\documentclass[
  reprint,
  prl,
  aps,
  superscriptaddress,
  longbibliography
]{revtex4-2}

\usepackage{graphicx}
\usepackage{amsmath,amssymb,amsfonts}
\usepackage{mathrsfs}
\usepackage{xcolor}
\usepackage{booktabs}
\usepackage{multirow}
\usepackage{algorithm}
\usepackage{algpseudocode}
\usepackage[english]{babel}

\begin{document}

\title{Microbiome association diversity reflects proximity to the edge of instability}

\author{Rub\'en Calvo}
\affiliation{Instituto Carlos I de F\'isica Te\'orica y Computacional, Universidad de Granada, Granada, Spain}
\affiliation{Departamento de Electromagnetismo y F\'isica de la Materia, Universidad de Granada, Granada, Spain}

\author{Adri\'an Roig}
\affiliation{Instituto Carlos I de F\'isica Te\'orica y Computacional, Universidad de Granada, Granada, Spain}
\affiliation{Departamento de Electromagnetismo y F\'isica de la Materia, Universidad de Granada, Granada, Spain}

\author{Roberto Corral L\'opez}
\affiliation{Instituto Carlos I de F\'isica Te\'orica y Computacional, Universidad de Granada, Granada, Spain}
\affiliation{Departamento de Electromagnetismo y F\'isica de la Materia, Universidad de Granada, Granada, Spain}

\author{Jos\'e Camacho-Mateu}
\affiliation{Institute of Functional Biology \& Genomics, CSIC - Universidad de Salamanca, Salamanca, Spain}

\author{Jos\'e A. Cuesta}
\affiliation{Universidad Carlos III de Madrid, Departamento de Matem\'aticas, Grupo Interdisciplinar de Sistemas Complejos (GISC), Legan\'{e}s, Spain}
\affiliation{Instituto de Biocomputaci\'on y F\'isica de Sistemas Complejos (BIFI), Universidad de Zaragoza, Zaragoza, Spain}

\author{Miguel A. Mu\~noz}
\email{mamunoz@onsager.ugr.es}
\affiliation{Instituto Carlos I de F\'isica Te\'orica y Computacional, Universidad de Granada, Granada, Spain}
\affiliation{Departamento de Electromagnetismo y F\'isica de la Materia, Universidad de Granada, Granada, Spain}

\date{\today}

\begin{abstract}
Recent advances in metagenomics have revealed macroecological patterns or "laws" describing robust statistical regularities across microbial communities. Stochastic logistic models (SLMs), which treat species as independent---akin to ideal gases in physics---and incorporate environmental noise, reproduce many single-species patterns but cannot account for the pairwise covariation observed in microbiome data. Here we introduce an interacting stochastic logistic model (ISLM) that minimally extends the SLM by sampling an ensemble of random interaction networks chosen to preserve these single-species laws. Using dynamical mean-field theory, we map the model's phase diagram ---stable, chaotic, and unbounded-growth regimes--- where the transition from stable fixed-point to chaos is controlled by network sparsity and interaction heterogeneity via a May-like instability line. Going beyond mean-field theory to account for finite communities, we derive an estimator of an effective stability parameter that quantifies distance to the edge of instability and can be inferred from the width of the distribution of pairwise covariances in empirical species-abundance data. Applying this framework to synthetic data, environmental microbiomes, and human gut cohorts indicates that these communities tend to operate near the edge of instability. Moreover, gut communities from healthy individuals cluster closer to this edge and exhibit broader, more heterogeneous associations, whereas dysbiosis-associated states shift toward more stable regimes---enabling discrimination across conditions such as Crohn's disease, inflammatory bowel syndrome, and colorectal cancer. Together, our results connect macroecological laws, interaction-network ensembles, and May's stability theory, suggesting that complex communities may benefit from operating near a dynamical phase transition.
\end{abstract}


\maketitle

\begin{figure*}[!t]
\centering
\includegraphics[width=1.0\textwidth]{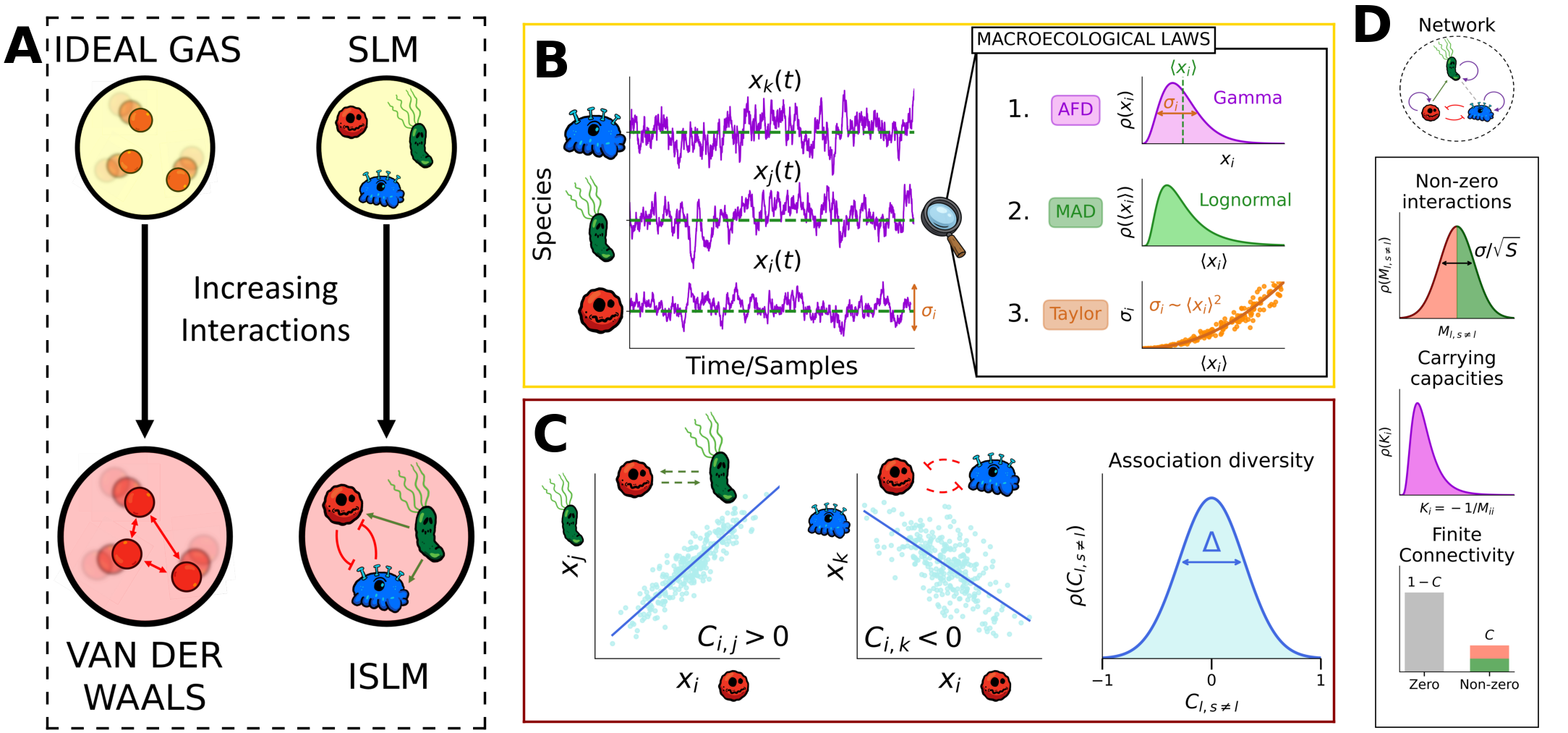}
\caption{\textbf{From ideal-gas-like descriptions to interacting ensembles.} Non-interacting models such as the SLM capture marginal statistical patterns of empirical data (Grilli's laws) but fail to reproduce interspecific associations, while Bayesian inference can fit both at the cost of losing interpretability. \textbf{A}. Analogy between physics and ecology: the ideal gas corresponds to the uncoupled SLM, whereas the Van der Waals gas plays the role of the interacting SLM (ISLM) that explicitly includes interactions among species. \textbf{B}. Three macroecological laws observed in empirical microbiomes: (i) for each species, temporal abundances follow a Gamma distribution (abundance-fluctuation distribution, AFD); (ii) across species, mean abundances are approximately lognormally distributed (mean-abundance distribution, MAD); and (iii) variance scales as the square of the mean (Taylor's law with exponent $\approx 2$). These patterns concern marginal statistics and ignore correlations between species’ relative abundances. \textbf{C}. Pairwise associations: two species can be positively correlated ($i$ and $j$) or negatively correlated ($i$ and $k$). Empirical microbiomes display a broad histogram of pairwise correlations, whose width we denote by $\Delta$, much wider than predicted by uncoupled models. \textbf{D}. Statistics of the synthetic ISLM. First row: schematic visualization of the interaction network. Second row: histogram of the non-zero entries of $G_{ij}$ (see Materials and Methods). Third row: distribution of carrying capacities, set as in \eqref{eq: ensemble 2}. Fourth row: link sparsity; a fraction $1-C$ of links is zero, while a fraction $C$ is non-zero.}\label{fig: 1}
\end{figure*}

From the seventeenth to nineteenth centuries, cumulative work by physicists and chemists culminated in the ideal-gas law, capturing the universal behavior of dilute gases under simple conditions \cite{Atkins}. In the mid-nineteenth century, Maxwell, Boltzmann, and others provided a microscopic basis via kinetic theory: neglecting intermolecular forces, they derived macroscopic observables from microscopic dynamics, establishing the ideal-gas law. Later experiments showed that gases and liquids form a single continuum separated by a phase transition, revealing the limits of the non-interacting ideal-gas description. Van der Waals then introduced minimal corrections---effective cohesive attractions between particles and an impenetrable particle core---yielding an equation of state that accounts for different phases, phase coexistence, and criticality \cite{Pathria,Kardar}. An analogous progression is now emerging in ecology. High-throughput metagenomics have generated a wealth of data that has revealed robust macroecological regularities in the composition and variability of microbiomes across taxonomic scales \cite{Shoemaker2017, Grilli2020, Descheemaeker2020, Ji2020, Wolff2023}. In particular, Grilli \cite{Grilli2020} identified three empirical ``laws'' that appear universal across different microbiomes (Fig.~\ref{fig: 1}B): (i) for each species, temporal abundance fluctuations follow a Gamma distribution (the abundance-fluctuation distribution, AFD); (ii) across species, mean abundances are log-normally distributed (the mean-abundance distribution, MAD); and (iii) abundance variance scales as a power of the mean (Taylor's law with exponent close to $2$). These laws probe single-species statistics and thus do not address interspecific couplings. Theoretically, they are successfully captured by a stochastic logistic model (SLM)---the ecological analogue of an ideal gas (Fig.~\ref{fig: 1}A)---that reproduces marginal abundance statistics in the absence of interactions.

Yet species abundances are not independent. A simple way to quantify interdependence, or ``association,'' is through pairwise correlations (Fig.~\ref{fig: 1}C): the relative abundance of a given species can be positively, negatively, or negligibly associated with that of another, reflecting shared external drivers, direct interactions, or both \cite{LIMITS,Martini2024,Carr2019,Pinto2022}. The width of the distribution of pairwise correlations (denoted $\Delta$ in Fig.~\ref{fig: 1}C) measures the heterogeneity of these associations, and in empirical microbiomes it is much broader than predicted for uncoupled species \cite{Grilli2020, Ho2022, Pinto2022}. By construction, the SLM does not account for the breadth of correlations, as it ignores ``ecological forces'' or interactions that shape species associations. Thus, much as Van der Waals introduced minimal interaction terms to extend the ideal-gas law, ecology calls for an extension of the uncoupled SLM that preserves its success at the single-species level while explaining the emergence of rich correlation patterns across communities \cite{Ho2022,Camacho,Sireci2023}.

From a broader perspective, recent work has sought to reconstruct the underlying networks of species interactions from time-series and cross-sectional microbiome data \cite{Faust2012, Stein2013, Venturelli2018, Coyte2015, LIMITS}. However, such inverse problems are notoriously ill-posed: high-dimensional ecological  models (such as the GLV) are over-parameterized, inference is strongly affected by noise and limited sampling, and even Bayesian approaches can yield broad posteriors compatible with quantitatively different interaction networks \cite{Masuda2025, Castro-INFERENCE, Pinto2022, zapien-campos_stochastic_2024}. These difficulties have motivated---following the strategy of statistical physics---a shift toward ``ensemble descriptions'' that focus on families of interaction networks able to reproduce empirically constrained macroecological patterns---such as diversity, abundance distributions, and correlation structures---rather than attempting to infer a unique, fully specified network of microscopic interactions and ascribe it a literal mechanistic meaning  \cite{Camacho} (see also \cite{Pasqualini2025}). However, while this ensemble framework  can successfully reproduce the richness of empirical associations \cite{Camacho}, it often does so at the cost of interpretability: posteriors may concentrate on networks that are hard to parse mechanistically, and inference may start from simple Gaussian-matrix perturbations of the SLM yet converge to highly non-Gaussian interaction structures, hampering further analyses and conceptual advances. 

In parallel with this network-inference program, a complementary line of theoretical work on complex microbiomes focuses directly on mapping the possible collective dynamical regimes that may emerge from large random-interaction networks \cite{Roy-chaos}. May’s pioneering analysis of the stability of large, complex ecological communities \cite{May1972} opened the door to a body of work showing that generic models with random, typically non-reciprocal interactions can exhibit both stable coexistence and regimes of persistent, intrinsically chaotic fluctuations, especially when spatial structure, migration, and resource dynamics are included \cite{Biroli2018, Altieri2021, Fisher2020, Fisher2023, Roy-chaos, Bunin-chaos-CR, Kessler, Mehta-PRL2024, Gore2025, Aguade24, Corral}. These regimes are separated by a sharp transition line---an ``edge of instability'' (or ``edge of chaos'') \cite{Bunin-chaos}. By analogy with other biological and artificial information-processing systems \cite{Munoz-RMP,MoraBialek2011,Chialvo,Langton1990,BerNat,Plenz-functional,Hidalgo}, one might hypothesize that real microbiomes are naturally poised near this edge. If so, operating close to the transition could have important consequences for stability, resilience, and responses to perturbations \cite{Amor,Zamponi,Pasqualini2025,Corral}, a hypothesis that deserves deeper, systematic scrutiny.

In this work, we connect these two perspectives: ensemble-based inference and dynamical-phase structure. We adopt the ensemble framework \cite{Camacho,Castro-INFERENCE,Pasqualini2025} and ask which generic classes of random interaction networks are both compatible with observed macroecological patterns and sufficiently constrained to retain mechanistic interpretability while keeping analytical tractability. We show that empirical microbiome data can be reproduced by a minimal interacting-SLM  built from Gaussian ensembles of random interactions. The only relevant parameter is the distance to instability, $g=\sigma \sqrt{C}$, which controls how broadly species covary (i.e., the spread of association strengths). Knowing $g$ ---and thus where a community sits in the phase diagram--- directly constrains the implied interaction architecture, the dynamical phase consistent with the data, and the community’s proximity to a critical transition. As we show in what follows, in human gut microbiomes, $g$ clearly discriminates healthy individuals ---whose communities appear closer to  instability, with wider associations--- from patients with dysbiosis-associated conditions. Our results suggest that a minimal ``Van der Waals-like'' correction to the ecological ideal gas not only reproduces observed correlation patterns, but also yields a biologically interpretable order parameter that places real microbiomes in parameter space and quantifies the deterioration of ecosystem function as they move away from a “healthy” state toward a “diseased” one.

\section*{Model building and analyses}\label{sec: 3}

We consider an interacting stochastic logistic model (ISLM), i.e. a generalized Lotka-Volterra (gLV) system with multiplicative environmental noise, representing species-dependent stochastic growth rates. The abundance of species $i=1,\dots,S$ evolves according to the (It\^o) Langevin equation
\begin{equation}
    \label{eq: ISLM}
    \dot{x}_i(t)
    = \frac{x_i(t)}{\tau_i}\left[ 1 + \sum_{j=1}^S A_{ij}\,x_j(t) \right]
      + x_i(t)\,\xi_i(t),
\end{equation}
where $\tau_i$ sets the characteristic time scale of growth/decay of species $i$, $A_{ij}$ is the interaction matrix, where diagonal entries are the single-species carrying capacities in the absence of interactions, as in the SLM (see below), and $\xi_i(t)$ is a stochastic term describing the effect of environmental fluctuations on species $i$. For most of our analysis, we assume temporally white and uncorrelated environmental noise,
\begin{equation}
    \langle \xi_i(t)\,\xi_j(t')\rangle = \sigma_0^2\,\delta(t-t')\,\delta_{ij},
\end{equation}
so that cross-species covariances are generated entirely by the interaction matrix $A$. Note that, more generally, one could consider a non-trivial correlation matrix $W$ encoding shared responses to latent environmental drivers or fields. A structured $W$ could by itself induce broad distributions of pairwise correlations, even in the absence of biotic interactions, and both $A$ and $W$ can in principle be derived from more mechanistic consumer-resource models ~\cite{Sireci2023}. Here we deliberately focus solely on biotic interactions and set $W_{ij}=\delta_{ij}$, as in previous work some of us found that heterogeneity in $A$ shapes the correlation distribution more strongly than noise correlations~\cite{Camacho}. Neglecting noise correlations is thus a modeling choice that yields a minimal proof-of-concept for interaction-driven associations.

We parameterize the interaction matrix as
\begin{equation}
    A = D + \sigma M,
    \label{matrix}
\end{equation}
where $D$ is a diagonal matrix corresponding to the uncoupled SLM part (self-regulation terms that fix the carrying capacities in the absence of interactions), $M$ is a random matrix whose statistics will be chosen to match empirical patterns, and $\sigma$ controls the magnitude of the perturbation away from the SLM. In this way, we interpolate between an ``ideal-gas'' limit of uncoupled species ($\sigma = 0$) and an interacting regime ($\sigma > 0$), analogous to a Van der Waals correction. 
Our goal is to understand how the statistics of $A$ shape community-level dynamics.

\paragraph{Balanced Gaussian interaction ensemble}
To define the synthetic ISLM ensemble, we proceed in three steps (illustrated in Fig.\ref{fig: 1}D). 

First, we consider the uncoupled SLM ($\sigma=0$) and, for simplicity,  assume identical time scales ($\tau_i=\tau$ for all $i$). The fixed-point solution of Eq.~(\ref{eq: ISLM}), denoted by $\mathbf{x}^\ast$, satisfies
\begin{equation}
\label{eq: FP condition}
A\mathbf{x}^\ast = D\mathbf{x}^\ast = -\mathbf{1},
\end{equation}
where $\mathbf{1}$ is the all-ones vector (see SI Section~1). Feasibility requires $x_i^\ast>0$ for all $i$.
 Thus, in the uncoupled limit,  $D_{ii}x_i^\ast=-1$, i.e. $x_i^\ast=-1/D_{ii}$.
 Following Grilli's work \cite{Grilli2020}, we impose that the fixed points (equivalently, the carrying capacities $K_i=-1/D_{ii}$) are positive and log-normally distributed to match empirical distribution of mean abundances (MAD):
\begin{equation}
    \label{eq: ensemble 2}
K_i = -\frac{1}{D_{ii}} \sim \mathrm{LogNormal}(\mu_{K},\sigma_{K}),
\end{equation}
where $\mu_{K}$, and $ \sigma_{K}$ are the mean and variance of the distribution. 

Secondly, we introduce interactions as a perturbation of the uncoupled SLM by writing the interaction matrix as in \eqref{matrix}. Rather than sampling $M$ directly, we first construct an intermediate random matrix $G$ with zero row sum
and then define $M$ from $G$ and the fixed point $x^\ast$. Concretely, we draw a sparse Gaussian matrix $G$ with connectivity $C$ by setting a random fraction $1-C$ of the off-diagonal
entries to zero and sampling the remaining ones as independent Gaussian variables with zero mean and variance
$\mathrm{Var}(G_{ij}) = 1/S$ (see Methods and SI Section 3.B). The diagonal of $G$ is set to zero, and for each
row $i$ we subtract the mean over the nonzero entries so that
\begin{equation}
    \sum_{j=1}^S G_{ij} = 0 \quad \text{for all } i,
    \label{1M}
\end{equation}
ensuring that the net interaction received by each species is centered. This row-balance condition corresponds to orthogonality with respect to the homogeneous vector, $G \mathbf{1} = \mathbf{0}$. We then define $M = - G D$. By construction, $M$ has zero diagonal and, because of \eqref{eq: FP condition}, satisfies 
\begin{equation}
    M \mathbf{x}^\ast = -GD\mathbf{x}^\ast = G \mathbf{1} = \mathbf{0},
     \label{2M}
\end{equation}
so the balance (row-orthogonality) condition holds and the fixed-point vector is left unchanged when interactions are
turned on. The full interaction matrix is then
\begin{equation}
    A = D + \sigma M = D - \sigma G D,
     \label{3M}
\end{equation}
which automatically ensures $A \mathbf{x^*} = -\mathbf{1}$ for any $\sigma$.

This procedure generates an ensemble of sparse, balanced, random interaction matrices with log-normally distributed
mean abundances and single-species statistics unchanged, while allowing us to tune the typical number of associations
per species via the connectance $C$ and the overall interaction heterogeneity via $\sigma$ to match empirical
correlation patterns.

In what follows, using techniques from the physics of disordered systems and random matrix theory, as well as computational analyses, we characterize the phase diagram of the ISLM and show that the connectivity, diversity, and strength of interactions play a crucial role in shaping the dynamics and stability of \textit{in silico} ecological communities. 

\section*{Results}\label{sec3}

\paragraph{Random-matrix stability analysis.}

Because the ISLM involves a large number of coupled degrees of freedom, stability is controlled by collective properties of the interaction matrix rather than by any single entry. Random matrix theory provides a natural framework to quantify these collective effects: in particular, it predicts the typical location and spread of the eigenvalue spectrum for large disordered networks, and thus yields analytic criteria for when fixed-point solutions lose stability as interaction strength and heterogeneity are increased.

\begin{figure*}[!t]
\centering
\includegraphics[width=1.0\textwidth]{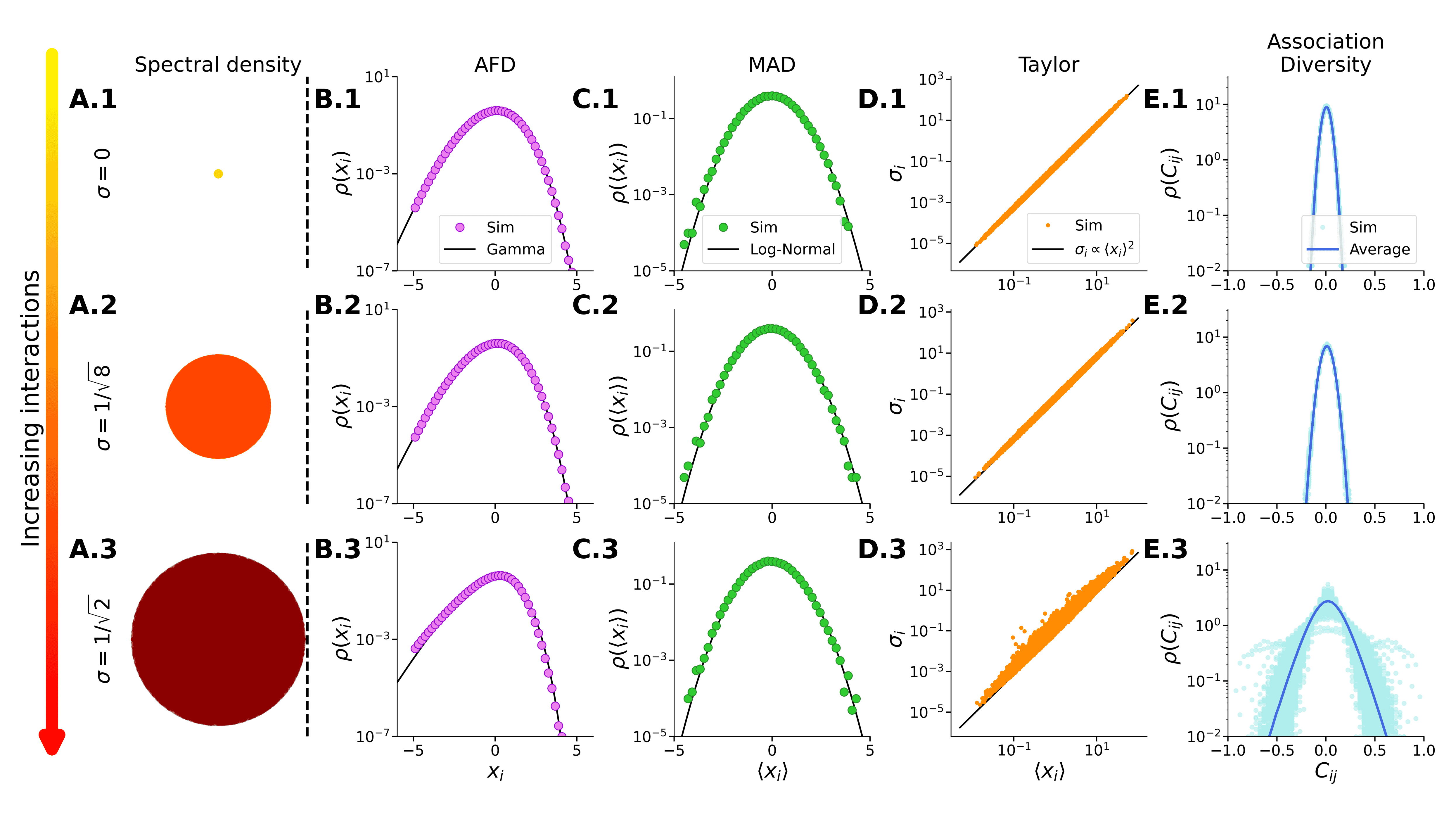}
\caption{\textbf{Grilli's laws are preserved in the interacting ISLM model, while approaching the edge of instability broadens the correlation histogram and enhances association diversity.}  
\textbf{B.1--B.3} Spectral density of the effective interaction matrix in \eqref{eq: Aeff} for three values of the interaction strength: B.1 $\sigma = 0$ (pure SLM), B.2 $\sigma = 1/\sqrt{8}$, and B.3 $\sigma = 1/\sqrt{2}$. The connectivity is fixed at $C = 1/2$, so panel B.3 lies exactly at the edge of instability. We use $\mu_{\mathrm{LN}} = 0$ and $\sigma_{\mathrm{LN}} = 1$ for the log-normal parameters.
\textbf{C.1--C.3} Abundance fluctuation distribution (AFD) for the same values of $\sigma$. Points show simulation results for a community with $S=200$ species, integrated up to $t_{\max} = 100$ with time step $h = 0.01$, noise variance $\sigma_0 = 1$, and growth time scale $\tau = 1$, averaged over $M = 500$ independent realizations. The black line is the best-fit Gamma distribution.
\textbf{D.1--D.3} Mean abundance distribution (MAD). Green dots are simulation results; the black curve is the log-normal distribution from which the values of $-1/D_{ii}$ are drawn.
\textbf{E.1--E.3} Taylor's law. Points show the empirical relationship between mean and standard deviation from simulations; the black line is the theoretical scaling $\sigma_i \propto \langle x_i\rangle^2$ (plotted as $y = x^2$).
\textbf{H.1--H.3} Association diversity for the three values of $\sigma$. For small and intermediate interaction strengths (H.1--H.2) the correlation histograms are narrow, while at the edge of instability (H.3) they become much broader. Light points correspond to different realizations of the quenched disorder, and the dark line shows their average, highlighting the multitude of possible correlation patterns compatible with the same single-species macroecological laws.}\label{fig: 3}
\end{figure*}

More specifically, dividing \eqref{eq: ISLM} by $x_i(t)$ and linearizing around $\mathbf{x}^\ast$ yields a multivariate Ornstein--Uhlenbeck process with an effective matrix $A_{\mathrm{eff}}$ given by (see Methods and SI--Section 1.B for details)
\begin{equation}
\label{eq: Aeff}
\tau A_{\mathrm{eff}} = -AD^{-1} =-I-\sigma MD^{-1} = -I+\sigma G.
\end{equation}
Importantly, as noted in the previous section, the balance condition implies $G\mathbf{1}=\mathbf{0}$, so $\mathbf{1}$ is an eigenvector with eigenvalue $0$. Since this is a rank-one constraint, it pins just one eigenvalue and does not affect the bulk spectrum as $S\to\infty$. In that limit, the eigenvalues of $G$ follow the circular law \cite{Tao,Allesina,Rajan+Abbott}, forming (to leading order) a disk of radius $\sqrt{C}$ in the complex plane, so the spectrum of $\tau A_{\mathrm{eff}}=-I+\sigma G$ is the same disk rescaled by $\sigma$ and shifted to be centered at $-1$. Linear stability is lost when the disk first reaches the origin, i.e. when $g \equiv \sigma\sqrt{C}=1$, which is the classic May criterion for random ecosystems (see SI--Section 1.D). As a particular example, Fig. \ref{fig: 3} panels A.1-A.3 show the spectral density of matrix $G$, confined to a disk whose radius increases with $\sigma$. In this particular case, since $C=1/2$, the instability threshold occurs at $\sigma=1/\sqrt{2}$.

\paragraph{Dynamical mean-field theory approach}
In order to go beyond linear stability analysis, we carried out a dynamical mean-field theory (DMFT) analysis of the ISLM. As previously discussed, this extends the standard generalized Lotka-Volterra framework by incorporating three key ingredients: a balanced, random ensemble of interactions, log-normally distributed diagonal elements, and multiplicative environmental noise. 
In the thermodynamic limit $S\to\infty$, DMFT becomes exact, reducing the full $S$-dimensional stochastic dynamics to an effective one-dimensional process for a single ``representative'' species with abundance $x(t)$ \cite{SCS,Bunin-chaos}. As derived in the SI, the effective equation is (SI, Eq.~(131))
\begin{equation}
    \dot x(t) = x(t)\big[1 - x(t) + \eta(t) + \xi(t)\big],
\label{MF-eq}
\end{equation}
where $\xi(t)$ is the original multiplicative environmental noise and $\eta(t)$ is an emergent effective Gaussian noise generated by the random interactions with the rest of the network. The statistics of $\eta(t)$ are fixed by the DMFT self-consistency condition, which relates its covariance to the autocorrelation of the effective trajectory (see SI--Section 3)
\begin{equation}
    \langle \eta(t)\eta(t') \rangle =
\sigma^2\Big(\langle x(t)x(t')\rangle - \langle x(t)\rangle \langle x(t')\rangle\Big),
\label{noise}
\end{equation}
 so that interaction-induced fluctuations feedback onto the dynamics of the focal species.

As in generalized Lotka-Volterra models \cite{Bunin-chaos,Roy-chaos,DeMonte-chaos,Altieri2021,Biroli2018}, the effective process displays distinct dynamical phases as parameters vary: a stable fixed-point regime with finite stationary abundances, a chaotic regime with persistent irregular fluctuations that can drive extinctions, and an unstable regime with diverging populations (see SI--Section 3).

Because in our case environmental fluctuations enter multiplicatively, the term ``fixed point'' must be understood in a statistical sense: in this phase the noise-averaged abundance (and the full stationary distribution) is time-independent, but individual trajectories still wander stochastically according to the stochastic logistic model (SLM). By contrast, in the chaotic phase even the noise-averaged abundance exhibits irregular temporal dynamics, and chaos must be characterized at the level of these underlying mean trajectories, on top of which environmental noise adds an additional layer of randomness (see SI--Section 5).

Quite remarkably—as we show analytically in SI Section 3.D/E—the orthogonality constraint forces a self-consistent cancellation of interaction-generated noise across the entire stable fixed-point phase. This implies that, for the balanced random ensemble, the noise term vanishes in the large-community limit, $\eta(t)\equiv 0$ as $S\to\infty$. The effective dynamics therefore reduce \emph{exactly} to a stochastic logistic equation driven only by multiplicative environmental noise, recovering the single-species SLM previously used to explain universal ``macroecological laws'' \cite{Grilli2020}. In this way, we provide a microscopic, analytically controlled derivation of Grilli's one-species description from a high-dimensional interacting community with log-normal carrying capacities and balanced interactions—one of the central results of this work.

\begin{figure}[!t]
\centering
\includegraphics[width=0.5\textwidth]{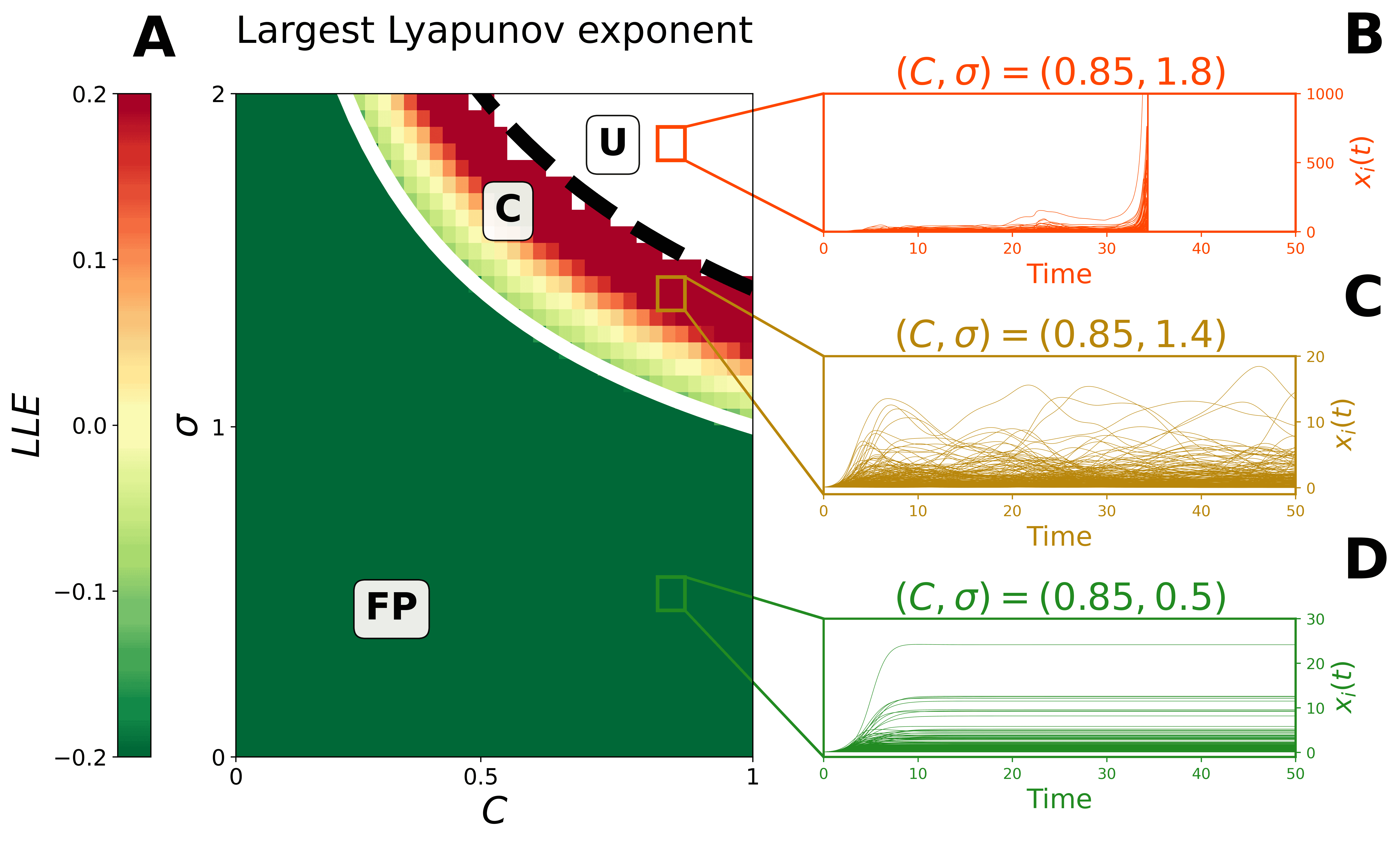}
\caption{\textbf{The synthetic model exhibits fixed-point, chaotic, and unbounded dynamical phases.}
\textbf{A}. Heat map of the largest Lyapunov exponent (LLE) as a function of $\sigma$ and $C$ (negative values are plotted as zero for clarity). Three regimes can be distinguished: a fixed-point phase (FP), a chaotic phase (C), and an unbounded phase (U) in which trajectories blow up. The white solid line and the black dashed line delimit the FP--C and C--U boundaries, respectively. Simulations were performed for communities with $S = 10^3$ species (to suppress finite-size effects), without environmental noise ($\sigma_0 = 0$, the effect of noise is further explored in the SI--Section 5.B), and averaged over $M = 50$ independent realizations of the quenched interactions.
\textbf{B--D}. Representative time series illustrating the dynamical behavior in each regime shown in panel A.}
\label{fig: 2}
\end{figure}

\paragraph{Computational validation of dynamical phases and chaos}
Before proceeding, we test the dynamical phases predicted by DMFT by simulating finite communities with interactions drawn from our synthetic ensemble and estimating the largest Lyapunov exponent (LLE), whose positivity is the standard hallmark of chaos. 
We simulate a community with $S=1000$ species up to $t_{\max}=10^2$ (time step $h=0.01$), for varying interaction widths and connectivities. The phase diagram in Fig.~\ref{fig: 2}A reveals the same three regimes. For small interaction heterogeneity $\sigma$, the LLE is negative across the $(C,\sigma)$ plane and trajectories relax to a unique fixed point (region FP), consistent with the stable DMFT solution. Beyond the fixed-point solution limit of instability $\sigma = 1/\sqrt{C}$ (white curve), a thin band with positive LLE appears (region C), marking the onset of chaotic dynamics. Owing to finite size and finite observation time, this chaotic band is slightly shifted inward and narrower than the DMFT prediction; in the large-$S$ limit we expect the boundaries to coincide (see Figure S3 included in the SI). As $\sigma$ increases further (keeping $C$ fixed), trajectories cross a second boundary (dashed line) beyond which they diverge and leave the biologically realistic domain, corresponding to the unbounded phase $U$ in DMFT. The time series in Fig.~\ref{fig: 2}B--D illustrate these behaviors at fixed connectivity $C=0.85$: below the stability line, all species relax to stationary abundances (FP, panel D); in the intermediate band, abundances show bounded irregular fluctuations with positive LLE (chaotic phase C, panel C); and for larger $\sigma$, trajectories blow up in finite time (unbounded phase U, panel B). Together, these simulations provide a direct dynamical confirmation of the fixed-point, chaotic, and unbounded regimes predicted by DMFT.

\paragraph{Computational verification of macroecological laws in the synthetic ISLM model}

Having characterized the dynamical phases of the ISLM in the infinite-size (mean-field) limit, we next use simulations to test whether the macroecological laws of \cite{Grilli2020} are preserved when interactions are turned on in finite communities.

Because of the constraint in \eqref{eq: FP condition}, the perturbation $M$ leaves the fixed points unchanged, so the MAD is satisfied by construction (see Figure \ref{fig: 3} panels C.1-C.3). The remaining macroecological laws are likewise robust, as shown in the other panels of Fig.~\ref{fig: 3}. We simulate a community with $S=100$ species up to $t_{\max}=10^2$ (time step $h=0.01$), collecting $M=10^3$ independent measurements, for three interaction strengths $\sigma$ ($\sigma=0$ in the first row, $\sigma=1/\sqrt{8}$ in the second, and $\sigma=1/\sqrt{2}$ in the third) at fixed connectivity $C=0.5$.  Remarkably, approaching this edge leaves the AFD and MAD essentially unchanged and preserves Taylor's law on average---the points scatter around the ideal straight line, as in real microbiome data (see e.g. \cite{Pasqualini2025, Camacho-PRE2024} and Panels B, C, and D)---while association diversity is strongly enhanced (Panels E.1--E.3). Near the instability, a wide range of correlation-histogram shapes emerges across network realizations (E.3), indicating many distinct coupling patterns that reproduce the same single-species laws yet differ markedly in interspecific structure.

\paragraph{Beyond mean-field theory: from model to data}

As already discussed, DMFT is exact in the infinite-community limit $S\to\infty$, where it reduces the full interacting dynamics to an effective single-species process. In this limit, pairwise covariances---and thus inter-species correlations---vanish by self-averaging over many interaction partners. Empirical microbiome data from diverse environments, however, exhibit nontrivial pairwise associations, so reproducing their distribution requires going beyond mean field. For large but finite communities, we capture these residual covariances by including the leading $1/S$ finite-size corrections around the DMFT solution; we develop this beyond-DMFT framework below (see Materials and Methods and SI Section 4). \footnote{Related beyond-mean-field approaches have been developed for neural networks \cite{dahmen_second_2019, clark_dimension_2023, clark_connectivity_2025, dick_linking_2024}, where they show that the width of the distribution of (long-time window) correlation pairs diverges as the DMFT fixed-point solution approaches the edge of a transition.}

This divergence can be exploited to construct a practical estimator of an effective stability parameter $g$, directly linking it to simple observables such as the empirical covariance distribution and the community size $S$. Here we extend this framework to the ISLM ensemble and, because microbiome time series are typically short, adapt it to use short-time covariances \cite{Robust} rather than asymptotic long-time ones (see SI Section 4.L for details).

This yields an explicit expression relating the normalized width of covariances, $\Delta$, to the stability parameter $g$ (with the edge of instability for the fixed-point solution at $g=1$) thereby allowing us to place actual microbiome data in the phase diagram. The implicit expression, derived in detail in the SI, reads:
\begin{equation}
\label{eq: g estimator}
    S \Delta^2
    = \frac{1}{8}\left(
        -6 - \frac{2}{\lambda}
        + \sqrt{\lambda}
        + \frac{4}{\sqrt{\lambda}}
        + \frac{3}{\lambda^{3/2}}
    \right), 
\end{equation}
where $S$ is the number of species, $\lambda=1-g^2$ and $\Delta$ is the normalized width of covariances:
\begin{equation}
    \Delta^2 = \frac{1}{S}\frac{\sum_{i,j\neq i}^S C_{ij}^2}{\left(\sum_i^S C_{ii}\right)^2},\\
    \label{Delta}
\end{equation}
where $C_{ij}$ denotes the short-time covariance matrix estimated from empirical data.

\paragraph{Fitting empirical biome association patterns at the edge of instability}

We now use our estimator \eqref{eq: g estimator} to infer an effective stability parameter $g$ from empirical correlation patterns and the community size. In the model, the edge of instability corresponds to $g=\sigma\sqrt{C}=1$. An inferred value of $g$ therefore does not pin down a unique point in the $(C,\sigma)$ plane, but rather a one-dimensional manifold defined by $g=\sigma\sqrt{C}$. This degeneracy reflects that, aside from finite-size effects, the correlation width is controlled only by the total interaction disorder, which depends on $\sigma$ and $C$ through their combination $\sigma\sqrt{C}$. Consequently, one parameter (e.g., $C$) can be chosen freely; given $C$ and the inferred $g$, one then fixes $\sigma=g/\sqrt{C}$ and samples many realizations of the ISLM interaction network at these values.

For each realization, we set $S$ equal to the empirical number of species, draw an interaction network from the balanced ensemble, simulate the dynamics, and compute the resulting covariance (association) histogram. We then retain those networks whose association statistics best match the empirical distribution. This two-step procedure explicitly acknowledges that associations are non-universal and realization-dependent in finite networks, and uses them to select representative networks within an ensemble that already captures the robust one-point statistics.

\begin{figure*}[!t]
\centering
\includegraphics[width=1.0\textwidth]{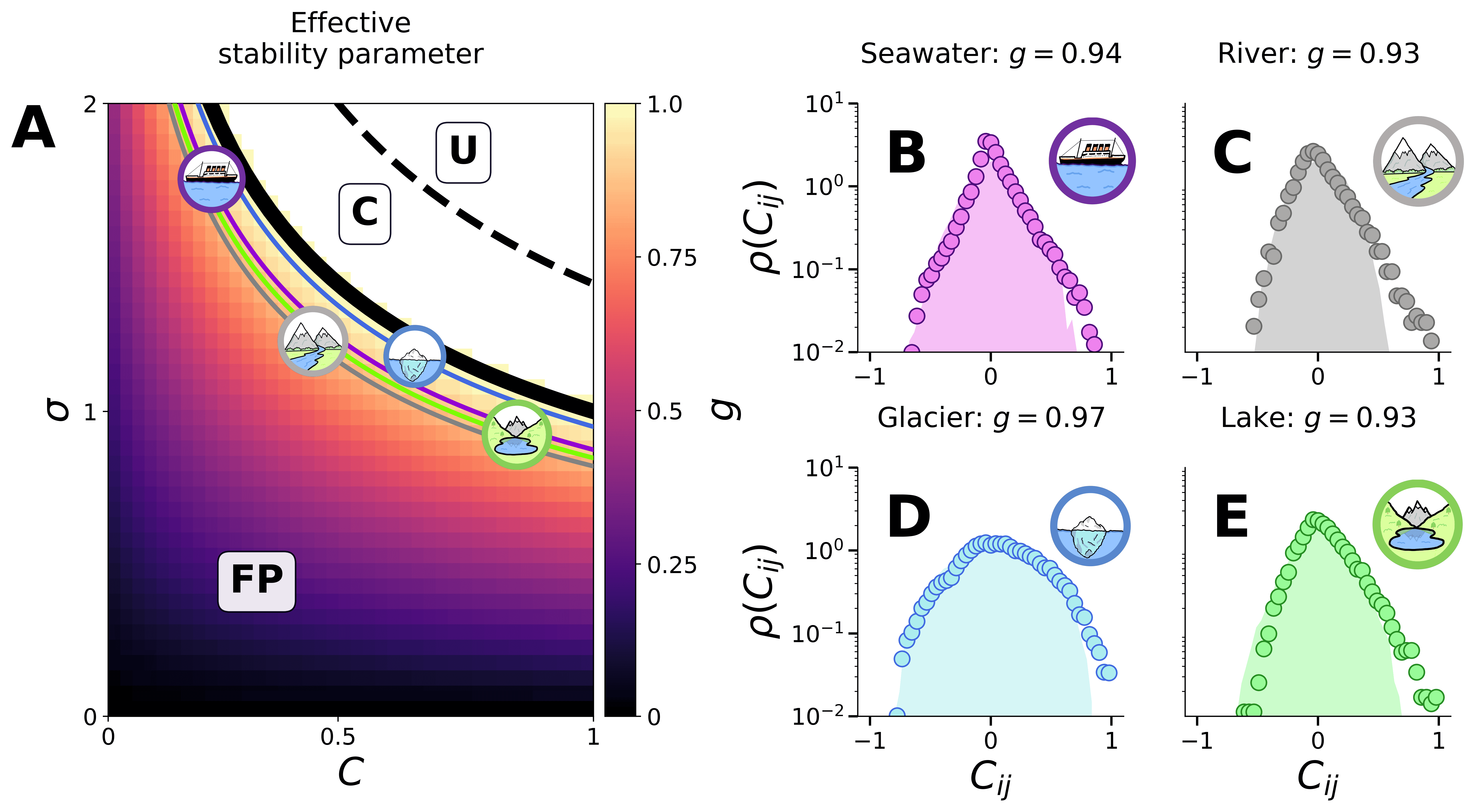}
\caption{\textbf{Empirical biomes are found more likely at the edge of instability}. \textbf{A}. Distance to the edge of instability, measured via the estimator of \eqref{eq: g estimator}. The color map shows the value of $g$. The black line shows the boundary of instability ($\sigma\sqrt{C}=1$). Below this boundary, the fixed point solution is stable and, above it, it becomes linearly unstable. Four different lines are shown in violet, grey, blue and green, indicating each of the four microbiomes analyzed: Seawater (Panel B), River (Panel C), Glacier (Panel D) and Lake (Panel E), respectively. Lines depict the possible combinations of $(\sigma, C)$ giving rise to the reported values of $g$, measured for each biome. \textbf{B-E} Histograms of correlations for the four microbiomes, each with their own inferred value of $g$. Points show empirical values; shades show the best fit found by simulating the dynamics for $S=200$ and $M=10^3$ different realizations of the quenched disorder at the value of $\sigma$ needed to reproduce the empirical $g$ (with fixed $C=0.5$).}\label{fig: 4}
\end{figure*}

Applying this inference scheme first to synthetic data generated with the ISLM model, we verified that the estimator behaves as expected: the inferred stability parameter $ g(C,\sigma)$ closely follows the theoretical prediction and increases towards $1$ as $(C,\sigma)$ approaches the DMFT stability boundary $\sigma = 1/\sqrt{C}$ (Fig.~\ref{fig: 4}A, colormap and black line). We then applied the same procedure to empirical time series from seawater, river, glacier, and lake microbiomes (see Methods). Their distance to the edge of instability was quantified and found to lie in the range $g \in [0.93,0.97]$ (icons and labels in Fig.~\ref{fig: 4}B--E). Thus, all analyzed communities appear to operate very close to the edge of instability, with systematic but modest biome-to-biome variation. Part of this variation is captured by differences in the inferred stability parameter $g$: consistent with the beyond-mean-field theory, biomes with larger $g$ (such as glacier microbiomes) display broader empirical correlation histograms (Fig.~\ref{fig: 4}D), reflecting increased association heterogeneity, whereas those with slightly smaller $g$ have narrower distributions (Fig.~\ref{fig: 4}B,C,E). In addition, the theory predicts that, even at fixed $g$, the detailed shape of the correlation histogram depends on the specific realization of the interaction and noise networks. Consequently, variability across biomes reflects both genuine differences in $g$ and sample-to-sample fluctuations arising from network-realization. 

\paragraph{Dysbiosis is linked to a retreat from the edge of instability}

Our analysis of environmental communities suggests that microbiomes often operate near the edge of instability, a regime frequently associated with enhanced responsiveness, adaptability, and resilience in complex systems \cite{Munoz-RMP,MoraBialek2011}. The human gut microbiome is a particularly relevant case because it plays a central role in host health \cite{deVos22,Hou,Blaser-review}. Dysbiosis---a persistent departure from a healthy microbial state, often accompanied by loss of key taxa and functions---has been linked to a wide range of diseases \cite{Blaser-review, Huang, buchanan2018understanding, Jovel, Lozupone2012, Seppi, Pasqualini2024, Corral}.

Despite its clinical relevance, the ecological mechanisms underlying the transition from healthy to dysbiotic states remain unclear, limiting mechanistically grounded interventions \cite{Schmidt,deVos12}. While recent work has argued that dysbiosis reflects a shift in the global balance of interactions toward more positive coupling \cite{Corral}, such claims typically rely on reconstructing full interaction networks from short and noisy time series---a difficult inverse problem with well-known caveats \cite{Faust2012,Faust2021}.

Here we take a complementary approach and test whether healthy and dysbiotic microbiomes differ in their proximity to the ISLM edge of instability. Importantly, this distance can be inferred directly from the breadth of short-time pairwise correlations, without reconstructing microscopic interaction networks. We therefore apply the same pipeline used above for environmental microbiomes to estimate the stability parameter 
g from human gut metagenomic datasets, comparing healthy individuals with multiple dysbiosis-associated conditions, such as inflammatory bowel disease (IBD), irritable bowel syndrome (IBS), colorectal cancer (CRC), and Clostridioides difficile infection (CDI).

\begin{figure*}[!t]
\centering
\includegraphics[width=1.0\textwidth]{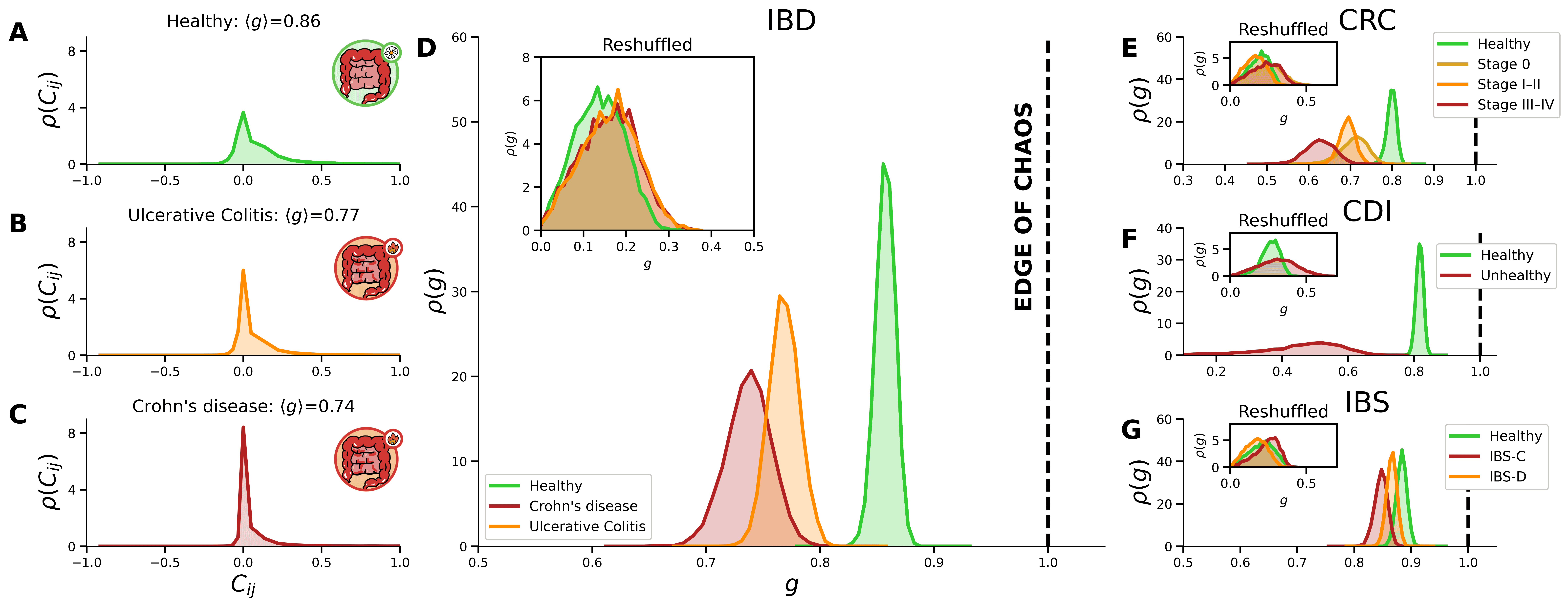}
\caption{\textbf{Dysbiosis-associated conditions are farther from the critical point than healthy cohorts}. \textbf{(A--C)} Distributions of pairwise species--species correlations
  $\rho(C_{ij})$ for Healthy controls (A), Ulcerative Colitis (B), and Crohn's
  disease (C) gut microbiomes. For each cohort we generated $M=1000$ independent subsamples; each
  subsample contains $T_2=200$ randomly selected samples (without replacement within a subsample). An occupancy filter of $0.2$ was applied so that only taxa present
  in at least $20\%$ of the selected samples were retained, yielding $\sim$100 taxa
  per dataset. \textbf{(D)} Distributions of the estimated distance to
  criticality ${g}$ for the four groups (two healthy cohorts (group 1 and group 2); two
  disease cohorts (UC, CD). Brackets report the AUC p-value between (i) the two disease cohorts (U--U), (ii) the two healthy cohorts (H--H), and
  (iii) pooled Unhealthy versus pooled Healthy (U--H). Small $p$-values indicate
  evidence for distributional differences; the U--H contrast shows clear separation (at least for the available data), whereas H--H is not significant and U--U shows a modest shift.
  As an interpretable effect size, the \emph{AUC} equals the probability that a randomly chosen value from one group exceeds a value from the
  other; in this setting it quantifies separability of both distributions.
  The inset (\emph{Reshuffled}) shows $g$ after random reshuffling that
  destroys the association between taxa and samples; the distinctive cohort
  structure disappears, confirming that the observed separations arise from
  genuine ecological organization rather than sampling artifacts.}
\label{fig: 5}
\end{figure*}

The correlation histograms (Fig.~\ref{fig: 5}A--C) already suggest a systematic difference between
groups: healthy subjects display slightly broader, more heterogeneous associations and a larger mean
stability estimate, $\langle g\rangle \approx 0.86$, whereas ulcerative colitis and Crohn's disease microbiomes show progressively narrower histograms and smaller values, $\langle g\rangle \approx 0.77$ and $\langle g\rangle \approx 0.74$, respectively. Pooling subjects within each group (Fig.~\ref{fig: 5}D) yields significantly separated $g$ distributions: the two independent healthy cohorts overlap and lie markedly closer to the edge of instability ($g=1$) than the inflammatory bowel disease (IBD) cohorts (ulcerative colitis and Crohn’s disease). To quantify this separation, we treat the individual $g$ estimates as samples and compute the area under the receiver operating characteristic (ROC) curve between pairs of diagnoses (see SI Section~6.B/C for methodological details and SI Section~6.E for the resulting AUC values).

As a null check, reshuffling bacterial abundances across samples (randomly reassigning each
species’ abundance profile to different individuals) destroys the correlation structure and
collapses these differences in $g$ (inset to Fig.~\ref{fig: 5}D), confirming that the signal in $g$
originates from genuine dynamical correlations rather than from potential confounders such as
sampling noise. Across additional disease cohorts analyzed with the same pipeline (IBS, CRC, CDI;
Fig.~\ref{fig: 5}E--G), the same qualitative pattern holds: healthy microbiomes systematically sit
closer to the critical regime of the ISLM ensemble, whereas dysbiosis is associated with a robust
shift toward more strongly stable states with reduced dynamical diversity and reactivity to
perturbations. The sampling scheme used to construct the covariance matrices and $g$
distributions in Fig.~\ref{fig: 5} (subsampling parameters and cohort-specific
filters) is described in SI Section~6.B/D, and its robustness to changes in these choices is examined
in SI Section~6.F. Finally, as a negative control, we analyzed healthy subjects from two independent
studies (SI Section~6.G): over a broad range of sampling parameters, these two groups are not
separable by our estimator, indicating that the procedure does not generate spurious differences
between comparable healthy cohorts.

\section*{Discussion}\label{sec12}

Microbial macroecology poses an apparent puzzle. On the one hand, a simple effective stochastic logistic model (SLM) without inter-species interactions already reproduces the single-species macroecological laws of Grilli \emph{et al.}. On the other hand, experiments and clinical evidence show that interactions in microbiomes are central to stability, function, and disease: for example, they have been implicated in gut disorders such as Crohn’s disease and inflammatory bowel conditions \cite{Newell2014,Khanna2017,Frank2007}, and several current treatments explicitly harness competitive interactions among bacteria \cite{DeDios2017,Palmer2022}.

Reconciling these two facts requires models in which interactions can be strong and structurally diverse without destroying the robust one-point statistics captured by the SLM. In this work, we address this tension by constructing a minimal, analytically tractable interacting stochastic logistic model (ISLM) that preserves the SLM backbone at the level of single-species statistics while embedding species in a high-dimensional interaction ensemble, where interaction networks are encoded by a balanced Gaussian matrix built on log-normally distributed self-interaction terms and subject to environmental noise.

Dynamic mean-field theory (DMFT) shows that this ensemble exhibits a phase diagram with a stable fixed-point phase, a chaotic phase, and an unbounded phase, separated by an instability line that obeys a May-like linear-stability criterion. Remarkably, throughout the stable phase the DMFT effective process reduces exactly to the one-species stochastic logistic equation, so the three macroecological laws (AFD, MAD, Taylor's law) remain intact even as interactions strongly reshape community-level dynamics. This provides a natural explanation for why a non-interacting model such as the stochastic logistic model can successfully capture macroecological patterns of empirical (interacting) microbiomes. In this sense, the ISLM acts as a ``Van der Waals correction’’ to the ecological ideal gas: it preserves the successful mean-field description of marginal abundances while enriching the dynamical and associational structure.

Finite-community-size (beyond-mean-field) corrections around the DMFT solution generate nonzero species-species covariances, whose distribution broadens as the instability line is approached. Extending previous theoretical frameworks \cite{dahmen_second_2019, clark_dimension_2023, clark_connectivity_2025, dick_linking_2024} to our ISLM ensemble and to short-time covariances, we derive an estimator for a single stability parameter $g$ that can be inferred directly from empirical covariance distributions and community size. In our parametrization, $g=1$ marks the edge of instability, so $g$ acts as an effective order parameter: it determines the width of the correlation histogram and places communities in the $(C,\sigma)$ phase diagram without reconstructing the full microscopic interaction network.

Applying this framework to synthetic data confirmed that the estimator behaves as expected and that, near the edge, the model generates a wide variety of correlation patterns across realizations while preserving single-species statistics. We then used the same pipeline on transversal and longitudinal microbiome data from several natural environments, including aquatic, glacial, and others. All these biomes are found to operate close to the instability line ($ g \approx 0.93$--$0.97$), 
with systematic but modest differences between environments. Biomes with larger $g$ display broader empirical correlation histograms, consistent with the theoretical prediction that association diversity increases as communities approach the edge of instability. 
Relatedly, our conclusion that microbiomes operate near the edge of instability echoes recent work on ecological “collectivity,” quantified by the spectral radius of the interaction matrix normalized by self-regulation, which increases as indirect effects propagate more broadly through the community \cite{Zelnik24,Aguade25}. In this sense, their collectivity metric and our distance to the May-like instability line capture the same underlying mechanism: as the dominant spectral scale approaches its critical threshold, community-wide fluctuations and heterogeneous covariation are amplified.

Focusing next on the human gut, we asked whether health and disease differ  in their distance to instability \cite{Relman2012,Zheng2020,Sorbara2022}. Across multiple cohorts, healthy microbiomes consistently show broader, more heterogeneous association patterns and larger values of $ g$ than dysbiosis-associated states. In particular, inflammatory bowel disease cohorts (ulcerative colitis and Crohn’s disease) exhibit narrower correlation histograms and lower $ g$ than healthy controls \cite{Frank2007,Khanna2017,Lee2021}, indicating a retreat from the edge of instability. The same qualitative shift is observed in additional conditions commonly linked to dysbiosis (IBS, CRC, CDI; see Fig. \ref{fig: 5}), and it is robust to changes in filtering and inference parameters (see SI Section 6.F). Null reshuffling tests, which destroy data structure while preserving marginal abundances, largely collapse these differences, confirming that they arise from genuine dynamical organization rather than from static sampling artifacts. 

Thus, our results suggest that $g$ may offer a simple and robust dynamical marker of microbiome health. Distinguishing healthy from dysbiotic communities remains a long-standing challenge \cite{VanHul2024, Rinninella}, with many proposed signals---most notably reduced diversity---proving non-universal across diseases \cite{Magne, Gupta, Halfvarson, Ma}. In contrast, $g$ consistently separates healthy from diseased individuals while showing no significant differences between independent healthy cohorts, indicating that it captures a reproducible and disease-relevant signal of the underlying ecological dynamics.  Moreover, its computation relies solely on abundance covariances, making it fast, data-efficient, and independent of any explicit inference of the interaction network.

Conceptually, our work connects several research lines. It embeds ensemble-based approaches to microbiome inference within the dynamical phase structure of high-dimensional ecological models, extending May’s stability criterion to a setting that preserves realistic single-species macroecology. It shows that a simple, random-interaction extension of the SLM is sufficient to reconcile universal one-point statistics with the observed breadth and variability of pairwise associations. And it translates these ideas into a practical inference tool: the scalar parameter $g$, estimated from short time series, provides a coarse but interpretable summary of community dynamics that can stratify health states and potentially track progression.
In doing so, it brings the analysis of microbiomes closer to the classical framework of community ecology based on generalized Lotka-Volterra models, and highlights microbial communities as an exceptional testbed for stability-diversity theory and other ecological laws that are difficult to probe in macroscopic ecosystems.

At the same time, our framework is merely a first step, with important limitations that future work will need to address. First, it rests on strong but transparent modeling assumptions: interactions are treated as pairwise and random with Gaussian statistics and constrained only by balance and sparsity; environmental noise is taken as unstructured and multiplicative; and spatial structure, migration, and explicit resource dynamics are neglected.  Real microbiomes undoubtedly display more structured interactions and stronger phylogenetic and metabolic constraints, and---in the human gut, for example---are further shaped by host-mediated feedbacks \cite{Zheng2020}. Moreover, correlations between species can arise not only from direct interactions but also from shared environmental responses or phylogenetic similarity, which we do not model explicitly here \cite{Sireci2023}. Our demonstration that pairwise direct interactions drawn from a simple ensemble can reproduce broad association patterns should therefore be viewed as a proof of concept rather than a claim that this is the sole mechanism behind observed correlations.

Second, we have restricted in our parsimonious modeling approach to pairwise interactions of Lotka-Volterra type. There is growing evidence that higher-order interactions, which naturally emerge in coarse-grained consumer-resource models, may play an important role in community stability and in the relationship between diversity and robustness ~\cite{Grilli2017,Sireci2023,Stouffer}. 
Let us underline, however, that notably, when tuned close to the edge of instability, this pairwise model also generates non-trivial higher-order correlations \cite{Pasqualini2025,Rosas-Mediano}, which could be directly compared with empirically measured multi-species statistics as these become available.

Also, demographic noise and immigration are omitted. While demographic noise appears subdominant to environmental fluctuations for the macroecological laws we consider, and immigration mainly affects dynamics far from steady state, both ingredients might be relevant in open, spatially extended ecosystems and could modify the effective position of communities in the phase diagram.

These caveats point to natural extensions. On the theoretical side, it will be important to incorporate additional structure into the interaction ensemble---such as consumer-resource constraints, trophic or metabolic layers, or phylogenetic dependence---and to explore the impact of higher-order interactions, space, and migration. On the empirical side, controlled perturbation experiments and longer longitudinal datasets could test more directly whether communities with larger $g$ indeed display stronger and more diverse responses to perturbations, and whether therapeutic interventions that restore health (e.g., diet, probiotics, fecal microbiota transplantation) drive $g$ back toward the critical regime identified here.

In summary, our study suggests that association diversity in microbiomes is not an incidental byproduct of noise and sampling, but a quantitative reflection of where communities sit in a dynamical phase diagram. By linking broad, realization-dependent correlation patterns to a simple stability parameter and showing that this parameter stratifies microbiome health states, we provide a bridge between macroecological laws, interaction ensembles, and the broader idea that some aspects of living systems may have evolved to exploit functional advantages by operating near the edge of a phase transition---without necessarily being tuned to an exact critical point.

\section*{Materials and Methods}

\subsection*{Numerical integration of stochastic dynamics and Lyapunov exponents}

We simulate the interacting stochastic logistic model (ISLM) by numerically integrating the
stochastic differential equations in Eq.~(1) using a Milstein scheme \cite{Toral2014}. 
Time is discretized in steps of size $h$ (typically $h = 10^{-2}$, with smaller values checked for convergence), and at each step both the deterministic drift and the multiplicative noise term are updated using the standard Milstein scheme for scalar stochastic differential equations with linear multiplicative noise, which ensures positivity of the solutions. Unless otherwise stated, we integrate up to a maximum time $t_{\max}$, discard an initial transient, and sample species
abundances at regular intervals to estimate macroecological observables (abundance-fluctuation
distributions, mean-abundance distributions, Taylor's law) and covariance matrices. 

To characterize dynamical regimes and locate the onset of chaos, we computed the largest Lyapunov
exponent (LLE) using the classical Benettin algorithm \cite{benettin_lyapunov_1980}.
For a given realization of the quenched disorder and noise, we evolve in parallel a reference
trajectory and a perturbed trajectory, initialized with an infinitesimal separation $\delta x(0)$
and driven by the \emph{same} realization of the stochastic noise. At fixed time intervals
$\Delta t$, we measure the norm of the separation, renormalize it to a fixed small value, and
accumulate the finite-time growth rates. The LLE is then estimated as the long-time average of
these growth rates. A positive LLE indicates chaotic dynamics, while a negative LLE corresponds
to a stable fixed-point regime; the transition between these regimes defines the edge of instability in
the $(C,\sigma)$ parameter space.

\subsection*{Construction of the random Gaussian interaction ensemble}

The random interaction ensemble is constructed in two steps: we first draw a sparse, row-balanced
Gaussian matrix $G$ and then obtain the interaction matrix $M$ and the full community matrix $A$
by rescaling with the fixed point $x^\ast$.

Given the number of species $S$ and connectance $C$, we start from a dense Gaussian matrix
$G^{(0)}$ with independent entries $G^{(0)}_{ij} \sim \mathcal{N}(0,1/S)$. We then draw an
independent Bernoulli mask $B_{ij}$ with $\mathbb{P}(B_{ij}=1)=C$ and set $B_{ii}=0$, so that only
a fraction $C$ of the off-diagonal entries are retained. The masked matrix is
$G_{ij} = B_{ij}\,G^{(0)}_{ij}$,
with $G_{ii}=0$. To enforce row balance, we subtract from each non-zero off-diagonal entry in row
$i$ the mean over the kept entries in that row:
\begin{equation}
    \mu_i = \frac{\sum_j G_{ij}}{\sum_j B_{ij}}, \qquad
G_{ij} \leftarrow G_{ij} - \mu_i \quad \text{for all } j \text{ with } B_{ij}=1.
\label{1}
\end{equation}
After this operation, each row satisfies $\sum_j G_{ij} = 0$, i.e. $G \mathbf{1} = 0$, and the
diagonal remains zero. The diagonal matrix $D$ is fixed by the uncoupled stochastic logistic model: we prescribe a
deterministic fixed point vector $x^\ast$ and set
$D = -\mathrm{diag}(1/x^\ast)$,
so that $\mathrm{diag}(x^\ast) = -D^{-1}$, as in \eqref{eq: FP condition}. The interaction matrix $M$ is then
obtained by rescaling $G$ with the components of $x^\ast$,
$M = -G D = G\,\mathrm{diag}(1/x^\ast)$.
By construction, $M$ has zero diagonal and satisfies \eqref{2M},
   $ M x^\ast = G\,\mathrm{diag}(1/x^\ast)\,x^\ast = G \mathbf{1} = 0,$
so that turning on interactions does not shift the fixed point. Finally, interactions are introduced as a perturbation of the uncoupled SLM via \eqref{3M}
$A = D + \sigma M = D - \sigma G D,$
and the effective stability matrix entering the linearized dynamics around $x^\ast$ reads as in \eqref{eq: Aeff}, $A_{\mathrm{eff}}=\frac{1}{\tau}\left(-I + \sigma G\right),$
which directly links the Gaussian ensemble for $G$ to the spectrum of the effective Jacobian
(see SI Section 1 for further details and analytical results).

\subsection*{Estimation of the distance to criticality from covariance matrices}

As shown in the SI (see Section 4 for details), in the fixed-point phase of the ISLM, the distance to
criticality can be quantified by a single parameter $g$ defined through the leading eigenvalue
$\lambda$ of the effective stability matrix, $\lambda = 1 - g^2$. Within dynamical mean-field
theory, the normalized width of the distribution of short-time covariances provides an explicit
relation between $\lambda$ and the empirical covariance matrix.

Given a covariance (or correlation) matrix $C$ of size $S \times S$, we define the normalized
width $\Delta$ as in \eqref{Delta}. This quantity can be linked to the short-time limits of the four-point observable, $\Psi^\text{short}$, and the average variance of the effective single-species process, $C_y^\text{short}$, in the DMFT description (see SI Section 4 for explicit definitions):
\begin{equation}
    S\Delta^2
:= \frac{\Psi^\text{short}}{\big(C_y^\text{short}\big)^2} - 1.
\label{delta2}
\end{equation}
In the ISLM this quantity is related to the leading eigenvalue
$\lambda$ by \eqref{eq: g estimator},
with $\lambda \in (0,1]$ in the stable phase.

To estimate the distance to criticality from data, we proceed as follows. For each community or
cohort, we first compute a short-time covariance matrix $C$ by subsampling the time series (or
individuals) with the procedure detailed in the SI (Section 6.B). From $C$ we compute
$\Delta$ and the scalar quantity $A = S \Delta^2$. \eqref{eq: g estimator} is then solved numerically for $\lambda$ by introducing
$x = \sqrt{\lambda}$, which yields a quartic equation
\begin{equation}
    x^4 - (8A + 6) x^3 + 4 x^2 - 2 x + 3 = 0.
\end{equation}
Among its real positive roots, we select the physically admissible solution $x \in (0,1]$,
corresponding to $\lambda = x^2 \in (0,1]$. The distance to criticality is finally obtained as
$g = \sqrt{1 - \lambda}$,
which coincides with the interaction parameter in our parametrization ($g = \sigma\sqrt{C}$) in the
mean-field limit.

\subsection*{ Data Availability}
All microbiome data analyzed in this study were previously published.  For  the analysis of different diseases and the impact on gut-microbiome dynamics, we used the following datasets:
\begin{itemize}
    \item \textbf{IBD}: Metagenomic data from the Inflammatory Bowel Disease Multi'omics Database
    (IBDMDB, \url{https://ibdmdb.org/}) and Ref.~\cite{LloydPrice}.
    \item \textbf{IBS}: Metagenomic data from Ref.~\cite{Mars}.
    \item \textbf{CDI}: Metagenomic data from Ref.~\cite{Ferretti}.
    \item \textbf{CRC}: Metagenomic data from Ref.~\cite{Yachida}.
    \item \textbf{Independent healthy cohorts}: Two independent healthy-control cohorts based on
    16S rRNA sequencing from Ref.~\cite{Park}.
\end{itemize}
The environmental and human-gut microbiome datasets used to characterize macroecological patterns
and association diversity are the same as in Ref.~\cite{Grilli2020}.

\subsection*{Data processing}

\paragraph*{Environmental microbiomes (Fig.~\ref{fig: 4})}

For the biome analysis in Fig.~\ref{fig: 4} we use the same environmental
microbiome datasets as in \cite{Grilli2020}, comprising glacier, lake, river and
seawater communities, all of them cross-sectional. For each biome we start from the species–by–sample relative–abundance table of Ref.\cite{Grilli2020} and apply
an occupancy filter to remove extremely rare taxa. Specifically, for a given
occupancy threshold $o$ we retain only the most prevalent species, i.e. those
present in at least a fraction $o$ of the samples in that biome. This yields an
effective number of species $S$, together with a number of samples $T$ for each
choice of $o$, and we define the units-to-samples ratio $r = S/T$. We explore several
occupancy thresholds (e.g. $o = 0.5, 0.95, 0.99$) and, for each biome,
select the threshold that minimizes $r$ under the constraint $S \ge 100$. The
resulting values of $o$, $S$, $T$ and $r$ for all biomes are summarized in
Table \ref{Tab: 1}.

\begin{table}[h]
\centering
\begin{tabular}{lcccc}
\toprule
Biome    & chosen $o$ & $S$ & $T$ & $r = S/T$ \\
\midrule
Glacier  & 0.99 & 260 &  30 &  8.67 \\
Lake     & 0.95 & 132 & 198 &  0.67 \\
River    & 0.95 & 147 & 188 &  0.78 \\
Seawater & 0.50 & 199 & 474 &  0.42 \\
\bottomrule
\end{tabular}
\caption{Occupancy thresholds $o$ selected for the environmental biomes in
Fig.~\ref{fig: 4}, together with the resulting number of species $S$, number of
samples $T$, and aspect ratio $r = S/T$ (selection criterion: minimize $r$ with
$S \ge 100$).}
\label{Tab: 1}
\end{table}

\paragraph*{Human gut cohorts and disease datasets (Fig. \ref{fig: 5})}

For the analysis of human gut microbiome data under different diseases (IBD, IBS, CRC, CDI and independent healthy cohorts)
we follow the preprocessing and subsampling pipeline described in detail in
SI Section~6.B. Briefly, for each study we construct a taxa–by–sample table of
relative abundances, apply the same basic filtering steps as in the biome
datasets (removing extremely low-occupancy taxa and samples with insufficient
coverage), and restrict the analysis to taxa passing a study–specific
occupancy threshold. Disease cohorts are typically cross-sectional: each sample corresponds to one individual, and subsamples used
for estimating $g$ are constructed by randomly drawing groups of individuals
within each diagnosis category. 

\paragraph*{Disease separation from distance to criticality}

After filtering, the abundance of each retained taxon is standardized across
samples within a given biome or cohort (subtracting the sample mean and
dividing by the sample standard deviation). We then compute the empirical covariance matrix $C$ of these standardized variables. The width of
the covariance distribution is summarized by the normalized second moment
$\Delta$ defined in \eqref{Delta}. We then use the dynamical mean-field relation of
\eqref{eq: g estimator} to solve for the leading eigenvalue
$\lambda$ and infer the distance to criticality $g = \sqrt{1-\lambda}$.
Repeating this procedure over many subsamples within each biome or diagnosis
group yields the distributions of $g$ reported in Figs.~\ref{fig: 4} and
\ref{fig: 5}; full details, including the choice of subsample size and the
number of repetitions per cohort, are given in SI Sections~6.C–6.F.

For each biome or diagnosis group, the pipeline described above and in
SI Section~6.B yields a distribution of distance-to-instability estimates
$g$ obtained from many short-time subsamples. To quantify how well $g$ separates healthy and diseased microbiomes, we
treat the individual $g$ values as classification scores and perform a
receiver operating characteristic (ROC) analysis between pairs of groups.
For a given pair (e.g.\ healthy vs.\ ulcerative colitis), we compute the area under the curve (AUC) as a non-parametric measure of separation (SI Section 6.C). An AUC of $1$ (or $0$) corresponds to perfect
separation by $g$, whereas $\mathrm{AUC} = 0.5$ indicates no discriminative
power. In the main text we report AUC values and associated confidence
intervals (estimated by resampling $g$ values within each group) as
our summary statistics for the degree of separation between diagnoses
(SI Section~6.E).

As controls, we compare independent healthy cohorts from different
studies, for which AUC values remain close to $0.5$ over a broad range of
sampling parameters, indicating that our estimator does not generate
spurious differences between comparable healthy groups (SI Section~6.G).

\vspace{0.5cm}
\begin{acknowledgments}
This work has been supported by Grant No. PID2023-149174NB-I00
financed by the Spanish Ministry and Agencia Estatal de Investigaci\'on MICIU/AEI/10.13039/501100011033 and EDRF funds.
We are thankful to Jacopo Grilli, William Shoemaker, Samir Suweis and Aniello Lampo  for enlightening discussions.
\end{acknowledgments}

\bibliography{Biomes+edge}

\end{document}


\maketitle

\section{The Interacting Stochastic Logistic Model (ISLM):} \label{sec: The ISLM}
In the main text, we focus on a model of interacting species where the abundance of each species $i$, with $i=1,2,\dots,S$, evolves according to the Langevin equation
\begin{equation}
    \label{eq: GLV}
    \dot{x}_i(t)
    = \frac{x_i(t)}{\tau}\left[ 1 + \sum_{j=1}^S A_{ij}\,x_j(t) \right]
      + \sigma_0 x_i(t)\,\xi_i(t),
\end{equation}
where $\xi_i(t)$ is a zero-mean, white noise term with $\langle \xi_i(t)\xi_j(s)\rangle = \delta(t-s)\delta_{ij}$. This stochastic differential equation is only well defined after we set a particular interpretation and, in the main text, we mainly discussed Itô's \cite{Gardiner}. In this interpretation (i.e. using Itô's calculus),  the change of variables $u_i = \log x_i$, leads to
\begin{equation}
    \label{eq: log-space GLV}
    \frac{d}{dt}\log x_i(t)=\frac{\dot{x_i}}{x_i}-\frac{\sigma_0^2}{2}=
    \frac{1}{\tau}\left[ \left(1 - \frac{\sigma_0^2\tau}{2}\right) + \sum_{j=1}^S A_{ij}\,x_j(t) \right]+\sigma_0 \xi_i(t).
\end{equation}
Let us analyze this stochastic differential equation (SDE) in detail.

\subsection{Mean abundance of each species}
Taking averages over noise realizations (denoted $\langle \cdot  \rangle$)  on both sides of the equation gives
\begin{equation}
    \frac{d}{dt}\langle \log x_i \rangle = \frac{1}{\tau}\left( 1+\sum_{j=1}^S A_{ij} \langle x_j\rangle \right)-\frac{\sigma_0^2}{2}.
\end{equation}
At stationary states the left-hand side of the previous equation vanishes, i.e. $\dot{u}_i(t)=0$, and one has
\begin{equation}
    \frac{\sigma_0^2 \tau}{2} = 1+\sum_{j=1}^S A_{ij} \langle x_j \rangle,
\end{equation}
which, in vector form, may be written as
\begin{equation}
    A\langle \mathbf{x}\rangle = -\left(1-\frac{\sigma_0^2\tau}{2}\right) \mathbf{1}.
\end{equation}
Denoting by $\mathbf{x}^*$ the fixed point of the deterministic limit of \eqref{eq: GLV} (i.e., the limit as $\sigma_0\rightarrow 0$), then
\begin{equation}
    A\mathbf{x^*}=-\mathbf{1} \quad \Rightarrow \quad \langle \mathbf{x} \rangle = \left(1-\frac{\sigma_0^2\tau}{2}\right)\mathbf{x}^*.
\end{equation}
That is, the addition of the noise term only shifts the position of the fixed point. In particular, if the fixed points are log-normally distributed, so will be the mean abundances of each species after the noise is turned on ($\sigma\neq 0$). Interestingly, this puts an upper limit to how large the noise can be, namely
\begin{equation}
    \sigma_0 < \sqrt{\frac{2}{\tau}}.  
\end{equation}

\subsection{Linear approximation of the dynamics}\label{section: linear app}
Linearizing Equation \eqref{eq: log-space GLV} around the fixed point leads to
\begin{equation}
    \label{eq: effective A}
    \frac{d}{dt}u_i(t) \approx A_{\text{eff}} u_i(t) +\sigma_0 \xi_i(t), \;\; \text{with}\;\; A_{\text{eff}}=\frac{1}{\tau}A\:\text{diag}\left(e^{u_1^*}, ..., e^{u_S^*}\right),
\end{equation}
where the fixed point $\mathbf{u}^*$ (let us denote also $\tilde{\mathbf{x}}$, such that $\exp\left( \mathbf{u}^* \right) = \tilde{\mathbf{x}}$) is defined (in Itô's interpretation) as
\begin{equation}
    \frac{1}{\tau}\left[ \left(1 - \frac{\sigma_0^2\tau}{2}\right) + \sum_{j=1}^S A_{ij}\,\tilde{x}_j(t) \right] = 0 \quad \Rightarrow \quad \mathbf{\tilde{x}}=\langle \mathbf{x} \rangle.
\end{equation}
In particular, this means that $\mathbf{u}^*=\log \langle \mathbf{x} \rangle$. Substituting into \eqref{eq: effective A} leads to
\begin{equation}
    A_{\text{eff}}=\frac{1}{\tau}A\:\text{diag}\left(\langle x_1\rangle, ..., \langle x_S \rangle\right)=\frac{k}{\tau} A \:\text{diag}\left(x^*_1, ..., x^*_S \right)=\frac{k}{\tau}\left( D \cdot \text{diag}(\mathbf{x}^*) + \sigma M \cdot \text{diag}(\mathbf{x}^*)\right),
\end{equation}
where we have used the decomposition $A=D+\sigma M$, and $k \equiv (1-\sigma_0^2\tau/2)$. Now, because $D$ and $M$ were explicitly built such that $D\cdot \mathbf{x^*}=-\mathbf{1}$ and $M\cdot \mathbf{x^*}=\mathbf{0}$, one gets
\begin{equation}
\label{eq: Aeff final}
    A_{\text{eff}}=\frac{k}{\tau}\left( - \text{I}+\sigma G\right), \qquad G\equiv M \cdot \text{diag}(\mathbf{x}^*).
\end{equation}
As $M$ is, by construction, orthogonal to the fixed point, $G$ fulfills
\begin{equation}
\label{eq: G-rowsum}
G\,\mathbf{1} \;=\; M\,\mathrm{diag}(x^\ast)\,\mathbf{1}
\;=\; M\,x^\ast \;=\; 0,
\end{equation}
which imposes the condition that the homogeneous eigenmode $\mathbf{1}$ is an eigenvector of $G$ with eigenvalue $0$.

\subsection{Correlations in the linearized model}
Once the dynamics are reduced, close to the mean abundances, to a set of linear coupled Ornstein-Uhlenbeck equations, the (short-time) covariance matrix $\hat{\Sigma}$ can be expressed as the solution to the Lyapunov equation (see e.g. \cite{Robust} and references therein)
\begin{equation}
    \label{eq: Lyapunov equation}
    A_{\text{eff}} \hat{\Sigma}+ \hat{\Sigma}A_{\text{eff}}^{\mathsf{T}}=\sigma_0^2 \text{I}.
\end{equation}
This equation does not admit, in general, a closed-form solution for $\hat{\Sigma}$ in terms of $A_{\text{eff}}$. Let us remark that $\hat{\Sigma}$ is the covariance matrix for the linearized process in log space $u_i=\log(x_i/\langle x_i \rangle)$. If environmental fluctuations are small (i.e., if $\sigma_0$ is small) then $u_i\approx x_i/\langle x_i\rangle -1 = (x_i-\langle x_i \rangle)/\langle x_i\rangle$. This means that the covariance, $\Sigma$, of the original process and the covariance of its linearized counterpart are related as
\begin{equation}
    \label{eq: relating the two covariances}
    \hat{\Sigma}_{ij} = \frac{\Sigma_{ij}}{\langle x_i\rangle \langle x_j \rangle} = \frac{1}{k^2}\frac{\Sigma_{ij}}{ x^*_i x^*_j } \quad \Rightarrow \quad \hat{\Sigma} = D^{-1} \Sigma D^{-1}.
\end{equation}
On the other hand, the (Pearson's) correlation matrix associated with $\Sigma$ is
\begin{equation}
  C_{ij}
  \;=\;
  \frac{\Sigma_{ij}}{\sqrt{\Sigma_{ii} \Sigma_{jj}}},
\end{equation}
so that:
\begin{equation}
    \tilde C_{ij}\equiv 
  \frac{\tilde \Sigma_{ij}}{\sqrt{\tilde \Sigma_{ii}\,\tilde \Sigma_{jj}}}= \frac{\Sigma_{ij}/(x_i^* x_j^*)}
{\sqrt{\dfrac{\Sigma_{ii}}{(x_i^*)^2} \dfrac{\Sigma_{jj}}{(x_j^*)^2}}}=
\frac{\Sigma_{ij}/(x_i^* x_j^*)}
       {\sqrt{\Sigma_{ii} \Sigma_{jj}}/(x_i^* x_j^*)}=
  \frac{\Sigma_{ij}}{\sqrt{\Sigma_{ii} \Sigma_{jj}}}=
  C_{ij}.
\end{equation}
Therefore the Pearson's correlation matrix is invariant under this diagonal rescaling: 
$\tilde C_{ij} = C_{ij}$ for all $i,j$.

\subsection{Condition of linear stability}
If the matrix elements $G_{ij}$, with $i,j \in \{1,\dots,S\}$, are i.i.d., drawn from a distribution with finite first and second moments and such that each row is orthogonal to the vector $\mathbf{1} = (1,\dots,1)$, then in the limit $S \to \infty$ the spectral density of $G$ converges to a uniform distribution bounded in a disk in the complex plane centered at $0$ with radius \cite{Tao,Allesina}
\begin{equation}
    R = \sqrt{S\,\mathrm{Var}(G_{ij})}.
\end{equation}
In the main text we choose $G$ such that $\mathrm{Var}(G_{ij}) = C/S$ and therefore the limiting spectral radius is
\begin{equation}
    R = \sqrt{C}.
\end{equation}
The effective linear response matrix can be written as
\[
\tau A_{\mathrm{eff}} = -\mathrm{I} + \sigma G,
\]
so its eigenvalues are contained in a disk of radius $\sigma \sqrt{C}$ centered at $-1$ in the complex plane.

The condition for linear stability is
\begin{equation}
    \max_{\lambda \in \mathrm{sp}(A_{\mathrm{eff}})} \Re(\lambda) < 0.
\end{equation}
Combining this condition with the bounds on the spectrum above implies
\[
-1 + \sigma \sqrt{C} < 0,
\]
which is the resulting stability criterion.

\section{Grilli's laws for the stochastic logistic model without interactions}

We consider the stochastic logistic model (SLM) for the abundance $x_i(t)$ of
species $i$,
\begin{equation}
    \dot x_i(t)
    = \frac{x_i(t)}{\tau_i}\left[1 + \sum_{j=1}^S A_{ij} x_j(t)\right]
      + \sigma_0 x_i(t)\,\xi_i(t),
    \label{eq:SLM_general}
\end{equation}
where $\xi_i(t)$ is a Gaussian white noise with
$\langle \xi_i(t)\rangle = 0$ and
$\langle \xi_i(t)\xi_j(t')\rangle = \delta_{ij}\,\delta(t-t')$.
In the absence of interactions we take
\begin{equation}
    A_{ij} = -\frac{1}{K_i}\,\delta_{ij},
\end{equation}
and assume that all time scales are equal, $\tau_i=\tau$.  The dynamics of
each species is then uncoupled and obeys
\begin{equation}
    \dot x_i(t)
    = \frac{x_i(t)}{\tau}\left(1 - \frac{x_i(t)}{K_i}\right)
      + \sigma_0 x_i(t)\,\xi_i(t).
    \label{eq:SLM_single}
\end{equation}
In the following we drop the species index $i$ for clarity and work with a
single generic species.

\subsection{Fokker--Planck equation and stationary distribution}

Interpreting \eqref{eq:SLM_single} in the It\^o sense, we can write
\begin{equation}
    \dot{x}(t)
    = f(x) + g(x)\xi(t),
\end{equation}
with drift and noise amplitude
\begin{equation}
    f(x) = \frac{x}{\tau}\left(1 - \frac{x}{K}\right),
    \qquad
    g(x) = \sigma_0 x.
\end{equation}
The corresponding Fokker--Planck equation for the probability density
$P(x,t)$ is
\begin{equation}
    \frac{\partial P(x,t)}{\partial t}
    = -\frac{\partial}{\partial x}\bigl[f(x)P(x,t)\bigr]
      + \frac{1}{2}\frac{\partial^2}{\partial x^2}
        \bigl[g^2(x)P(x,t)\bigr]
    = -\frac{\partial}{\partial x}
        \left[\frac{x}{\tau}\left(1-\frac{x}{K}\right)P\right]
      + \frac{\sigma_0^{2}}{2}\frac{\partial^2}{\partial x^2}
        \left[x^2 P\right].
    \label{eq:FP}
\end{equation}
The stationary distribution $P_\ast(x)$ satisfies
$\partial_t P_\ast = 0$.  For a one--dimensional It\^o process, the
stationary solution of \eqref{eq:FP} can be written as
\begin{equation}
    P_\ast(x)
    \propto \frac{1}{g^2(x)}
    \exp\!\left(
        \int^x \mathrm{d}y\,\frac{2f(y)}{g^2(y)}
    \right).
    \label{eq:FP_stationary_general}
\end{equation}
In our case $g^2(x) = \sigma_0^{2} x^2$, and
\begin{equation}
    \frac{2f(y)}{g^2(y)}
    = \frac{2}{\sigma_0^{2}}\,
      \frac{1}{\tau}\,\frac{y(1-y/K)}{y^2}
    = \frac{2}{\sigma_0^{2} \tau}
      \left(\frac{1}{y} - \frac{1}{K}\right).
\end{equation}
Integrating this expression we obtain
\begin{equation}
    \int^x \mathrm{d}y\,\frac{2f(y)}{g^2(y)}
    = \frac{2}{\sigma_0^{2} \tau}
      \left(\log x - \frac{x}{K}\right) + \text{const}.
\end{equation}
Substituting into Eq.~\eqref{eq:FP_stationary_general} gives
\begin{equation}
    P_\ast(x)
    \propto \frac{1}{\sigma_0^{2} x^2}
    \exp\!\left[
        \frac{2}{\sigma_0^{2} \tau}
        \left(\log x - \frac{x}{K}\right)
    \right]
    = \text{const}\times
      x^{\frac{2}{\sigma_0^{2} \tau}-2}
      \exp\!\left(
        -\frac{2x}{\sigma_0^{2} \tau K}
      \right).
\end{equation}
This is a Gamma distribution
\begin{equation}
    P_\ast(x)
    = \frac{1}{\Gamma(\alpha)\,\theta^\alpha}
      x^{\alpha-1}\exp\!\left(-\frac{x}{\theta}\right),
    \qquad x>0,
\end{equation}
with shape and scale parameters
\begin{equation}
    \alpha = \frac{2}{\sigma_0^{2} \tau} - 1, \qquad
    \theta = \frac{\sigma_0^{2} \tau K}{2}.
    \label{eq:gamma_params}
\end{equation}
Thus, in the absence of interactions, the abundance fluctuation
distribution (AFD) of each species is Gamma-distributed.

\subsection{Taylor's law}
For a Gamma distribution with parameters $(\alpha,\theta)$, the mean and
variance are
\begin{equation}
    \langle x \rangle = \alpha \theta, \qquad
    \mathrm{Var}(x) = \alpha \theta^2.
\end{equation}
Eliminating $\theta$ we obtain
\begin{equation}
    \mathrm{Var}(x)
    = \frac{1}{\alpha}\,\langle x \rangle^2.
    \label{eq:Taylor}
\end{equation}
\eqref{eq:Taylor} is precisely Taylor's law with exponent $2$:
across species, the variance of abundances scales as the square of the
mean, with a prefactor $1/\alpha$ that is independent of the species
identity,
\begin{equation}
    \mathrm{Var}(x_i)
    = \frac{1}{\alpha}\,\langle x_i \rangle^2,
    \qquad i=1,\dots,S.
\end{equation}
In the SLM, $\alpha$ depends only on the noise amplitude $\sigma_0$ and
the common time scale $\tau$ (see \eqref{eq:gamma_params}) and is
therefore the same for all species (provided they share the same
environmental noise statistics). Differences in carrying capacities
$K_i$ change only the scale parameter $\theta_i$ and hence the mean
abundance $\langle x_i \rangle$, but not $\alpha$. This hence proves that Taylor's 
law is fulfilled in the SLM.

\subsection{Mean abundance distribution (MAD)}
In the absence of interactions, each species follows the stochastic
logistic dynamics in \eqref{eq:SLM_single}.  As shown above, the
stationary abundance distribution of a single species is Gamma with
parameters
\begin{equation}
    \alpha = \frac{2}{\sigma_0^{2} \tau} - 1,
    \qquad
    \theta_i = \frac{\sigma_0^{2} \tau K_i}{2},
    \label{eq:gamma_params_MAD}
\end{equation}
where $K_i$ is the carrying capacity of species $i$. In It\^o's calculus, the stationary mean abundance of species $i$ is therefore
\begin{equation}
    \langle x_i \rangle
    = \alpha \theta_i
    = \left(\frac{2}{\sigma_0^{2} \tau} - 1\right)
      \frac{\sigma_0^{2} \tau K_i}{2}
    = c\,K_i,
    \qquad
    c \equiv 1 - \frac{\sigma_0^{2} \tau}{2}.
    \label{eq:mean_x_i_MAD}
\end{equation}
Hence the mean abundance is simply a constant dilation of the
deterministic fixed point $x_i^\ast = K_i$.

To obtain the mean abundance distribution (MAD) across species, we now
assume that the carrying capacities are independently drawn from a
log-normal distribution,
\begin{equation}
    K_i \sim \mathrm{LogNormal}(\mu_K,\sigma_K^2),
\end{equation}
i.e.
\begin{equation}
    \log K_i \sim \mathcal{N}(\mu_K,\sigma_K^2).
\end{equation}
Using \eqref{eq:mean_x_i_MAD}, the mean abundance of species $i$ is
\begin{equation}
    m_i \equiv \langle x_i \rangle = c K_i.
\end{equation}
Taking logarithms,
\begin{equation}
    \log m_i = \log c + \log K_i.
\end{equation}
Since $\log K_i$ is Gaussian and $\log c$ is a constant shift, we obtain
\begin{equation}
    \log m_i \sim \mathcal{N}(\mu_K + \log c,\sigma_K^2),
\end{equation}
which implies that the mean abundances are themselves log-normally
distributed,
\begin{equation}
    m_i \sim \mathrm{LogNormal}(\mu_K + \log c,\sigma_K^2).
\end{equation}
Therefore, choosing log-normally distributed carrying capacities
directly yields a log-normal mean abundance distribution (MAD) across
species.  The multiplicative effect of environmental noise in the
It\^o interpretation appears only as a global dilation factor $c$ and
does not change the log-normal form of the MAD.


\section{Dynamical Mean Field Derivation of the Effective process}\label{secApend: DMFT}
\subsection{Dynamical model}

Here we address the problem of disentangling ecological forces from a statistical-physics perspective. For this, we first focus on a the interacting stochastic logistic model (ISLM). The abundance of species $i = 1,\dots,S$ evolves according to the set of coupled stochastic differential equations
    \begin{equation}
        \label{eq:GLV}
        \dot{x}_i(t)
        = \frac{x_i(t)}{\tau_i} \left[ 1 + \sum_{j=1}^S A_{ij}\,x_j(t) \right]
        + x_i(t)\,\xi_i(t),
    \end{equation}
    where $\tau_i$ sets the characteristic timescale of growth or decline for species $i$, $A_{ij}$ is the interaction matrix, and $\xi_i(t)$ is an uncorrelated Gaussian white noise with
    \begin{equation}
        \langle \xi_i(t) \rangle = 0,
        \qquad
        \langle \xi_i(t)\,\xi_j(t') \rangle = \sigma_0^2\,\delta_{ij}\,\delta(t-t').
    \end{equation}

We decompose the interaction matrix as
 $  A = D + \sigma M$,
where:
\begin{itemize}
    \item $D$ is a diagonal matrix that encodes the intrinsic regulation (effective carrying capacity) of each species in the absence of interactions.
    \item $M$ is a random matrix whose statistics will be chosen to reproduce empirical ecological patterns.
    \item $\sigma$ controls the overall strength of interactions, continuously interpolating between the uncoupled SLM limit ($\sigma = 0$, ideal-gas-like) and a strongly interacting regime ($\sigma \gg 0$, Van der Waals-like).
\end{itemize}

\subsection{Construction of the interaction matrix}
\label{subsec:interaction-matrix}

We now specify the statistical ensemble used to generate the interaction matrix $A$.

\subsubsection*{Distribution of intrinsic terms}

The diagonal elements of $D$ are sampled independently from a log-normal distribution with negative support:
\begin{equation}
    D_{ii} \sim -\log\mathcal{N}(\hat{\mu}, \hat{\sigma}),
\end{equation}
so that $D_{ii}<0$ and each species is self-limiting in the absence of interactions.

For $\sigma=0$, the fixed point $x^*$ of Eq.~\eqref{eq:GLV} is determined by
\begin{equation}
    1 + D_{ii}\,x_i^* = 0 
    \quad\Rightarrow\quad
    x_i^* = -\frac{1}{D_{ii}},
\end{equation}
which we collect into the diagonal matrix
\begin{equation}
    \mathrm{diag}(x^*) = -D^{-1}.
\end{equation}

\subsubsection*{Random connectivity and weights}

To introduce sparse random interactions, we draw two independent random matrices $B$ and $Z$:
\begin{align}
    Z_{ij} &\sim \mathcal{N}\!\left(0,\frac{1}{S}\right), \\
    B_{ij} &\sim \mathrm{Bernoulli}(C),
\end{align}
where the connectivity $C \in [0,1]$ is the probability that an interaction between $i$ and $j$ is present. Equivalently, the distribution of each entry of $B$ is
\begin{equation}
    P(B_{ij}) = C\,\delta(B_{ij}-1) + (1-C)\,\delta(B_{ij}).
\end{equation}

We then construct an intermediate random matrix $G$ with zero row-sum, ensuring that the net interaction of each species is centered. For each row $i$ we define
\begin{equation}
    G_{ij} = B_{ij}\left[Z_{ij} - \frac{1}{S}\sum_{k=1}^S B_{ik}Z_{ik}\right],
\end{equation}
so that
\begin{equation}
    \sum_{j=1}^S G_{ij} = 0 \quad \text{for all } i.
\end{equation}
By construction, all entries $B_{ij},B_{ik}$ and $Z_{ij},Z_{ik}$ are independent for $j \neq k$.

\subsubsection*{Rescaling by the deterministic fixed point}

We define the random matrix $M$ by rescaling $G$ with the typical abundances $x^*$ of the uncoupled system:
\begin{equation}
    M = G \cdot \mathrm{diag}^{-1}(x^*),
    \qquad
    \mathrm{diag}(x^*) = -D^{-1},
\end{equation}
so that
\begin{equation}
    M = -G D.
\end{equation}
Hence the full interaction matrix
\begin{equation}
    A = D + \sigma M = D - \sigma G D
\end{equation}
has components
\begin{equation}
    A_{ij} =
    \begin{cases}
        D_{ii}, & i = j, \\[4pt]
        -\sigma D_{jj}\,G_{ij}, & i \neq j.
    \end{cases}
\end{equation}
The sum of all entries of $A$ can therefore be expressed as
\begin{equation}
    \sum_{i,j} A_{ij}
    = \sum_{i} D_{ii}
      - \sigma \sum_{i\neq j} D_{jj}\,G_{ij}.
\end{equation}

Our strategy is to construct the generating functional of the stochastic dynamics and then integrate over the random degrees of freedom associated with $Z$ and $B$.  
Since the diagonal entries $D_{ii}$
are independent and do not induce direct coupling between different species, we keep them explicit until the end of the calculation.
At that stage, the entries $D_{ii}$ will be encoded as an effective quenched randomness, distributed according to the prescribed log-normal law.

\subsection{Generating Functional analysis}\label{subsecA2}

We now perform a generating functional analysis in order to derive a dynamical mean-field theory (DMFT) for the generalized Lotka--Volterra dynamics defined in Eq.~\eqref{eq:GLV}. Throughout this section the number of species is denoted by $S$, and we work in the thermodynamic limit $S\to\infty$.

\subsubsection{Definition of the generating functional}

We consider connectivity $C=1$ for this analysis. For a given realization of the quenched disorder $D$ and $Z$, we introduce the Martin--Siggia--Rose--Janssen--De~Dominicis (MSRJD) generating functional
\begin{equation}
    Z[\boldsymbol{j}]
    = \int \mathcal{D}D\,\mathcal{P}(D)
      \int \mathcal{D}[\underline{\boldsymbol{x}},\hat{\underline{\boldsymbol{x}}}]
      \int \mathcal{D}Z\,\mathcal{P}(Z)\,
      \exp\!\left( i S_{\text{MSR}}[\underline{\boldsymbol{x}},\hat{\underline{\boldsymbol{x}}};D,Z]
                 + i\sum_i \int \!dt\, j_i(t)\,x_i(t)\right),
\end{equation}
where $j_i(t)$ are source fields for the abundances and
\begin{equation}
    S_{\text{MSR}}
    = \sum_i \int\!dt\,\hat{x}_i(t)\left[
         \frac{\dot{x}_i(t)}{x_i(t)}
        -\left(1+\sum_j A_{ij} x_j(t)\right)
    \right].
\end{equation}
Here $A_{ij}$ is the interaction matrix defined in the previous subsection,
\begin{equation}
    A_{ij} =
    \begin{cases}
        D_{ii}, & i=j,\\[4pt]
        -\sigma D_{jj} G_{ij}, & i\neq j,
    \end{cases}
\end{equation}
with $G_{ij} = Z_{ij} - \frac{1}{S}\sum_k Z_{ik}]$.

It is convenient to separate the different contributions in the action:
\begin{align}
    S_{\text{MSR}}
    &= \sum_i \int\!dt\,\hat{x}_i(t)\left[\frac{\dot{x}_i(t)}{x_i(t)} - 1\right]
       - \sum_{ij}\int\!dt\,\hat{x}_i(t) A_{ij} x_j(t)\\
    &= \sum_i \int\!dt\,\hat{x}_i(t)\left[\frac{\dot{x}_i(t)}{x_i(t)} - 1\right]
       - \sum_i \int\!dt\,\hat{x}_i(t) D_{ii} x_i(t)
       - \sigma \sum_{i\neq j}\int\!dt\,\hat{x}_i(t) D_{jj} G_{ij} x_j(t).
\end{align}

Introducing the shorthand
\begin{equation}
    R_{ij} \equiv \int\!dt\,\hat{x}_i(t)\,x_j(t),
\end{equation}
we can rewrite
\begin{align}
    S_{\text{MSR}}
    &= \sum_i \int\!dt\,\hat{x}_i(t)\left[\frac{\dot{x}_i(t)}{x_i(t)} - 1\right]
       - \sum_i D_{ii} R_{ii}
       - \sigma \sum_{i\neq j} D_{jj} G_{ij} R_{ij}.
\end{align}

Thus the generating functional takes the form
\begin{align}
    Z[\boldsymbol{j}]
    &= \int \mathcal{D}D\,\mathcal{P}(D)
       \int \mathcal{D}[\underline{\boldsymbol{x}},\hat{\underline{\boldsymbol{x}}}]
       \int \mathcal{D}Z\,\mathcal{P}(Z)\,
       \exp\Bigg\{
            i\sum_i\int\!dt\, j_i(t)x_i(t)
            \\
            &
            + i\sum_i\int\!dt\,\hat{x}_i(t)\Big[\frac{\dot{x}_i(t)}{x_i(t)}-1\Big]
            - i\sum_i D_{ii}R_{ii}
            - i\sigma\sum_{i\neq j}D_{jj}G_{ij}R_{ij}
       \Bigg\}.
       \label{eq:GF_with_G}
\end{align}

\subsubsection{Disorder average over the Gaussian weights $Z$}

We now average over the Gaussian weights $Z_{ij}$, drawn independently as
\begin{equation}
    Z_{ij} \sim \mathcal{N}\!\left(0,\frac{1}{S}\right).
\end{equation}
Using the definition
\begin{equation}
    G_{ij} = Z_{ij} - \frac{1}{S}\sum_k Z_{ik},
\end{equation}

we write the interaction term as
\begin{align}
    -i\sigma \sum_{i\neq j} D_{jj} G_{ij} R_{ij}
    &= -i\sigma \sum_{i\neq j} D_{jj}  Z_{ij} R_{ij}
     + i \frac{\sigma}{S}\sum_{i\neq j}\sum_k D_{jj}  Z_{ik} R_{ij}.
\end{align}
In the second term we relabel the dummy indices $(j,k)$ and reorganize the sums so that both contributions take the form of a linear coupling to $Z_{ij}$. More explicitly,
\begin{align}
    -i\sigma \sum_{i\neq j} D_{jj} G_{ij} R_{ij}
    &= -i\sigma\sum_{i\neq j} D_{jj}  R_{ij}\,Z_{ij}
       + i\frac{\sigma}{S}\sum_{i\neq k}\sum_{j\neq i} D_{kk}  R_{ik}\,Z_{ij}\\
    &= -\sum_{i\neq j} H_{ij}\,Z_{ij},
\end{align}
where we have defined the (in general complex) coefficients
\begin{equation}
    H_{ij}
    \equiv i\sigma \left[
        D_{jj}  R_{ij}
        - \frac{1}{S}\sum_k D_{kk}  R_{ik}
    \right].
    \label{eq:H_ij_def}
\end{equation}

The part of the action depending on $Z_{ij}$ is therefore
\begin{equation}
    S_Z = -\sum_{i\neq j} H_{ij} Z_{ij}.
\end{equation}
The Gaussian average over $Z$ is straightforward:
\begin{align}
    \left\langle
        \exp\Big(-\sum_{i\neq j} H_{ij} Z_{ij}\Big)
    \right\rangle_Z
    &= \prod_{i\neq j}
       \int\ dZ_{ij}\,
       \sqrt{\frac{S}{2\pi}}
       \exp\!\left(-\frac{S}{2}Z_{ij}^2 - H_{ij}Z_{ij}\right)\\
    &\propto \exp\!\left(\frac{1}{2S}\sum_{i\neq j} H_{ij}^2\right).
\end{align}
Substituting \eqref{eq:H_ij_def}, we note that $H_{ij}\propto i\sigma$, so that
\begin{equation}
    H_{ij}^2 = (i\sigma)^2 \left[\cdots\right]^2 = -\sigma^2 \left[\cdots\right]^2,
\end{equation}
and the exponential becomes
\begin{equation}
    \left\langle
        \exp\Big(-\sum_{i\neq j} H_{ij} Z_{ij}\Big)
    \right\rangle_Z
    \propto
    \exp\!\left(
        -\frac{\sigma^2}{2S}
        \sum_{i\neq j}
        \left[
            D_{jj}  R_{ij}
            - \frac{1}{S}\sum_k D_{kk}  R_{ik}
        \right]^2
    \right).
\end{equation}

Using this result in Eq.~\eqref{eq:GF_with_G}, the disorder-averaged generating functional over $Z$ reads
\begin{align}
    \langle Z[\boldsymbol{j}] \rangle_Z
    &= \int \mathcal{D}D\,\mathcal{P}(D)
       \int \mathcal{D}[\underline{\boldsymbol{x}},\hat{\underline{\boldsymbol{x}}}]
       \int 
       \exp\Bigg\{
            i\sum_i\int\!dt\, j_i(t)x_i(t)
            \nonumber\\[-2pt]
            &\hspace{2.9cm}
            + i\sum_i\int\!dt\,\hat{x}_i(t)\Big[\frac{\dot{x}_i(t)}{x_i(t)}-1\Big]
            - i\sum_i D_{ii}R_{ii}
            \nonumber\\
            &\hspace{2.9cm}
            -\frac{\sigma^2}{2S}\sum_{i\neq j}
                \left[
                    D_{jj} B_{ij} R_{ij}
                    - \frac{1}{S}\sum_k D_{kk} B_{ij} R_{ik}
                \right]^2
       \Bigg\}.
       \label{eq:GF_after_Z}
\end{align}

Expanding the square, one obtains 
\begin{align}
    &\sum_{i\neq j}
        \left[
            D_{jj} R_{ij}
            - \frac{1}{S}\sum_k D_{kk}  R_{ik}
        \right]^2
    \nonumber\\
    &\quad
    = \sum_{i\neq j} D_{jj}^2 R_{ij}^2
      - \frac{2}{S}\sum_{i\neq j}\sum_k D_{jj}D_{kk}  R_{ij} R_{ik}
    \nonumber\\
    &\quad\quad
      + \frac{1}{S^2}\sum_{i\neq j}\sum_{k_1\neq j}\sum_{k_2\neq j}
         D_{k_1k_1}D_{k_2k_2} R_{ik_1} R_{ik_2}.
\end{align}

\subsubsection{Macroscopic order parameters}

Guided by the structure of the terms appearing after the disorder average, we introduce the following order parameters:
\begin{align}
    M(t) &= \frac{1}{S}\sum_i D_{ii} x_i(t), \label{eq:OrderParameterM}\\[4pt]
    C_D(t,t') &= \frac{1}{S}\sum_i D_{ii}^2 x_i(t)x_i(t'), \label{eq:OrderParameterC}\\[4pt]
    L(t,t') &= \frac{1}{S}\sum_i \hat{x}_i(t)\hat{x}_i(t'). \label{eq:OrderParameterL}
\end{align}
The object $M(t)$ is a disorder-weighted mean abundance, $C_D(t,t')$ is a disorder-weighted two-time correlation function, and $L(t,t')$ encodes correlations between response fields.  

After some algebra (reorganizing the sums over species indices and keeping only terms that scale as $\mathcal{O}(S)$), the exponent in \eqref{eq:GF_after_Z} can be written as
\begin{align}
    &i\sum_i\int\!dt\,\hat{x}_i(t)\Big[\frac{\dot{x}_i(t)}{x_i(t)}-1\Big]
    - i\sum_i D_{ii}R_{ii}
    \nonumber\\
    &\quad
    - \frac{\sigma^2}{2}\int\!dt\,dt'\,\Bigg\{
        S\left(1-\frac{1}{S}\right)^2 C_D(t,t') L(t,t')
        - 2(S-1) M(t)M(t') L(t,t')
        \nonumber\\
    &\quad\quad
        + M(t)M(t')L(t,t')
        + \frac{1}{S}C_D(t,t')L(t,t')
        + \mathcal{O}\!\left(\frac{1}{S}\right)
    \Bigg\}.
\end{align}
In the thermodynamic limit $S\to\infty$, keeping the leading $\mathcal{O}(S)$ terms and defining the connected correlator
\begin{equation}
    \tilde{C}(t,t') = C_D(t,t') - M(t)M(t'),
\end{equation}
this simplifies to
\begin{align}
    &i\sum_i\int\!dt\,\hat{x}_i(t)\Big[\frac{\dot{x}_i(t)}{x_i(t)}-1\Big]
    - i\sum_i D_{ii}R_{ii}
    - S\,\frac{\sigma^2}{2}\int\!dt\,dt'\,\tilde{C}(t,t')\,L(t,t')
    + \mathcal{O}(1).
    \label{eq:GF_order_S}
\end{align}

Therefore, up to subleading corrections in $S$, the disorder-averaged generating functional can be written as
\begin{align}
    \langle Z[\boldsymbol{j}] \rangle_{B,Z}
    &= \int \mathcal{D}D\,\mathcal{P}(D)
       \int \mathcal{D}[\underline{\boldsymbol{x}},\hat{\underline{\boldsymbol{x}}}]
       \exp\Bigg\{
            i\sum_i\int\!dt\, j_i(t)x_i(t)
            \nonumber\\[-2pt]
            &\hspace{2.9cm}
            + i\sum_i\int\!dt\,\hat{x}_i(t)\Big[\frac{\dot{x}_i(t)}{x_i(t)}-1 + D_{ii} x_i(t)\Big]
            \nonumber\\
            &\hspace{2.9cm}
            - S\,\frac{\sigma^2}{2}\int\!dt\,dt'\,\tilde{C}(t,t')\,L(t,t')
            + \mathcal{O}(1)
       \Bigg\}.
       \label{eq:GF_after_disorder_leading}
\end{align}

To treat the order parameters at the saddle point level, we enforce their definitions via functional delta constraints. For $M(t)$,
\begin{align}
    1
    &= \int \mathcal{D}\underline{M}\,\underline{\delta}\!\left(
        M(t)-\frac{1}{S}\sum_i D_{ii} x_i(t)
    \right)\\
    &= \int \mathcal{D}[\underline{\hat{M}},\underline{M}]\,
       \exp\left(
           i S\int\!dt\, \hat{M}(t)\left[
               M(t) - \frac{1}{S}\sum_i D_{ii}x_i(t)
           \right]
       \right).
    \label{eq:identityM}
\end{align}
Similarly, for $C_D(t,t')$,
\begin{align}
    1
    &= \int \mathcal{D}\underline{\underline{C_D}}\,
       \underline{\underline{\delta}}\!\left(
           C_D(t,t') - \frac{1}{S}\sum_i D_{ii}^2 x_i(t)x_i(t')
       \right)
    \nonumber\\
    &= \int \mathcal{D}[\underline{\underline{\hat{C}}},\underline{\underline{C_D}}]\,
       \exp\left(
           i S \int\!dt\,dt'\,\hat{C}(t,t')\left[
               C_D(t,t') - \frac{1}{S}\sum_i D_{ii}^2 x_i(t)x_i(t')
           \right]
       \right),
    \label{eq:identityC}
\end{align}
and for $L(t,t')$,
\begin{align}
    1
    &= \int \mathcal{D}\underline{\underline{L}}\,
       \underline{\underline{\delta}}\!\left(
           L(t,t') - \frac{1}{S}\sum_i \hat{x}_i(t)\hat{x}_i(t')
       \right)
    \nonumber\\
    &= \int \mathcal{D}[\underline{\underline{\hat{L}}},\underline{\underline{L}}]\,
       \exp\left(
           i S \int\!dt\,dt'\,\hat{L}(t,t')\left[
               L(t,t') - \frac{1}{S}\sum_i \hat{x}_i(t)\hat{x}_i(t')
           \right]
       \right).
    \label{eq:identityL}
\end{align}

Here, by $\mathcal{D}[\underline{\underline{\hat{\mathcal{N}}}},\underline{\underline{\mathcal{N}}}]$ we denote a functional measure over fields of two time arguments,
\begin{equation}
    \mathcal{D}[\underline{\underline{\hat{\mathcal{N}}}},\underline{\underline{\mathcal{N}}}]
    = \prod_{t,t'} \frac{d\hat{\mathcal{N}}(t,t')\,d\mathcal{N}(t,t')}{2\pi/S},
\end{equation}
and similarly for fields of a single time argument.

Inserting the identities \eqref{eq:identityM}--\eqref{eq:identityL} into Eq.~\eqref{eq:GF_after_disorder_leading} we obtain a functional integral of the form
\begin{equation}
    \langle Z[\boldsymbol{j}] \rangle
    = \int \mathcal{D}[\hat{M},M]\,
      \mathcal{D}[\hat{C},C_D]\,
      \mathcal{D}[\hat{L},L]\,
      \exp\left( S\,\mathcal{Q}[M,\hat{M},C_D,\hat{C},L,\hat{L}]
                 + \mathcal{O}(1)\right),
    \label{eq:SPZ}
\end{equation}
where the functional $\mathcal{Q}$ naturally splits into three contributions,
\begin{equation}
    \mathcal{Q} = \Psi + \Phi + \Omega.
\end{equation}

\paragraph{(i) Conjugate-order-parameter part \(\Psi\).}

The first contribution collects the direct products of order parameters and their conjugates:
\begin{equation}
    \Psi
    = i\int\!dt\, \hat{M}(t) M(t)
      + i\int\!dt\,dt'\,\big[
          \hat{C}(t,t') C_D(t,t') + \hat{L}(t,t') L(t,t')
      \big].
    \label{eq:Psi}
\end{equation}

\paragraph{(ii) Variance term \(\Phi\).}

The second contribution originates from the variance term proportional to
$\tilde{C}(t,t')L(t,t')$ in Eq.~\eqref{eq:GF_after_disorder_leading}:
\begin{equation}
    \Phi
    = -\frac{\sigma^2}{2}\int\!dt\,dt'\, L(t,t')\,\tilde{C}(t,t').
    \label{eq:Phi}
\end{equation}

\paragraph{(iii) Single-species contribution \(\Omega\).}

Finally, the third term $\Omega$ results from the remaining path integral over the microscopic dynamics of each species, which factorizes over $i$ in the thermodynamic limit:
\begin{align}
    \Omega
    &= \frac{1}{S}\sum_i \log
       \int \mathcal{D}\underline{x}_i\,\mathcal{D}\hat{\underline{x}}_i\,
           p_i(x_i(0))\,
       \exp\left(
           i\int\!dt\, j_i(t)x_i(t)
         + i\int\!dt\, \hat{j}_i(t)\hat{x}_i(t)
       \right)
       \nonumber\\
    &\quad\times
       \exp\left(
           i\int\!dt\,\hat{x}_i(t)\bigg[
               \frac{\dot{x}_i(t)}{x_i(t)} - \big(1 - D_{ii}x_i(t)\big)
           \bigg]
       \right)
       \nonumber\\
    &\quad\times
       \exp\left(
           -i\int\!dt\, \hat{M}(t) D_{ii} x_i(t)
       \right)
       \nonumber\\
    &\quad\times
       \exp\left(
           -i\int\!dt\,dt'\,\Big[
               \hat{C}_D(t,t') D_{ii}^2 x_i(t)x_i(t')
               + \hat{L}(t,t')\hat{x}_i(t)\hat{x}_i(t')
           \Big]
       \right).
    \label{eq:Omega}
\end{align}
Here $p_i(x_i(0))$ denotes the initial condition of species $i$, and we have assumed a factorized initial distribution over species,
$p_0(\boldsymbol{x}(0)) = \prod_i p_i(x_i(0))$.

The crucial point is that each of the three functionals $\Psi$, $\Phi$ and $\Omega$ contains contributions of at least order $\mathcal{O}(S^0)$ in the thermodynamic limit. Therefore the leading behaviour of $\langle Z\rangle$ is controlled by the saddle point of $\mathcal{Q}$.

\subsubsection{Saddle-point equations and interpretation of \(\Omega\)}

We now find the saddle point by taking functional derivatives of $\mathcal{Q}$ with respect to the order parameters and their conjugates, and setting them to zero. This yields a set of self-consistency relations.

From variations with respect to the ``non-hatted'' order parameters we obtain:
\begin{align}
    \frac{\delta\mathcal{Q}}{\delta M(t)} = 0
    &\quad\Rightarrow\quad
    \hat{M}(t) = 2\int\!dt'\, M(t') L(t,t'),
    \label{eq:M_hat_sp}\\[4pt]
    \frac{\delta\mathcal{Q}}{\delta C_D(t,t')} = 0
    &\quad\Rightarrow\quad
    i\hat{C}(t,t') = \frac{\sigma^2}{2} L(t,t'),
    \label{eq:C_hat_sp}\\[4pt]
    \frac{\delta\mathcal{Q}}{\delta L(t,t')} = 0
    &\quad\Rightarrow\quad
    i\hat{L}(t,t') = \frac{\sigma^2}{2} \tilde{C}(t,t').
    \label{eq:L_hat_sp}
\end{align}

On the other hand, variations with respect to the conjugate fields yield:
\begin{align}
    \frac{\delta\mathcal{Q}}{\delta \hat{M}(t)} = 0
    &\quad\Rightarrow\quad
    M(t) = -\lim_{S\to\infty}\frac{\delta\Omega}{\delta\hat{M}(t)},
    \label{eq:M_sp}\\[4pt]
    \frac{\delta\mathcal{Q}}{\delta \hat{C}(t,t')} = 0
    &\quad\Rightarrow\quad
    C(t,t') = -\lim_{S\to\infty} \frac{\delta\Omega}{\delta\hat{C}(t,t')},
    \label{eq:C_sp}\\[4pt]
    \frac{\delta\mathcal{Q}}{\delta \hat{L}(t,t')} = 0
    &\quad\Rightarrow\quad
    L(t,t') = -\lim_{S\to\infty} \frac{\delta\Omega}{\delta\hat{L}(t,t')}.
    \label{eq:L_sp}
\end{align}

Equations \eqref{eq:M_hat_sp}--\eqref{eq:L_sp} show that the conjugate fields $\hat{M}$, $\hat{C}$ and $\hat{L}$ play a role analogous to source fields for the macroscopic order parameters, with $\Omega$ acting as an effective ``free-energy'' functional at the level of single-species trajectories.

Using the saddle-point relations \eqref{eq:M_hat_sp}--\eqref{eq:L_hat_sp} and the definitions of the order parameters \eqref{eq:OrderParameterM}--\eqref{eq:OrderParameterL}, we can write $\Omega$ as
\begin{equation}
    \Omega
    = \frac{1}{S}\sum_i \log Z^{(i)}_{\text{eff}}[\underline{j}_i,\hat{\underline{j}}_i],
\end{equation}
where the effective single-species generating functional is
\begin{align}
    Z^{(i)}_{\text{eff}}[\underline{j}_i,\hat{\underline{j}}_i]
    &= \int \mathcal{D}\underline{x}_i\,\mathcal{D}\hat{\underline{x}}_i\,
       p_i(x_i(0))\,
       \exp\left(
           i\int\!dt\, \hat{x}_i(t)\left[
               \frac{\dot{x}_i(t)}{x_i(t)}
               -\left(1-D_{ii}x_i(t)\right)
           \right]
       \right)
       \nonumber\\
    &\quad\times
       \exp\left(
           i\int\!dt\, D_{ii} j_i(t)x_i(t)
           + i\int\!dt\,\hat{j}_i(t)\hat{x}_i(t)
           - i\int\!dt\, M(t)\hat{x}_i(t)
       \right)
       \nonumber\\
    &\quad\times
       \exp\left(
           -\frac{\sigma^2}{2}\int\!dt\,dt'\,\Big[
               L(t,t') x_i(t)x_i(t')
               + \tilde{C}(t,t')\hat{x}_i(t)\hat{x}_i(t')
           \Big]
       \right).
    \label{eq:Zeff_i}
\end{align}

This expression shows clearly how the effective single-species process is coupled to the macroscopic order parameters.

\subsubsection{Vanishing of pure response correlators and self-consistent order parameters}

The response sources $\hat{j}_i(t)$ have been introduced so that derivatives of $\langle Z\rangle$ with respect to $\hat{j}$ generate moments of the response fields $\hat{x}_i(t)$. However, physical averages contain only the $x_i(t)$; moments involving only response fields are purely auxiliary and, for a properly normalized generating functional, must vanish.

Indeed, imposing that the disorder-averaged generating functional is normalized for all $\hat{\boldsymbol{j}}$,
\begin{equation}
    \overline{Z[\boldsymbol{0},\hat{\boldsymbol{j}}]} = 1,
\end{equation}
implies that
\begin{equation}
    \big\langle \hat{x}_i(t_1)\cdots \hat{x}_i(t_n)\big\rangle_i
    = (-i)^n
      \frac{\delta^n}{\delta\hat{j}_i(t_1)\cdots\delta\hat{j}_i(t_n)}
      \overline{Z[\boldsymbol{0},\hat{\boldsymbol{j}}]}
      \Big|_{\hat{\boldsymbol{j}}=0}
    = 0
\end{equation}
for any $n\ge1$. In particular,
\begin{equation}
    L(t,t') = \frac{1}{S}\sum_i \langle \hat{x}_i(t)\hat{x}_i(t')\rangle_i = 0.
\end{equation}
Thus the saddle-point solution has $L=0$, but from \eqref{eq:L_hat_sp} the conjugate field $\hat{L}$ remains finite and encodes the noise kernel through $\tilde{C}(t,t')$.

Using the effective generating functional \eqref{eq:Zeff_i}, the order parameters can be expressed as averages over the effective process:
\begin{align}
    M(t)
    &= \lim_{S\to\infty}\frac{1}{S}\sum_i
       \big\langle D_{ii} x_i(t)\big\rangle_i,
    \\
    C(t,t')
    &= \lim_{S\to\infty}\frac{1}{S}\sum_i
       \big\langle D_{ii}^2 x_i(t)x_i(t')\big\rangle_i,
    \\
    L(t,t')
    &= \lim_{S\to\infty}\frac{1}{S}\sum_i
       \big\langle \hat{x}_i(t)\hat{x}_i(t')\big\rangle_i = 0.
    \label{eq:OrderZeff}
\end{align}
Here $\langle\cdot\rangle_i$ denotes an average with respect to $Z^{(i)}_{\text{eff}}$.

Equivalently, we can express these quantities as derivatives of $\Omega$ with respect to the sources:
\begin{align}
    -i\frac{\delta\Omega}{\delta j_i(t)}\Big|_{\boldsymbol{j}=\boldsymbol{0}}
        &= M(t),
    \\
    -\frac{\delta^2\Omega}{\delta j_i(t)\delta j_i(t')}\Big|_{\boldsymbol{j}=\boldsymbol{0}}
        &= C(t,t') - M(t)M(t'),
    \\
    -\frac{\delta^2\Omega}{\delta\hat{j}_i(t)\delta\hat{j}_i(t')}\Big|_{\boldsymbol{j}=\boldsymbol{0}}
        &= L(t,t') = 0.
\end{align}

\subsubsection{Effective single-species process}

After integrating out the degrees of freedom and enforcing the saddle-point conditions, the microscopic dynamics of different species become statistically independent. The macroscopic behaviour is fully encoded in the order parameters $M$ and $\tilde{C}$, while the index $i$ becomes irrelevant in the thermodynamic limit. We can therefore define an effective process for a representative species, with measure
\begin{align}
    \mathcal{M}[x,\hat{x}]
    &= p_0(x(0))\,
       \exp\left(
           i\int\!dt\,\hat{x}(t)\left[
               \frac{\dot{x}(t)}{x(t)}
               -\big(1 - D x(t)\big)
               + \hat{j}(t)
           \right]
       \right)
       \nonumber\\
    &\quad\times
       \exp\left(
           -\frac{\sigma^2}{2}\int\!dt\,dt'\,
               \tilde{C}(t,t')\,\hat{x}(t)\hat{x}(t')
       \right),
\end{align}
where $D$ is now a single random variable sampled from the log-normal distribution $\mathcal{P}(D)$ (the distribution of diagonal elements $D_{ii}$). Expectation values with respect to the effective process are defined as
\begin{equation}
    \langle f[\underline{x},\hat{\underline{x}}]\rangle_*
    = \frac{
        \int \mathcal{D}\underline{x}\,\mathcal{D}\hat{\underline{x}}\,
            f[\underline{x},\hat{\underline{x}}]\,
            \mathcal{M}[\underline{x},\hat{\underline{x}}]
    }{
        \int \mathcal{D}\underline{x}\,\mathcal{D}\hat{\underline{x}}\,
            \mathcal{M}[\underline{x},\hat{\underline{x}}]
    }.
\end{equation}

By performing a Hubbard--Stratonovich transformation, the quadratic term in $\hat{x}$ can be written as an average over a Gaussian noise $\eta(t)$ with
\begin{equation}
    \mathbb{E}[\eta(t)] = 0,
    \qquad
    \mathbb{E}[\eta(t)\eta(t')] = \sigma^2 \tilde{C}(t,t').
\end{equation}
Equivalently, we can write the effective measure as
\begin{equation}
    \mathcal{M}[x,\hat{x}]
    \propto
    \left\langle
        \exp\left(
            i\int\ dt\,\hat{x}(t)\left[
                \frac{\dot{x}(t)}{x(t)}
                -\big(1 - D x(t)\big)
                - \eta(t) + \hat{j}(t)
            \right]
        \right)
    \right\rangle_{\eta},
    \label{eq:EffMeasure}
\end{equation}
where the average is over realizations of $\eta$. The self-consistency conditions take the form
\begin{align}
    M(t) &= \langle D x(t)\rangle_*,
    \\
    \sigma^2 \tilde{C}(t,t')
    &= \sigma^2\big(\langle D^2 x(t)x(t')\rangle_* - \langle D x(t)\rangle_*\langle D x(t')\rangle_*\big)
     = \langle \eta(t)\eta(t')\rangle_*.
\end{align}

Finally, \eqref{eq:EffMeasure} shows that the effective dynamics of a representative species is governed by the one-dimensional stochastic differential equation
\begin{equation}
    \dot{x}(t)
    = x(t)\Big[1 - D x(t) + \eta(t) + \hat{j}(t) + \xi(t)\Big],
    \label{eq:Dyeff}
\end{equation}
where $\xi(t)$ denotes the original environmental noise (if present) and $\eta(t)$ is the emergent, self-consistent noise generated by the random interactions. The parameter $D$ is a quenched random variable drawn from the prescribed log-normal distribution of intrinsic rates.

We rename $x\to Dx$ such that we can write the effective process as follows

\begin{equation}
    \dot{x}(t)
    = x(t)\Big[1 - x(t) + \eta(t) + \hat{j}(t) + \xi(t)\Big],
    \label{eq:DyeffE}
\end{equation}
which is a random Generalized Lotka-Volterra effective process with temporal correlations of the effective noise given by $ \langle \eta(t)\eta(t')\rangle_*=\sigma^2\big(\langle  x(t)x(t')\rangle_* - \langle x(t)\rangle_*\langle  x(t')\rangle_*\big)$

\subsection{Static phase: self–consistent fixed--point distribution}
\label{subsec:static_phase}

In the stable phase the DMFT effective process for a representative species 
\begin{equation}
    \dot x(t) = x(t)\big[1 - x(t) + \eta(t)\big],
    \label{eq:eff_process_static_phase}
\end{equation}
where $\eta(t)$ is an effective Gaussian noise with zero mean and two time correlations
\begin{equation}
    \langle \eta(t)\eta(t')\rangle = \sigma^2 \ \tilde C(t,t').
\end{equation}
relax to a stationary distribution.
In the static phase, trajectories for single realizations converge to a fixed point and the effective noise becomes static,
\begin{equation}
    \eta(t) \equiv \eta^* = \sigma\sqrt{q - M^2}\,z,
    \qquad z\sim\mathcal{N}(0,1),
\end{equation}

Here and in the following, $\langle \cdot \rangle$ denotes an average over species and disorder in the thermodynamic limit.

For a fixed realization of $z$, the stationary condition $\dot x=0$ applied to~\eqref{eq:eff_process_static_phase} gives
\begin{equation}
    x^*(z)\Big[1 - x^*(z) + \eta^*(z)\Big] = 0.
\end{equation}
Besides the absorbing solution $x^*(z)=0$, a positive fixed point exists whenever
\begin{equation}
    1 - x^*(z) + \eta^*(z) = 0,
\end{equation}
i.e.
\begin{equation}
    x^*(z) = 1 + \eta^*(z)
    = 1 + \sigma\sqrt{q - M^2} \ z.
\end{equation}
It is convenient to parameterize the noise amplitude in terms of a dimensionless parameter
\begin{equation}
    \Delta \equiv \frac{1}{\sigma\sqrt{q - M^2}}.
    \label{eq:def_Delta}
\end{equation}
Then we can write
\begin{equation}
    x^*(z) =\left(1 + \frac{z}{\Delta}\right)
           = \frac{\Delta + z}{\Delta}.
    \label{eq:xz_fixed_point}
\end{equation}

The positivity constraint $x^*(z)>0$ implies
\begin{equation}
    1 + \frac{z}{\Delta} > 0
    \quad\Longleftrightarrow\quad
    z > -\Delta.
\end{equation}
Therefore, only those realizations of the static noise with $z>-\Delta$ correspond to surviving species; those with $z\leq -\Delta$ fall into the absorbing state $x^*=0$.

Let us obtain the distribution of $x^*$ induced by the Gaussian variable $z\sim\mathcal{N}(0,1)$. Denote the standard normal measure by
\begin{equation}
    \mathcal{D}z = \frac{dz}{\sqrt{2\pi}}\,\mathrm{e}^{-z^2/2}.
\end{equation}
For a given $D>0$ we define the following local moments of the stationary abundance:
\begin{align}
    \phi &= \int_{-\Delta}^{\infty} \mathcal{D}z,
            \label{eq:phiD_def}\\
    M &= \int_{-\Delta}^{\infty} \mathcal{D}z\,x^*(z),
         \label{eq:mD_def}\\
    q &=  \int_{-\Delta}^{\infty} \mathcal{D}z\,\big[x^*(z)\big]^2.
         \label{eq:qD_def}
\end{align}
Here $\phi$ is the survival probability (the fraction of species with $x^*>0$), while $M$ and $q$ are the first and second unweighted moments of $x^*$ among surviving species (unconditional on extinction, i.e. zero contribution for extinct ones).

Using~\eqref{eq:xz_fixed_point}, these integrals can be evaluated in closed form. Introduce the standard Gaussian pdf and cdf
\begin{equation}
    a_1(\Delta) = \frac{1}{\sqrt{2\pi}} \mathrm{e}^{-\Delta^2/2},
    \qquad
    a_0(\Delta) = \int_{-\infty}^{\Delta}\mathcal{D}z = \frac{1}{2}\left[1 + \operatorname{erf}\!\left(\frac{\Delta}{\sqrt{2}}\right)\right].
\end{equation}
Then
\begin{equation}
    \phi= \int_{-\Delta}^{\infty} \mathcal{D}z = a_0(\Delta).
    \label{eq:phiD_closed}
\end{equation}

For the first moment we obtain
\begin{align}
    M
    &= \int_{-\Delta}^{\infty} \mathcal{D}z\,\frac{\Delta + z}{\Delta} \\
    &= \frac{1}{\Delta}\left[
        \int_{-\Delta}^{\infty} \mathcal{D}z\,z
        + \Delta \int_{-\Delta}^{\infty} \mathcal{D}z
    \right]\equiv \Delta^{-1}w_1.
\end{align}
Using the standard Gaussian identities we find
\begin{equation}
    \int_{-\Delta}^{\infty} \mathcal{D}z = a_0(\Delta),\qquad
    \int_{-\Delta}^{\infty} \mathcal{D}z\,z = a_1(\Delta).
\end{equation}
Therefore
\begin{equation}
    M= \Delta^{-1}\left(a_0(\Delta) + \Delta\,a_1(\Delta)\right).
    \label{eq:mD_closed}
\end{equation}

For the second moment we proceed analogously:
\begin{align}
    q
    &= \int_{-\Delta}^{\infty} \mathcal{D}z\,\left(\frac{\Delta + z}{\Delta}\right)^2
     = \frac{1}{\Delta^2} \int_{-\Delta}^{\infty} \mathcal{D}z\,\big(z^2 + 2\Delta z + \Delta^2\big).
\end{align}
The remaining Gaussian integrals are

\begin{equation}
    \int_{-\Delta}^{\infty}\mathcal{D}z\,z^2 = a_0(\Delta) - \Delta\,a_1(\Delta).
\end{equation}
Combining everything,
\begin{align}
    \int_{-\Delta}^{\infty}\mathcal{D}z\,(z^2 + 2\Delta z + \Delta^2)
    &= \big[a_0(\Delta) - \Delta a_1(\Delta)\big]
       + 2\Delta\, a_1(\Delta)
       + \Delta^2\,a_0(\Delta) \\
    &= (1+\Delta^2)\,a_0(\Delta) + \Delta\,a_1(\Delta).
\end{align}
Hence
\begin{equation}
    q= \Delta^{-2}\left[(1+\Delta^2)\,a_0(\Delta) + \Delta\,a_1(\Delta)\right]\equiv \Delta^{-2}w_2.
    \label{eq:qD_closed}
\end{equation}

Equations~\eqref{eq:phiD_closed}, \eqref{eq:mD_closed} and~\eqref{eq:qD_closed} express the survival fraction and the first two moments of the abundance fixed point for a given $D$ in terms of the single dimensionless parameter $\Delta$ and the error function. All together

\begin{align}
    &\phi= a_0(\Delta) \\
    &M = \sigma\sqrt{q-M^2} \ \left(a_1(\Delta)+\Delta a_0(\Delta)\right)\equiv \sigma \sqrt{q-M^2} \ w_1 \label{eq:OrderStable2}\\
    & q= \sigma^2\left(q-M^2\right)\left[ (1+\Delta^2)\,a_0(\Delta) + \Delta\,a_1(\Delta)  \right]\equiv \sigma^2\left(q-M^2\right)w_2\label{eq:OrderStable3}
\end{align}

which is a closed system of equations. 

We want to show that in the stable phase the unique possible solutions that have finite abundance are restricted by the condition $\Delta^{-1}=0$. This implies $q=M^2$ and $\phi=1$ for $\sigma>0$. 

Subtracting the square of \eqref{eq:OrderStable2} from \eqref{eq:OrderStable3} leads to the condition

\begin{align}
    \left(q-M^2\right)\left[1-\sigma^2\left(w_2-w^2_1\right)\right]=0\label{eq: ConditionSelf}
\end{align}

Assume that $q>M^2$ and $1=\sigma^2\left(w_2-w^2_1\right)$, where $w_2,w_1$ are functions of $\Delta\geq 0$. However, for $\sigma<1$ the function $g(\Delta)=w_2(\Delta)-w_1(\Delta)^2-\frac{1}{\sigma^2}$ does not cross the x-axis; for $\sigma=1$ it tends to zero when $\Delta\to \infty$; and for $\sigma>1$ it has a finite $\Delta>0$ where it crosses. This means that for $\sigma\leq 1$ the stable phase is characterized by the solution $q-M^2=0$ in Eq. \ref{eq: ConditionSelf}, and whenever $\sigma>1$ there exists a solution where $\infty>\sigma\Delta^{-1}=q-M^2>0$. We show in the next section, the regime where $\sigma>1$ is already unstable. This leaves us with the unique choice $\sigma<1$ that characterizes the stable phase. Then, in all this phase it is verified that $\Delta^{-1}=0$, which implies that $q=M^2$ and the effective noise vanishes. Thus, the zero row-sum property of the interaction matrix $M$ implies that the effective noise vanishes in the thermodynamic limit, which, as desired, keeps the fixed points to be the inverse of the log-normal diagonal matrix $D$. 

\subsection{Grilli's laws in the stable phase} \label{Sec: Grilli DMFT}

As shown in the previous section, the stable phase is characterized by the condition $\Delta^{-1}=0$. Now, we add the environmental noise to the equation. For a small amplitude of the noise, the effect of it is that trajectories would fluctuate around fixed point solutions. Then condition $\Delta^{-1}=0$ still holds at first order approximation, this makes the effective process to be the following stochastic differential equation with multiplicative noise

\begin{align}
    \dot{x}(t) = x(t)\left(1-x(t) + \sigma_0 \xi(t) \right)
\end{align}
where $\xi(t)$ is white noise with variance $\sigma_0^2$. That is, the original interacting dynamics reduces in the thermodynamic limit due to the zero row-sum property to a decoupled single stochastic differential equation, encapsulating Grilli's decoupled logistic model and its consequent Grilli's laws for microbial communities. However, our model is dynamically richer: when the effective noise does not vanish, the model is able to exhibit chaotic behavior, a property that the decoupled model does not have.  

\subsection{Stability Analysis}

The next step is to deduce the combination of parameters for the instability line: the line that delimits when the dynamics goes from the stable phase to the chaotic phase. 

Let us assume that we run the dynamics and the system relaxes towards the static solution $(x^*,\eta^*)$, then we introduce a small perturbation and quantify if it decays towards the fixed point solution or not.

\begin{align}
    \dot{x} = x\left( 1-x + \eta +\epsilon \,\xi \right)
\end{align}
where $\epsilon$ is the small parameter that controls the intensity of the perturbation and $\xi$ is a white noise with unit variance and zero mean. 
That is, we define time dependent variables around the fixed point solution at order $\epsilon$ 

\begin{align}
    &\eta(t)= \eta^* + \epsilon \,\nu(t) \\ &
    x(t)= x^* + \epsilon \, y(t)
\end{align}

Linearizing the dynamics, we obtain
\begin{align}
    & O(\epsilon^0): \quad x^*\left(1-x^*+\eta^*\right)=0\\
    & O(\epsilon): \quad \dot{y} = x^*\left(-y+\nu+\xi\right)+ y\left(1-x^*+\eta^*\right) . 
\end{align}
In the stable phase, we know that we reach feasible solutions characterized by $x^*\neq 0$, then $1-x^*+\eta^*=0$ and the previous relation at order $\epsilon$ reduces to the equation

\begin{align}
    \dot{y} = x^*\left(-y+\nu+\xi\right) .
\end{align}

In Fourier space, we can write the first order differential equation as follows
\begin{align}
    \frac{i\omega}{x^*}\tilde{y}(\omega)=-\tilde{y}(\omega)+\tilde{\nu}(\omega)+\tilde{\xi}(\omega)\label{eq: FourierDyn}
\end{align}
where we denoted by tilded terms the corresponding Fourier transforms. 

We need relations between the moments of the dynamic variable $y$ and the effective noise $\nu$. Recurring to the self-consistent relations we obtain the following

\begin{align}
    & \langle \eta(t)\eta(t')\rangle =  \langle \eta^*\eta^*\rangle + \epsilon\left( \langle \eta^*\nu(t)\rangle + \langle \eta^* \nu(t')\rangle \right) + \epsilon^2 \langle \nu(t)\nu(t')\rangle\\
    & \langle x(t)x(t') \rangle = q + \epsilon \left(\langle x^* y(t') \rangle + \langle x^*y(t) \rangle\right) + \epsilon^2 \langle y(t) y(t') \rangle 
\end{align}
both are linked by the relation
\begin{align}
    \langle \eta(t)\eta(t')\rangle = \sigma^2\left[ \langle x(t)x(t')\rangle - \langle x(t)\rangle \langle x(t') \rangle  \right]
\end{align}

where averages are taken with respect to the static distribution and the effective process $\nu$ after the perturbation applied. Then

\begin{align}
    & O(\epsilon^0): \quad \langle \eta^*\eta^* \rangle = \sigma^2\left(q-M^2\right) \\
    & O(\epsilon): \quad \langle \eta^*\nu(t)\rangle + \langle \eta^* \nu(t')\rangle = \sigma^2\left[ \langle x^* y(t') \rangle-\langle x^*\rangle \langle y(t') \rangle + \langle x^*y(t) \rangle- \langle x^*\rangle \langle y(t)\rangle \right]\\
    & O(\epsilon^2):\quad \langle \nu(t)\nu(t')\rangle= \sigma^2\left(\langle y(t) y(t') \rangle -\langle y(t)\rangle \langle y(t')\rangle \right)
\end{align}

Computing $\langle |\tilde{y}(\omega)|^2\rangle$ from \eqref{eq: FourierDyn}, taking averages, using the previous self-consistent relations, and evaluating at $\omega=0$ we obtain the following result
\begin{align}
    \langle |\tilde{y}(0)|^2\rangle = \phi \left[\sigma^2\left(\langle |\tilde{y}(0)|^2\rangle-|\langle \tilde{y}(0)\rangle|^2\right)+1 \right]. \label{eq: RelationFourier}
\end{align}

From \eqref{eq: FourierDyn} we can directly take averages to show that $\langle \tilde{y}(0)\rangle =0$ since $\langle \nu\rangle = \langle \xi\rangle = 0$. Thus, clearing $\langle |\tilde{y}(0)|^2\rangle$ from \eqref{eq: RelationFourier}, we get
\begin{align}
    \langle |\tilde{y}(0)|^2\rangle = \frac{\phi}{1-\phi\sigma^2}.
\end{align}
For $1-\phi\sigma^2>0$ the system is in the stable phase, whereas for $1-\phi\sigma^2<0$ the term $\langle |\tilde{y}(0)|^2\rangle<0$ leads to an incongruence and therefore signals the unstable phase.


\section{Inferring the distance to the edge of chaos}
In the main text, we linked the distance to the edge of chaos to the width of the correlation histogram, following previous works \cite{dahmen_second_2019}. Here, we give a simple derivation for this methodology.

\subsection{The Ornstein-Uhlenbeck process}
In section \ref{section: linear app}, we proved that the population dynamics could be expressed (in log-space and after linearization) as a set of linear stochastic differential equations, an Ornstein-Uhlenbeck process \cite{uhlenbeck_theory_1930}. In vector form, this set of equations can be expressed as:
\begin{equation}
    \partial_t \mathbf{x}(t) = - \frac{k}{\tau}(\text{I} - \sigma\text{G})\,\mathbf{x}(t) + \sigma_0\boldsymbol{\xi}(t),
\end{equation}
where $G=M\cdot\diag(x^*)$, c.f~\eqref{eq: effective A}. Define a new time variable
\begin{equation}
    t' = \frac{k}{\tau}\,t
    \qquad\Rightarrow\qquad
    \partial_t = \frac{k}{\tau}\,\partial_{t'}.
\end{equation}
In terms of $t'$ the equation becomes
\begin{equation}
    \frac{k}{\tau}\,\partial_{t'} \mathbf{x}(t')
    = - \frac{k}{\tau}\bigl(I - \sigma G\bigr)\,\mathbf{x}(t')
      + \sigma_0\,\boldsymbol{\xi}(t),
\end{equation}
and dividing both sides by $k/\tau$ we obtain
\begin{equation}
    \partial_{t'} \mathbf{x}(t')
    = -\bigl(I - \sigma G\bigr)\,\mathbf{x}(t')
      + \frac{\tau}{k}\,\sigma_0\,\boldsymbol{\xi}(t).
\end{equation}
We now introduce a rescaled noise
\begin{equation}
    \boldsymbol{\eta}(t') \equiv \sqrt{\frac{\tau}{k}}\,\boldsymbol{\xi}(t),
\end{equation}
which satisfies
\begin{equation}
    \langle \eta_i(t') \rangle = 0,
    \qquad
    \langle \eta_i(t')\eta_j(s')\rangle = \delta_{ij}\,\delta(t'-s').
\end{equation}
In terms of $\boldsymbol{\eta}$, the dynamics reads
\begin{equation}
    \partial_{t'} \mathbf{x}(t')
    = -\bigl(I - \sigma G\bigr)\,\mathbf{x}(t')
      + \sigma_0 \sqrt{\frac{\tau}{k}}\,\boldsymbol{\eta}(t'),
\end{equation}
which is again an Ornstein--Uhlenbeck process, with an effective noise amplitude
\begin{equation}
    D = \sigma_0^2\,\frac{\tau}{k}.
\end{equation}
In what follows, in order to make notation more compact, we drop the primes to denote the time dilation, and just focus on the set of linear dynamics:
\begin{equation}
    \partial_{t} \mathbf{x}(t)
    = -\bigl(I - \sigma G\bigr)\,\mathbf{x}(t)
      + \sqrt{D}\,\boldsymbol{\xi}(t).
      \label{eq:dynamics_original}
\end{equation}
The aim of this section is to  derive the Dynamic Mean-Field expression for this set of linear dynamics.
\subsection{Change of basis to transversal and longitudinal modes}
It is convenient to work in a basis that is aligned with the homogeneous direction (i.e., the vector $\mathbf{1}=(1,...,1)$). To this end, we introduce an orthogonal matrix
$U \in \mathbb{R}^{S\times S}$,
\begin{equation}
    U^\top U = U U^\top = I,
\end{equation}
whose first column coincides with the normalized homogeneous vector,
\begin{equation}
    U_{i0} = \frac{1}{\sqrt{S}}
    \qquad \text{for all } i = 1,\dots,S.
\end{equation}
The remaining columns of $U$ form an orthonormal basis of the subspace
orthogonal to the homogeneous vector.

We then define the rotated variables
\begin{equation}
    \mathbf{X}(t) = U^\top \mathbf{x}(t),
    \qquad
    \boldsymbol{\xi}'(t) = U^\top \boldsymbol{\xi}(t),
    \qquad
    G' = U^\top G U .
\end{equation}
Because $U$ is orthogonal, the transformed noise
$\boldsymbol{\xi}'(t)$ has the same statistics as $\boldsymbol{\xi}(t)$,
\begin{equation}
    \bigl\langle \xi'_a(t)\,\xi'_b(s) \bigr\rangle
    = \delta_{ab}\,\delta(t-s),
    \qquad a,b = 0,\dots,S-1.
\end{equation}
Multiplying \eqref{eq:dynamics_original} on the left by $U^\top$,
and using the definitions above, we obtain the dynamics in the rotated basis,
\begin{equation}
    \partial_t \mathbf{X}(t)
    = -\bigl(I - \sigma G'\bigr)\,\mathbf{X}(t)
      + D\,\boldsymbol{\xi}'(t),
    \label{eq:dynamics_rotated}
\end{equation}
or, in components,
\begin{equation}
    \partial_t X_a(t)
    = - X_a(t)
      + \sigma \sum_{b=0}^{S-1} G'_{ab}\,X_b(t)
      + \sqrt{D}\,\xi'_a(t),
    \qquad a = 0,\dots,S-1.
    \label{eq: dynamics rotated}
\end{equation}
Here $a=0$ labels the longitudinal (homogeneous) mode, while
$a=1,\dots,S-1$ label the transverse modes.
\subsection{Moment generating functional for the linear dynamics}

We now formulate the dynamics \eqref{eq:dynamics_rotated} within the
Martin--Siggia--Rose--De~Dominicis--Janssen (MSRDJ) path–integral
formalism, which yields a generating functional of the
form:
\begin{equation}
    Z[\mathbf{j},\tilde{\mathbf{j}}]
    = \int \mathcal{D}\mathbf{X}\,\mathcal{D}\tilde{\mathbf{X}}\;
      \exp\Big(
        - S[\mathbf{X},\tilde{\mathbf{X}}]
        + \sum_{a=0}^{S-1}\int\!dt\,
          \big[ j_a(t)\,X_a(t) + \tilde{j}_a(t)\,\tilde{X}_a(t)\big]
      \Big),
    \label{eq:Z_MSRDJ}
\end{equation}
where the MSRDJ action $S$ reads
\begin{equation}
    S[\mathbf{X},\tilde{\mathbf{X}}]
    = \sum_{a=0}^{S-1}\int\!dt\,
      \bigg[
        \tilde{X}_a(t)\,
        \Big(
          \partial_t X_a(t)
          + X_a(t)
          - \sigma \sum_{b=0}^{S-1} G'_{ab}\,X_b(t)
        \Big)
        - \frac{D}{2}\,\tilde{X}_a(t)^2
      \bigg].
    \label{eq:action_MSRDJ_rotated}
\end{equation}
It is convenient to separate the bare (uncoupled) OU part of the action
from the interaction mediated by $G'$. We write
\begin{equation}
    S[\mathbf{X},\tilde{\mathbf{X}}]
    = S_0[\mathbf{X},\tilde{\mathbf{X}}]
      - S_{\mathrm{int}}[\mathbf{X},\tilde{\mathbf{X}}],
\end{equation}
where
\begin{equation}
    S_0[\mathbf{X},\tilde{\mathbf{X}}]
    = \sum_{a=0}^{S-1}\int\!dt\,
      \bigg[
        \tilde{X}_a(t)\big(\partial_t X_a(t) + X_a(t)\big)
        - \frac{D}{2}\,\tilde{X}_a(t)^2
      \bigg],
\end{equation}
and
\begin{equation}
    S_{\mathrm{int}}[\mathbf{X},\tilde{\mathbf{X}}]
    = \sigma \sum_{a,b=0}^{S-1}\int\!dt\,
      \tilde{X}_a(t)\,G'_{ab}\,X_b(t).
\end{equation}
The interaction term $S_{\mathrm{int}}$ contains the quenched disorder
through the random matrix $G'$, while $S_0$ describes $S$ independent
Ornstein--Uhlenbeck processes (one longitudinal and $S-1$ transverse
modes) in the absence of coupling.

\subsection{Average over the quenched disorder}
We now perform the average of the generating functional over the quenched
disorder in the connectivity. The random matrix $G$ is taken to have
independent Gaussian entries with zero mean and variance
$\mathrm{Var}(G_{ij}) = 1/S$, together with the constraint
$G \mathbf{1} = 0$, where $\mathbf{1}$ is the homogeneous vector. In the
rotated basis defined above, this constraint implies that the first
column of $G'$ vanishes,
\begin{equation}
    G'_{a0} = 0
    \qquad\text{for all } a=0,\dots,S-1,
\end{equation}
while the remaining entries
\begin{equation}
    G'_{\alpha\beta},\quad G'_{0\beta},
    \qquad \alpha,\beta = 1,\dots,S-1,
\end{equation}
are independent Gaussian random variables with
\begin{equation}
    \langle G'_{ab} \rangle = 0,
    \qquad
    \langle G'_{ab} G'_{cd} \rangle
    = \frac{1}{S}\,\delta_{ac}\,\delta_{bd}.
\end{equation}

The disorder-averaged generating functional is defined as
\begin{equation}
    \overline{Z[\mathbf{j},\tilde{\mathbf{j}}]}
    = \int \mathcal{D}G'\,P(G')\,
      Z[\mathbf{j},\tilde{\mathbf{j}}\,|\,G'],
\end{equation}
where
\begin{equation}
    P(G') \propto
    \exp\!\bigg[
        -\frac{S}{2}
        \sum_{a,b=0}^{S-1} G_{ab}'^{\,2}
    \bigg]
\end{equation}
implements the Gaussian distribution with variance $1/S$, and
$Z[\mathbf{j},\tilde{\mathbf{j}}\,|\,G']$ is the MSRDJ functional
\eqref{eq:Z_MSRDJ} for fixed $G'$.

It is convenient to split the rotated degrees of freedom into the
longitudinal mode $X_0$ and the $S-1$ transverse modes
$X_\alpha \equiv y_\alpha$, $\alpha = 1,\dots,S-1$. In this notation,
the interaction part of the MSRDJ action reads
\begin{equation}
    S_{\mathrm{int}}[\mathbf{X},\tilde{\mathbf{X}}]
    = \sigma \sum_{a,b=0}^{S-1}\int\!dt\,
      \tilde{X}_a(t)\,G'_{ab}\,X_b(t)
    = \sigma \int\!dt\,
      \biggl[
        \sum_{\alpha,\beta=1}^{S-1}
        \tilde{y}_\alpha(t)\,G'_{\alpha\beta}\,y_\beta(t)
        + \sum_{\beta=1}^{S-1}
        \tilde{X}_0(t)\,G'_{0\beta}\,y_\beta(t)
      \biggr],
\end{equation}
where the terms $G'_{a0}$ are absent because the first column of $G'$ is
identically zero.

The dependence of the generating functional on the quenched disorder
appears only through $S_{\mathrm{int}}$. The disorder average is thus a
Gaussian integral over the entries $G'_{\alpha\beta}$ and $G'_{0\beta}$:
\begin{equation}
    \overline{Z[\mathbf{j},\tilde{\mathbf{j}}]}
    =
    \int\mathcal{D}\mathbf{X}\,\mathcal{D}\tilde{\mathbf{X}}\;
    \exp\!\big( - S_0[\mathbf{X},\tilde{\mathbf{X}}]
                 + \mathbf{j}\cdot\mathbf{X}
                 + \tilde{\mathbf{j}}\cdot\tilde{\mathbf{X}} \big)\;
    \overline{\exp\big(-S_{\mathrm{int}}[\mathbf{X},\tilde{\mathbf{X}}]\big)}_{G'}.
\end{equation}
For a generic Gaussian variable $G$ with variance $1/S$ one has
\begin{equation}
    \int dG\,
    \exp\bigg( -\frac{S}{2}\,G^2 + \sigma G A \bigg)
    \propto
    \exp\bigg( \frac{\sigma^2}{2S}\,A^2 \bigg),
\end{equation}
so that the average over $G'_{\alpha\beta}$ and $G'_{0\beta}$ yields
quartic terms in the dynamical fields. Introducing the shorthand
\begin{equation}
    A_{\alpha\beta}
    := \int\!dt\,\tilde{y}_\alpha(t)\,y_\beta(t),
    \qquad
    A_{0\beta}
    := \int\!dt\,\tilde{X}_0(t)\,y_\beta(t),
\end{equation}
the average over the transverse--transverse block gives
\begin{equation}
    \overline{\exp\bigg[
        -\sigma \sum_{\alpha,\beta}\int\!dt\,
                 \tilde{y}_\alpha(t)\,G'_{\alpha\beta}\,y_\beta(t)
    \bigg]}_{G'_{\alpha\beta}}
    =
    \exp\bigg[
        \frac{\sigma^2}{2S}\sum_{\alpha,\beta=1}^{S-1} A_{\alpha\beta}^2
    \bigg],
\end{equation}
and the average over the longitudinal row produces
\begin{equation}
    \overline{\exp\bigg[
        -\sigma \sum_{\beta=1}^{S-1}\int\!dt\,
                 \tilde{X}_0(t)\,G'_{0\beta}\,y_\beta(t)
    \bigg]}_{G'_{0\beta}}
    =
    \exp\bigg[
        \frac{\sigma^2}{2S}\sum_{\beta=1}^{S-1} A_{0\beta}^2
    \bigg].
\end{equation}

The sums over indices can be expressed in terms of collective fields.
Writing
\begin{equation}
    \sum_{\alpha,\beta} A_{\alpha\beta}^2
    =
    \sum_{\alpha,\beta}\int\!dt\,ds\,
    \tilde{y}_\alpha(t)\tilde{y}_\alpha(s)\,
    y_\beta(t)y_\beta(s),
\end{equation}
and defining the transverse order parameters
\begin{equation}
    Q(t,s)
    := \frac{1}{S-1}\sum_{\alpha=1}^{S-1}
        \tilde{y}_\alpha(t)\,\tilde{y}_\alpha(s),
    \qquad
    C(t,s)
    := \frac{1}{S-1}\sum_{\beta=1}^{S-1}
        y_\beta(t)\,y_\beta(s),
\end{equation}
one obtains
\begin{equation}
    \sum_{\alpha,\beta} A_{\alpha\beta}^2
    = (S-1)^2 \int\!dt\,ds\,
      Q(t,s)\,C(t,s).
\end{equation}
Similarly,
\begin{equation}
    \sum_{\beta} A_{0\beta}^2
    =
    \sum_{\beta}\int\!dt\,ds\,
    \tilde{X}_0(t)\tilde{X}_0(s)\,y_\beta(t)y_\beta(s)
    = (S-1)\int\!dt\,ds\,
      \tilde{X}_0(t)\tilde{X}_0(s)\,C(t,s).
\end{equation}

Putting everything together, the quenched-disorder averaged generating
functional can be written as
\begin{equation}
\begin{aligned}
    \overline{Z[\mathbf{j},\tilde{\mathbf{j}}]}
    &=
    \int \mathcal{D}y\,\mathcal{D}\tilde{y}\,
          \mathcal{D}X_0\,\mathcal{D}\tilde{X}_0\;
    \exp\Bigg\{
        - S_0[y,\tilde{y}] - S_0[X_0,\tilde{X}_0]\\
        &\quad
        + \sum_{\alpha=1}^{S-1}\int\!dt\,
          \big[j_\alpha(t)\,y_\alpha(t)
              + \tilde{j}_\alpha(t)\,\tilde{y}_\alpha(t)\big]
        + \int\!dt\,
          \big[j_0(t)\,X_0(t)
              + \tilde{j}_0(t)\,\tilde{X}_0(t)\big]\\
        &\quad
        + \frac{\sigma^2}{2S}(S-1)^2
          \int\!dt\,ds\,Q(t,s)\,C(t,s)
        + \frac{\sigma^2}{2S}(S-1)
          \int\!dt\,ds\,
          \tilde{X}_0(t)\tilde{X}_0(s)\,C(t,s)
    \Bigg\},
\end{aligned}
\end{equation}
where $S_0[y,\tilde{y}]$ and $S_0[X_0,\tilde{X}_0]$ denote the bare
Ornstein--Uhlenbeck actions (without couplings) for the transverse and
longitudinal modes, respectively. The quenched average has generated an
effective quartic interaction between response and activity fields,
mediated by the collective correlators $Q(t,s)$ and $C(t,s)$.

\subsection{HS transformation}
We now introduce a Hubbard--Stratonovich (HS) transformation in order to
decouple the quartic interaction generated by the quenched disorder.
The quenched-averaged generating functional can be written as
\begin{equation}
\begin{aligned}
    \overline{Z[\mathbf{j},\tilde{\mathbf{j}}]}
    &=
    \int \mathcal{D}y\,\mathcal{D}\tilde{y}\,
          \mathcal{D}X_0\,\mathcal{D}\tilde{X}_0\;
    \exp\Bigg\{
        - S_0[y,\tilde{y}] - S_0[X_0,\tilde{X}_0]\\
        &\quad
        + \sum_{\alpha=1}^{S-1}\int\!dt\,
          \big[j_\alpha(t)\,y_\alpha(t)
              + \tilde{j}_\alpha(t)\,\tilde{y}_\alpha(t)\big]
        + \int\!dt\,
          \big[j_0(t)\,X_0(t)
              + \tilde{j}_0(t)\,\tilde{X}_0(t)\big]\\
        &\quad
        + \frac{\sigma^2}{2S}(S-1)^2
          \int\!dt\,ds\,Q(t,s)\,C(t,s)
        + \frac{\sigma^2}{2S}(S-1)
          \int\!dt\,ds\,
          \tilde{X}_0(t)\tilde{X}_0(s)\,C(t,s)
    \Bigg\},
\end{aligned}
\label{eq:Z_bar_QC}
\end{equation}
where
\begin{equation}
    Q(t,s)
    = \frac{1}{S-1}\sum_{\alpha=1}^{S-1}
      \tilde{y}_\alpha(t)\,\tilde{y}_\alpha(s),
    \qquad
    C(t,s)
    = \frac{1}{S-1}\sum_{\beta=1}^{S-1}
      y_\beta(t)\,y_\beta(s).
\end{equation}
The quartic interaction term proportional to $Q(t,s)\,C(t,s)$ couples
all transverse degrees of freedom through the empirical correlators. To
obtain a single-site description, we introduce a collective field
$C(t,s)$ together with a conjugate field $\tilde{C}(t,s)$, and enforce
the definition of $C(t,s)$ via a functional constraint.

We insert into \eqref{eq:Z_bar_QC} the following functional identity
\begin{equation}
\begin{aligned}
    1
    &=
    \int \mathcal{D}C\,\mathcal{D}\tilde{C}\;
    \exp\Bigg\{
        -\frac{(S-1)^2}{\sigma^2 S}
         \int\!dt\,ds\,C(t,s)\,\tilde{C}(t,s)\\
        &\qquad\qquad\quad
        + (S-1)\int\!dt\,ds\,
          \tilde{C}(t,s)\,
          \frac{1}{S-1}\sum_{\beta=1}^{S-1}
          y_\beta(t)\,y_\beta(s)
    \Bigg\},
\end{aligned}
\label{eq:HS_identity}
\end{equation}
which plays the role of a functional $\delta$-constraint selecting
$C(t,s)$ as the empirical two-time correlator of the transverse
variables. Up to an overall (irrelevant) normalization, the factor
\eqref{eq:HS_identity} implements
\begin{equation}
    C(t,s)
    = \frac{1}{S-1}\sum_{\beta=1}^{S-1} y_\beta(t)\,y_\beta(s)
\end{equation}
at the saddle point. The particular prefactor $(S-1)^2/(\sigma^2 S)$ is
chosen so that the resulting effective action acquires the standard
dynamical mean-field structure.

Upon inserting \eqref{eq:HS_identity} into \eqref{eq:Z_bar_QC} and
collecting all terms depending on $C$ and $\tilde{C}$, the quenched
average takes the form
\begin{equation}
\begin{aligned}
    \overline{Z[\mathbf{j},\tilde{\mathbf{j}}]}
    &=
    \int \mathcal{D}C\,\mathcal{D}\tilde{C}\;
    \exp\Bigg\{
        - \frac{(S-1)^2}{\sigma^2 S}
          \int\!dt\,ds\,C(t,s)\,\tilde{C}(t,s)
    \Bigg\}\\
    &\quad\times
    \int \mathcal{D}y\,\mathcal{D}\tilde{y}\,
          \mathcal{D}X_0\,\mathcal{D}\tilde{X}_0\;
    \exp\Bigg\{
        - S_0[y,\tilde{y}] - S_0[X_0,\tilde{X}_0]\\
        &\qquad
        + \sum_{\alpha=1}^{S-1}\int\!dt\,
          \big[j_\alpha(t)\,y_\alpha(t)
              + \tilde{j}_\alpha(t)\,\tilde{y}_\alpha(t)\big]
        + \int\!dt\,
          \big[j_0(t)\,X_0(t)
              + \tilde{j}_0(t)\,\tilde{X}_0(t)\big]\\
        &\qquad
        + \frac{\sigma^2}{2S}(S-1)^2
          \int\!dt\,ds\,Q(t,s)\,C(t,s)\\
        &\qquad
        + \frac{\sigma^2}{2S}(S-1)
          \int\!dt\,ds\,
          \tilde{X}_0(t)\tilde{X}_0(s)\,C(t,s)\\
        &\qquad
        + (S-1)\int\!dt\,ds\,
          \tilde{C}(t,s)\,
          \frac{1}{S-1}\sum_{\beta=1}^{S-1}
          y_\beta(t)\,y_\beta(s)
    \Bigg\}.
\end{aligned}
\label{eq:Z_with_C_tildeC}
\end{equation}
The exponent in \eqref{eq:Z_with_C_tildeC} is now \emph{quadratic} in
each transverse degree of freedom $(y_\alpha,\tilde{y}_\alpha)$, so that
the path integral over the transverse modes factorizes into a product of
identical single-site contributions. Writing the quadratic terms in
$\tilde{y}_\alpha$ and $y_\alpha$ explicitly, one finds that the
transverse part of the action can be rewritten as
\begin{equation}
\begin{aligned}
    S_{\mathrm{trans}}[y,\tilde{y};C,\tilde{C}]
    &=
    \sum_{\alpha=1}^{S-1}\int\!dt\,
    \bigg[
        \tilde{y}_\alpha(t)\big(\partial_t y_\alpha(t) + y_\alpha(t)\big)
        - \frac{D}{2}\,\tilde{y}_\alpha(t)^2
    \bigg]\\
    &\quad
    - \frac{1}{2}\sum_{\alpha=1}^{S-1}
      \int\!dt\,ds\,
      \tilde{y}_\alpha(t)\,C(t,s)\,\tilde{y}_\alpha(s)\\
    &\quad
    - \sum_{\alpha=1}^{S-1}
      \int\!dt\,ds\,
      y_\alpha(t)\,\tilde{C}(t,s)\,y_\alpha(s),
\end{aligned}
\end{equation}
so that the transverse contribution to
\eqref{eq:Z_with_C_tildeC} can be written as
\begin{equation}
\begin{aligned}
    &\int \mathcal{D}y\,\mathcal{D}\tilde{y}\;
    \exp\Big(
        - S_0[y,\tilde{y}]
        + \frac{1}{2}\sum_{\alpha=1}^{S-1}\tilde{y}_\alpha C \tilde{y}_\alpha
        + \sum_{\alpha=1}^{S-1} y_\alpha \tilde{C} y_\alpha
        + \dots
    \Big)\\
    &\qquad
    =
    \bigg[
      \int \mathcal{D}y\,\mathcal{D}\tilde{y}\;
      \exp\Big(
          - S_0[y,\tilde{y}]
          + \frac{1}{2}\,\tilde{y} C \tilde{y}
          + y \tilde{C} y
      \Big)
    \bigg]^{S-1},
\end{aligned}
\end{equation}
where in the last line $(y,\tilde{y})$ denote a single transverse mode,
and we have omitted the source terms for brevity. The longitudinal mode
contributes a separate factor
\begin{equation}
\begin{aligned}
    Z_{\mathrm{long}}[C]
    =
    \int \mathcal{D}X_0\,\mathcal{D}\tilde{X}_0\;
    \exp\Bigg\{
        - S_0[X_0,\tilde{X}_0]
        + \frac{\sigma^2}{2S}(S-1)
          \int\!dt\,ds\,
          \tilde{X}_0(t)\,C(t,s)\,\tilde{X}_0(s)
    \Bigg\}.
\end{aligned}
\end{equation}

The quenched-disorder averaged generating functional can therefore be
cast in the standard dynamical mean-field form
\begin{equation}
\begin{aligned}
    \overline{Z[\mathbf{j},\tilde{\mathbf{j}}]}
    &=
    \int \mathcal{D}C\,\mathcal{D}\tilde{C}\;
    \exp\Bigg\{
        - \frac{(S-1)^2}{\sigma^2 S}
          \int\!dt\,ds\,C(t,s)\,\tilde{C}(t,s)
    \Bigg\}\\
    &\qquad\times
    \Big[ Z_{\mathrm{trans}}[C,\tilde{C}] \Big]^{S-1}\,
    Z_{\mathrm{long}}[C],
\end{aligned}
\label{eq:Z_DMF_form}
\end{equation}
where
\begin{equation}
    Z_{\mathrm{trans}}[C,\tilde{C}]
    =
    \int \mathcal{D}y\,\mathcal{D}\tilde{y}\;
    \exp\Bigg\{
        - \int\!dt\,
          \big[
            \tilde{y}(t)\big(\partial_t y(t) + y(t)\big)
            - \tfrac{D}{2}\,\tilde{y}(t)^2
          \big]
        + \frac{1}{2}\,\tilde{y} C \tilde{y}
        + y \tilde{C} y
    \Bigg\},
\end{equation}
and $Z_{\mathrm{long}}[C]$ is given above. The fields $C(t,s)$ and
$\tilde{C}(t,s)$ play the role of dynamical order parameters, describing
the self-consistent two-time correlations and response functions of a
single effective transverse degree of freedom.

\subsection{Saddle-point approximation}

The disorder-averaged generating functional has the dynamical mean-field
representation
\begin{equation}
\begin{aligned}
    \overline{Z[\mathbf{j},\tilde{\mathbf{j}}]}
    &=
    \int \mathcal{D}C\,\mathcal{D}\tilde{C}\;
    \exp\Big\{- S_{\mathrm{eff}}[C,\tilde{C}]\Big\},\\[4pt]
    S_{\mathrm{eff}}[C,\tilde{C}]
    &=
    \frac{(S-1)^2}{\sigma^2 S}
    \int\!dt\,ds\,C(t,s)\,\tilde{C}(t,s)\\
    &\quad
    - (S-1)\,\ln Z_{\mathrm{trans}}[C,\tilde{C}]
    - \ln Z_{\mathrm{long}}[C],
\end{aligned}
\end{equation}
with
\begin{equation}
\begin{aligned}
    Z_{\mathrm{trans}}[C,\tilde{C}]
    &=
    \int \mathcal{D}y\,\mathcal{D}\tilde{y}\;
    \exp\Bigg\{
        - \int\!dt\,
          \Big[
            \tilde{y}(t)\big(\partial_t y(t) + y(t)\big)
            - \frac{D}{2}\,\tilde{y}(t)^2
          \Big]\\
    &\hspace{3.5cm}
        + \frac{1}{2}\int\!dt\,ds\,
          \tilde{y}(t)\,C(t,s)\,\tilde{y}(s)
        + \int\!dt\,ds\,
          y(t)\,\tilde{C}(t,s)\,y(s)
    \Bigg\},
\end{aligned}
\end{equation}
and
\begin{equation}
\begin{aligned}
    Z_{\mathrm{long}}[C]
    =
    \int \mathcal{D}X_0\,\mathcal{D}\tilde{X}_0\;
    \exp\Bigg\{
        - \int\!dt\,
          \Big[
            \tilde{X}_0(t)\big(\partial_t X_0(t) + X_0(t)\big)
            - \frac{D}{2}\,\tilde{X}_0(t)^2
          \Big]\\
        \hspace{2.7cm}
        + \frac{\sigma^2}{2S}(S-1)
          \int\!dt\,ds\,
          \tilde{X}_0(t)\,C(t,s)\,\tilde{X}_0(s)
    \Bigg\}.
\end{aligned}
\end{equation}

The dynamical mean-field (DMF) equations follow from a saddle-point
evaluation of the functional integral over $C$ and $\tilde{C}$ in the
limit of large system size $S$. The saddle-point fields
$C^\ast(t,s)$ and $\tilde{C}^\ast(t,s)$ are determined by the stationarity
conditions
\begin{equation}
    \frac{\delta S_{\mathrm{eff}}}{\delta C(t,s)}\bigg|_{C^\ast,\tilde{C}^\ast}
    = 0,
    \qquad
    \frac{\delta S_{\mathrm{eff}}}{\delta \tilde{C}(t,s)}\bigg|_{C^\ast,\tilde{C}^\ast}
    = 0.
\end{equation}

It is convenient to express these equations in terms of expectation values
with respect to the single-site effective measures. For the transverse
mode, expectation values are defined as
\begin{equation}
    \langle \mathcal{O}[y,\tilde{y}] \rangle_{\mathrm{trans}}
    :=
    \frac{1}{Z_{\mathrm{trans}}[C^\ast,\tilde{C}^\ast]}
    \int \mathcal{D}y\,\mathcal{D}\tilde{y}\;
    \mathcal{O}[y,\tilde{y}]\,
    e^{-S_{\mathrm{eff}}[y,\tilde{y};C^\ast,\tilde{C}^\ast]},
\end{equation}
where
\begin{equation}
\begin{aligned}
    S_{\mathrm{eff}}[y,\tilde{y};C^\ast,\tilde{C}^\ast]
    &=
    \int\!dt\,
    \Big[
        \tilde{y}(t)\big(\partial_t y(t) + y(t)\big)
        - \frac{D^2}{2}\,\tilde{y}(t)^2
    \Big]\\
    &\quad
    - \frac{1}{2}\int\!dt\,ds\,
      \tilde{y}(t)\,C^\ast(t,s)\,\tilde{y}(s)
    - \int\!dt\,ds\,
      y(t)\,\tilde{C}^\ast(t,s)\,y(s).
\end{aligned}
\end{equation}
For the longitudinal mode,
\begin{equation}
    \langle \mathcal{O}[X_0,\tilde{X}_0] \rangle_{\mathrm{long}}
    :=
    \frac{1}{Z_{\mathrm{long}}[C^\ast]}
    \int \mathcal{D}X_0\,\mathcal{D}\tilde{X}_0\;
    \mathcal{O}[X_0,\tilde{X}_0]\,
    e^{-S_{\mathrm{eff}}[X_0,\tilde{X}_0;C^\ast]},
\end{equation}
with
\begin{equation}
\begin{aligned}
    S_{\mathrm{eff}}[X_0,\tilde{X}_0;C^\ast]
    &=
    \int\!dt\,
    \Big[
        \tilde{X}_0(t)\big(\partial_t X_0(t) + X_0(t)\big)
        - \frac{D}{2}\,\tilde{X}_0(t)^2
    \Big]\\
    &\quad
    - \frac{\sigma^2}{2S}(S-1)
      \int\!dt\,ds\,
      \tilde{X}_0(t)\,C^\ast(t,s)\,\tilde{X}_0(s).
\end{aligned}
\end{equation}

Using the standard identities
\begin{equation}
    \frac{\delta \ln Z_{\mathrm{trans}}}{\delta \tilde{C}(t,s)}
    = \big\langle y(t)\,y(s)\big\rangle_{\mathrm{trans}},
    \qquad
    \frac{\delta \ln Z_{\mathrm{trans}}}{\delta C(t,s)}
    = \frac{1}{2}\big\langle \tilde{y}(t)\,\tilde{y}(s)\big\rangle_{\mathrm{trans}},
\end{equation}
and
\begin{equation}
    \frac{\delta \ln Z_{\mathrm{long}}}{\delta C(t,s)}
    = \frac{\sigma^2}{2S}(S-1)\,
      \big\langle \tilde{X}_0(t)\,\tilde{X}_0(s)\big\rangle_{\mathrm{long}},
\end{equation}
the saddle-point equations can be written explicitly.

The variation with respect to $\tilde{C}(t,s)$ gives
\begin{equation}
\begin{aligned}
    0
    &= \frac{\delta S_{\mathrm{eff}}}{\delta \tilde{C}(t,s)}\bigg|_{C^\ast,\tilde{C}^\ast}\\
    &= \frac{(S-1)^2}{\sigma^2 S}\,C^\ast(t,s)
       - (S-1)\,\big\langle y(t)\,y(s)\big\rangle_{\mathrm{trans}}.
\end{aligned}
\end{equation}
Therefore,
\begin{equation}
    C^\ast(t,s)
    = \frac{\sigma^2 S}{S-1}\,
      \big\langle y(t)\,y(s)\big\rangle_{\mathrm{trans}}=\frac{\sigma^2 S}{S-1}C_y^*(t,s),
    \label{eq:DMF_C_eq}
\end{equation}
where we have introduced the notation $C_y(t,s) = \big\langle y(t)\,y(s)\big\rangle_\text{trans}$ to denote the correlator of the variable $y(t)$ (the asterisk denotes just this correlator evaluated at the saddle).

The variation with respect to $C(t,s)$ yields
\begin{equation}
\begin{aligned}
    0
    &= \frac{\delta S_{\mathrm{eff}}}{\delta C(t,s)}\bigg|_{C^\ast,\tilde{C}^\ast}\\
    &= \frac{(S-1)^2}{\sigma^2 S}\,\tilde{C}^\ast(t,s)
       - (S-1)\,\frac{1}{2}\big\langle \tilde{y}(t)\,\tilde{y}(s)\big\rangle_{\mathrm{trans}}
       - \frac{\sigma^2}{2S}(S-1)\,
         \big\langle \tilde{X}_0(t)\,\tilde{X}_0(s)\big\rangle_{\mathrm{long}}.
\end{aligned}
\end{equation}
Dividing by $(S-1)$ and rearranging,
\begin{equation}
    \tilde{C}^\ast(t,s)
    = \frac{\sigma^2 S}{2(S-1)}
      \bigg[
        \big\langle \tilde{y}(t)\,\tilde{y}(s)\big\rangle_{\mathrm{trans}}
        + \frac{\sigma^2}{S}\,
          \big\langle \tilde{X}_0(t)\,\tilde{X}_0(s)\big\rangle_{\mathrm{long}}
      \bigg].
    \label{eq:DMF_Ctilde_eq}
\end{equation}

Equations \eqref{eq:DMF_C_eq} and \eqref{eq:DMF_Ctilde_eq} are the
self-consistent dynamical mean-field equations for the collective
correlator $C(t,s)$ and the conjugate kernel $\tilde{C}(t,s)$. In the
limit of large system size $S$, the factors $S/(S-1)$ approach unity, and
the longitudinal contribution in \eqref{eq:DMF_Ctilde_eq} is suppressed
by a factor $1/S$. To leading order in $1/S$ one obtains the familiar
DMF relations
\begin{equation}
    C^\ast(t,s)
    \simeq \sigma^2\,\big\langle y(t)\,y(s)\big\rangle_{\mathrm{trans}} = \sigma^2 C_y^*(t,s),
    \qquad
    \tilde{C}^\ast(t,s)
    \simeq \frac{\sigma^2}{2}\,
           \big\langle \tilde{y}(t)\,\tilde{y}(s)\big\rangle_{\mathrm{trans}}=0,
\end{equation}
which describe a single effective transverse degree of freedom
interacting with a self-consistent Gaussian bath. We have set $\tilde{C}^\ast(t,s)=0$ because the self-correlation of the fields $\tilde{y}$ must vanish \cite{helias_statistical_2020}.

\subsection{Effective single-particle dynamics and linear response}

At the saddle point, the transverse contribution to the
disorder-averaged generating functional factorizes into $(S-1)$
identical copies of a single effective degree of freedom. The dynamics
of a typical transverse mode $y(t)$ is encoded in the MSRDJ action
$S_{\mathrm{eff}}[y,\tilde y;C^\ast,\tilde C^\ast]$,
\begin{equation}
\begin{aligned}
    S_{\mathrm{eff}}[y,\tilde{y};C^\ast,\tilde{C}^\ast]
    &=
    \int\!dt\,
    \Big[
        \tilde{y}(t)\big(\partial_t y(t) + y(t)\big)
        - \frac{D}{2}\,\tilde{y}(t)^2
    \Big]\\
    &\quad
    - \frac{1}{2}\int\!dt\,ds\,
      \tilde{y}(t)\,C^\ast(t,s)\,\tilde{y}(s)
    - \int\!dt\,ds\,
      y(t)\,\cancelto{0}{\tilde{C}^\ast(t,s)}\,y(s),
\end{aligned}
\label{eq:Seff_transverse_general}
\end{equation}
where $C^\ast$ and $\tilde C^\ast$ are the saddle-point values of the
collective kernels. This is precisely the MSRDJ action of a linear Langevin equation with additive Gaussian noise. More explicitly, the effective single-particle dynamics of a transverse mode can be written as
\begin{equation}
    \partial_t y(t)
    = - y(t) + \eta(t) + \sqrt{D}\,\xi(t),
    \label{eq:effective_single_particle_dynamics}
\end{equation}
where $\xi(t)$ is a standard white noise with
$\langle \xi(t)\,\xi(s)\rangle = \delta(t-s)$, and $\eta(t)$ is a
zero-mean Gaussian ``mean-field'' input with covariance
\begin{equation}
    \big\langle \eta(t)\,\eta(s) \big\rangle
    = C^\ast(t,s).
    \label{eq:eta_covariance}
\end{equation}
The total effective noise $\zeta(t) := \eta(t) + D\,\xi(t)$ is therefore
Gaussian with covariance
\begin{equation}
    \big\langle \zeta(t)\,\zeta(s)\big\rangle
    = C^\ast(t,s) + D\,\delta(t-s),
\end{equation}
and the DMF equation \eqref{eq:DMF_C_eq} enforces the self-consistency
between this noise covariance and the autocorrelation of the solution
$y(t)$.

\paragraph{Linear response function.}
To compute the response function of the effective single-particle
process, we add a small external field $h(t)$ to the dynamics
\eqref{eq:effective_single_particle_dynamics},
\begin{equation}
    \partial_t y(t)
    = - y(t) + \eta(t) + \sqrt{D}\,\xi(t) + h(t).
\end{equation}
The (causal) linear response function is defined as
\begin{equation}
    R^\ast(t,s)
    := \frac{\delta \langle y(t)\rangle}{\delta h(s)}\bigg|_{h=0},
\end{equation}
and is related to the MSRDJ fields by
\begin{equation}
    R^\ast(t,s)
    = \big\langle y(t)\,\tilde{y}(s)\big\rangle_{\mathrm{eff}}.
\end{equation}
Because the equation is linear and the noise has zero mean, the average
$\langle y(t)\rangle$ obeys the deterministic equation
\begin{equation}
    \partial_t \langle y(t)\rangle
    = - \langle y(t)\rangle + h(t),
\end{equation}
irrespective of the detailed noise statistics. Writing the formal
solution as
\begin{equation}
    \langle y(t)\rangle
    = \int_{-\infty}^{\infty} ds\; R^\ast(t,s)\,h(s),
\end{equation}
and inserting this into the deterministic equation, one finds that
$R^\ast$ must satisfy
\begin{equation}
    \frac{\partial}{\partial t} R^\ast(t,s)
    = - R^\ast(t,s) + \delta(t-s),
    \qquad
    R^\ast(t,s) = 0 \quad \text{for } t < s,
\end{equation}
where causality enforces $R^\ast(t,s)=0$ for $t<s$. The unique solution
is
\begin{equation}
    R^\ast(t,s)
    = \theta(t-s)\,e^{-(t-s)},
\end{equation}
with $\theta$ the Heaviside step function. In terms of the time
difference $\tau = t-s$, we write
\begin{equation}
    R^\ast(\tau)
    := R^\ast(t,t-\tau)
    = \theta(\tau)\,e^{-\tau}.
\end{equation}

The response function in frequency space is defined as the Fourier
transform of $R^\ast(\tau)$,
\begin{equation}
    \chi(\omega)
    := \int_{-\infty}^{\infty} d\tau\;
       e^{i\omega\tau}\,R^\ast(\tau)
    = \int_0^{\infty} d\tau\;
      e^{-(1 - i\omega)\tau}
    = \frac{1}{1 - i\omega}.
    \label{eq:chi_omega_OU}
\end{equation}
Thus, for the linear network considered here, the effective transverse
degree of freedom has the same response function as a bare
Ornstein--Uhlenbeck process: all the dependence on the coupling
$\sigma$ enters through the self-consistent noise covariance $C^\ast$,
while the linear response \eqref{eq:chi_omega_OU} remains fixed.

\subsection{Longitudinal mode at the DMFT saddle}

The longitudinal single-site MSRDJ action at the saddle reads
\begin{equation}
\begin{aligned}
    S_{\mathrm{long}}[X_0,\tilde X_0;C^\ast]
    &=
    \int\!dt\,
    \Big[
        \tilde X_0(t)\big(\partial_t X_0(t) + X_0(t)\big)
        - \frac{D^2}{2}\,\tilde X_0(t)^2
    \Big]\\
    &\quad
    - \frac{\sigma^2}{2S}(S-1)
      \int\!dt\,ds\,
      \tilde X_0(t)\,C^\ast(t,s)\,\tilde X_0(s).
\end{aligned}
\end{equation}
This is of the standard MSRDJ form for a linear Langevin equation with
additive Gaussian noise. Introducing the effective noise kernel
\begin{equation}
    \Gamma_0(t,s)
    :=
    D^2\,\delta(t-s)
    + \frac{\sigma^2}{S}(S-1)\,C^\ast(t,s),
\end{equation}
we can rewrite
\begin{equation}
    S_{\mathrm{long}}[X_0,\tilde X_0;C^\ast]
    =
    \int\!dt\,\tilde X_0(t)\big(\partial_t X_0(t) + X_0(t)\big)
    - \frac{1}{2}\int\!dt\,ds\,
      \tilde X_0(t)\,\Gamma_0(t,s)\,\tilde X_0(s).
\end{equation}
By the usual MSRDJ correspondence, this action describes the effective
longitudinal dynamics
\begin{equation}
\label{eq:X0_dmft_eom} 
    \partial_t X_0(t)
    = -\,X_0(t) + \zeta_0(t),
\end{equation}
where $\zeta_0(t)$ is a zero-mean Gaussian process with covariance
\begin{equation}
\label{eq:zeta_cov}
    \big\langle \zeta_0(t)\,\zeta_0(s)\big\rangle
    =
    D^2\,\delta(t-s)
    + \frac{\sigma^2}{S}(S-1)\,C^\ast(t,s).
\end{equation}
The dynamical mean-field equations for the order parameters $C(t,s)$ and
$\tilde C(t,s)$ in the large-$S$ limit yield:
\begin{equation}
    C^\ast(t,s)
    \simeq \sigma^2\,C_y^\ast(t,s).
\end{equation}
Crucially, $C^\ast(t,s)$ in this expression is the transverse DMFT
order parameter determined self-consistently;
it does not depend on the longitudinal mode. Therefore, at the DMFT saddle the longitudinal mode is \emph{not} self-consistent: it is a simple Ornstein--Uhlenbeck process with linear drift $-X_0$ driven by a Gaussian noise whose colored component is entirely set by the transverse
dynamics. All nontrivial self-consistency and critical behavior reside in the transverse sector, while the longitudinal mode behaves as a passive readout of the network fluctuations.

On the other hand, the longitudinal mode can be related with the mean activity of the network:
\begin{equation}
    M(t) := \frac{1}{\sqrt{S}}\,X_0(t),
\end{equation}
so that $X_0(t) = \sqrt{S}\,M(t)$. Differentiating with respect to time
and using \eqref{eq:X0_dmft_eom} gives
\begin{equation}
\begin{aligned}
    \partial_t M(t)
    &= \frac{1}{\sqrt{S}}\,\partial_t X_0(t)
     = \frac{1}{\sqrt{S}}\big[-X_0(t) + \zeta_0(t)\big]\\[4pt]
    &= -M(t) + \frac{1}{\sqrt{S}}\,\zeta_0(t).
\end{aligned}
\label{eq:M_stochastic_eom}
\end{equation}
Thus, for finite $S$ the mean activity satisfies
\begin{equation}
    \partial_t M(t)
    = - M(t) + \eta_M(t),
\end{equation}
with an effective noise
\begin{equation}
    \eta_M(t) := \frac{1}{\sqrt{S}}\,\zeta_0(t),
\end{equation}
whose covariance, by \eqref{eq:zeta_cov}, is
\begin{equation}
\begin{aligned}
    \big\langle \eta_M(t)\,\eta_M(s)\big\rangle
    &= \frac{1}{S}\big\langle \zeta_0(t)\,\zeta_0(s)\big\rangle\\[4pt]
    &= \frac{1}{S}
       \left[
         D^2\,\delta(t-s)
         + \frac{\sigma^2}{S}(S-1)\,C^\ast(t,s)
       \right]\\[4pt]
    &= \mathcal{O}\!\left(\frac{1}{S}\right).
\end{aligned}
\end{equation}
Therefore, in the thermodynamic limit $S\to\infty$ the mean activity converges to the deterministic solution of
\begin{equation}
    \partial_t M(t) = -M(t),
\end{equation}
while the stochastic fluctuations around this mean are suppressed as
$\mathrm{Var}[M(t)] = \mathcal{O}(1/S)$. In contrast, the rescaled
longitudinal mode $X_0(t)=\sqrt{S}M(t)$ remains an $\mathcal{O}(1)$
fluctuation variable governed by the Ornstein--Uhlenbeck dynamics
\eqref{eq:X0_dmft_eom} with colored noise set by the transverse DMFT
kernel $C^\ast(t,s)$. This decoupling of the mean with respect to the quenched disorder is precisely what we observed when performing DMFT directly into the ISLM, a condition made possible by the row-sum normalization of the matrix $G_{ij}$.

\subsection{Gaussian fluctuations and Hessian around the saddle point}

The dynamical mean-field equations are obtained from the saddle-point
conditions of the effective action
\begin{equation}
    S_{\mathrm{eff}}[C,\tilde{C}]
    =
    \frac{(S-1)^2}{\sigma^2 S}
    \int\!dt\,ds\,C(t,s)\,\tilde{C}(t,s)
    - (S-1)\,\ln Z_{\mathrm{trans}}[C,\tilde{C}]
    - \ln Z_{\mathrm{long}}[C].
\end{equation}
Beyond mean-field, the leading corrections are governed by the Gaussian
fluctuations of the collective fields $C$ and $\tilde{C}$ around their
saddle-point values $C^\ast$ and $\tilde{C}^\ast$.

\paragraph{Fluctuation fields and expansion.}
Fluctuations of the order parameter (of the external field created by the quenchde disorder) are parametrized as
\begin{equation}
    C(t,s) = C^\ast(t,s)
             + \frac{1}{\sqrt{S-1}}\,\delta C(t,s),
    \qquad
    \tilde{C}(t,s) = \tilde{C}^\ast(t,s)
             + \frac{1}{\sqrt{S-1}}\,\delta\tilde{C}(t,s).
\end{equation}
The factor $1/\sqrt{S-1}$ is chosen so that the quadratic fluctuation
action remains of order one as $S \to \infty$.

The effective action is expanded to second order as
\begin{equation}
    S_{\mathrm{eff}}[C,\tilde{C}]
    =
    S_{\mathrm{eff}}[C^\ast,\tilde{C}^\ast]
    + S^{(1)}[\delta C,\delta\tilde{C}]
    + S^{(2)}[\delta C,\delta\tilde{C}]
    + \dots,
\end{equation}
where $S^{(1)}$ vanishes by construction (saddle-point equations), and
$S^{(2)}$ is the quadratic form
\begin{equation}
\begin{aligned}
    S^{(2)}[\delta C,\delta\tilde{C}]
    &=
    \frac{1}{2}
    \int\!dt_1 ds_1 dt_2 ds_2\;
    \begin{pmatrix}
        \delta C(t_1,s_1) \\
        \delta\tilde{C}(t_1,s_1)
    \end{pmatrix}^{\!\!\top}
    \mathcal{H}(t_1,s_1; t_2,s_2)
    \begin{pmatrix}
        \delta C(t_2,s_2) \\
        \delta\tilde{C}(t_2,s_2)
    \end{pmatrix},
\end{aligned}
\end{equation}
with the Hessian kernel
\begin{equation}
    \mathcal{H}
    =
    \begin{pmatrix}
        H_{CC} & H_{C\tilde{C}} \\
        H_{\tilde{C}C} & H_{\tilde{C}\tilde{C}}
    \end{pmatrix},
\end{equation}
defined by the second functional derivatives of $S_{\mathrm{eff}}$ at
the saddle point,
\begin{equation}
\begin{aligned}
    H_{CC}(t_1,s_1; t_2,s_2)
    &= \left.
       \frac{\delta^2 S_{\mathrm{eff}}}
            {\delta C(t_1,s_1)\,\delta C(t_2,s_2)}
       \right|_{C^\ast,\tilde{C}^\ast},\\[4pt]
    H_{\tilde{C}\tilde{C}}(t_1,s_1; t_2,s_2)
    &= \left.
       \frac{\delta^2 S_{\mathrm{eff}}}
            {\delta \tilde{C}(t_1,s_1)\,\delta \tilde{C}(t_2,s_2)}
       \right|_{C^\ast,\tilde{C}^\ast},\\[4pt]
    H_{C\tilde{C}}(t_1,s_1; t_2,s_2)
    &= \left.
       \frac{\delta^2 S_{\mathrm{eff}}}
            {\delta C(t_1,s_1)\,\delta \tilde{C}(t_2,s_2)}
       \right|_{C^\ast,\tilde{C}^\ast},\\[4pt]
    H_{\tilde{C}C}(t_1,s_1; t_2,s_2)
    &= \left.
       \frac{\delta^2 S_{\mathrm{eff}}}
            {\delta \tilde{C}(t_1,s_1)\,\delta C(t_2,s_2)}
       \right|_{C^\ast,\tilde{C}^\ast}.
\end{aligned}
\end{equation}

\paragraph{Second derivatives and connected four-point functions.}
The collective term
\begin{equation}
    \frac{(S-1)^2}{\sigma^2 S}\int\!dt\,ds\,C(t,s)\,\tilde{C}(t,s)
\end{equation}
contributes only to the mixed block $H_{C\tilde{C}}$:
\begin{equation}
    \frac{\delta^2}{\delta C(1)\,\delta \tilde{C}(2)}
    \bigg[
        \frac{(S-1)^2}{\sigma^2 S}\int\!C\tilde{C}
    \bigg]
    =
    \frac{(S-1)^2}{\sigma^2 S}\,
    \delta(t_1-t_2)\,\delta(s_1-s_2),
\end{equation}
while its $CC$ and $\tilde{C}\tilde{C}$ derivatives vanish.

The contributions from the single-site partition functions are expressed
in terms of connected four-point functions. For the transverse mode, the identities
\begin{equation}
\begin{aligned}
    \frac{\delta \ln Z_{\mathrm{trans}}}{\delta \tilde{C}(t_1,s_1)}
    &= \big\langle y(t_1)\,y(s_1) \big\rangle_{\mathrm{trans}},
    \qquad
    \frac{\delta \ln Z_{\mathrm{trans}}}{\delta C(t_1,s_1)}
    &= \frac{1}{2}\big\langle
        \tilde{y}(t_1)\,\tilde{y}(s_1)
    \big\rangle_{\mathrm{trans}},
\end{aligned}
\end{equation}
imply
\begin{equation}
\begin{aligned}
    \frac{\delta^2 \ln Z_{\mathrm{trans}}}
         {\delta \tilde{C}(t_1,s_1)\,\delta \tilde{C}(t_2,s_2)}
    &=
    \big\langle
        y(t_1)\,y(s_1)\,
        y(t_2)\,y(s_2)
    \big\rangle_{\mathrm{trans},c},\\[4pt]
    \frac{\delta^2 \ln Z_{\mathrm{trans}}}
         {\delta C(t_1,s_1)\,\delta C(t_2,s_2)}
    &=
    \frac{1}{4}\big\langle
        \tilde{y}(t_1)\,\tilde{y}(s_1)\,
        \tilde{y}(t_2)\,\tilde{y}(s_2)
    \big\rangle_{\mathrm{trans},c},\\[4pt]
    \frac{\delta^2 \ln Z_{\mathrm{trans}}}
         {\delta C(t_1,s_1)\,\delta \tilde{C}(t_2,s_2)}
    &=
    \frac{1}{2}\big\langle
        \tilde{y}(t_1)\,\tilde{y}(s_1)\,
        y(t_2)\,y(s_2)
    \big\rangle_{\mathrm{trans},c},
\end{aligned}
\end{equation}
where $\langle\cdots\rangle_{\mathrm{trans},c}$ denotes connected
correlation functions with respect to the effective single-site action
of the transverse mode. Similarly, the longitudinal partition function
$Z_{\mathrm{long}}[C]$ generates connected four-point functions of the
longitudinal response field $\tilde{X}_0$,
\begin{equation}
    \frac{\delta^2 \ln Z_{\mathrm{long}}}
         {\delta C(t_1,s_1)\,\delta C(t_2,s_2)}
    \propto
    \big\langle
        \tilde{X}_0(t_1)\,\tilde{X}_0(s_1)\,
        \tilde{X}_0(t_2)\,\tilde{X}_0(s_2)
    \big\rangle_{\mathrm{long},c},
\end{equation}
with analogous relations for mixed derivatives involving $C$.

\paragraph{Explicit form of the Hessian blocks.}
Collecting the contributions from the collective term and from
$Z_{\mathrm{trans}}$ and $Z_{\mathrm{long}}$, the Hessian blocks are
\begin{equation}
\begin{aligned}
    H_{\tilde{C}\tilde{C}}(t_1,s_1; t_2,s_2)
    &=
    - (S-1)\,
      \big\langle
        y(t_1)\,y(s_1)\,
        y(t_2)\,y(s_2)
      \big\rangle_{\mathrm{trans},c},
    \\[4pt]
    H_{CC}(t_1,s_1; t_2,s_2)
    &=
    - (S-1)\,\frac{1}{4}\,
      \big\langle
        \tilde{y}(t_1)\,\tilde{y}(s_1)\,
        \tilde{y}(t_2)\,\tilde{y}(s_2)
      \big\rangle_{\mathrm{trans},c}
      \;-\;
      \left.
      \frac{\delta^2 \ln Z_{\mathrm{long}}}
           {\delta C(t_1,s_1)\,\delta C(t_2,s_2)}
      \right|_{C^\ast},
    \\[4pt]
    H_{C\tilde{C}}(t_1,s_1; t_2,s_2)
    &=
    \frac{(S-1)^2}{\sigma^2 S}\,
    \delta(t_1-t_2)\,\delta(s_1-s_2)
    - (S-1)\,\frac{1}{2}\,
      \big\langle
        \tilde{y}(t_1)\,\tilde{y}(s_1)\,
        y(t_2)\,y(s_2)
      \big\rangle_{\mathrm{trans},c}
      \;-\;
      \left.
      \frac{\delta^2 \ln Z_{\mathrm{long}}}
           {\delta C(t_1,s_1)\,\delta \tilde{C}(t_2,s_2)}
      \right|_{C^\ast},
\end{aligned}
\end{equation}
and, by symmetry,
\begin{equation}
    H_{\tilde{C}C}(t_1,s_1; t_2,s_2)
    =
    H_{C\tilde{C}}(t_2,s_2; t_1,s_1).
\end{equation}
The longitudinal contributions (second derivatives of
$\ln Z_{\mathrm{long}}[C]$) are of order $1$ in $S$, whereas the
transverse contributions scale as $(S-1)$, which means that the quadratic action $S^{(2)}$ is dominated, for large $S$, by the
transverse sector. The longitudinal corrections enter $H_{CC}$ and
$H_{C\tilde{C}}$ with an extra factor $1/(S-1)$ relative to the
transverse four-point functions and therefore vanish in the
thermodynamic limit. To leading order in $1/S$, the Hessian kernel
reduces to
\begin{equation}
\begin{aligned}
    H_{\tilde{C}\tilde{C}}(t_1,s_1; t_2,s_2)
    &\simeq
    - (S-1)\,
      \big\langle
        y(t_1)\,y(s_1)\,
        y(t_2)\,y(s_2)
      \big\rangle_{\mathrm{trans},c},\\[4pt]
    H_{CC}(t_1,s_1; t_2,s_2)
    &\simeq
    - (S-1)\,\frac{1}{4}\,
      \big\langle
        \tilde{y}(t_1)\,\tilde{y}(s_1)\,
        \tilde{y}(t_2)\,\tilde{y}(s_2)
      \big\rangle_{\mathrm{trans},c},\\[4pt]
    H_{C\tilde{C}}(t_1,s_1; t_2,s_2)
    &\simeq
    \frac{(S-1)^2}{\sigma^2 S}\,
    \delta(t_1-t_2)\,\delta(s_1-s_2)
    - (S-1)\,\frac{1}{2}\,
      \big\langle
        \tilde{y}(t_1)\,\tilde{y}(s_1)\,
        y(t_2)\,y(s_2)
      \big\rangle_{\mathrm{trans},c}.
\end{aligned}
\end{equation}

The Gaussian theory defined by $S^{(2)}$ determines the fluctuations of the correlations $\delta C(t,s)$ around its mean-field value
$C^\ast(t,s)$. Since all the Hessian elements contain an overall factor $(S-1)$, it is convenient to define an intensive Hessian $\mathcal{H}_{AB}=H_{AB}/(S-1)$ and work just with $\mathcal{H}$ for now on.

\paragraph{The Hessian in Fourier space}

Since the effective process is Gaussian, Wick’s theorem applies and the four–point functions factorize into products of two–point functions.
For instance,
\begin{equation}
    \big\langle
        y(t_1)y(s_1)y(t_2)y(s_2)
    \big\rangle_{\text{eff},c}
    =
    C^\ast(t_1,t_2)\,C^\ast(s_1,s_2)
    + C^\ast(t_1,s_2)\,C^\ast(s_1,t_2),
    \label{eq:wick_yyyy_again}
\end{equation}
where \(C^\ast(t,s)=\sigma^2 \langle y(t)y(s)\rangle_{\text{eff}}\) is the DMF two–point function. Similar Wick decompositions hold for the mixed correlators \(\langle\tilde y\tilde y y y\rangle_c\) and
\(\langle\tilde y\tilde y\tilde y\tilde y\rangle_c\) in terms of
\(C^\ast\) and the response functions.

At this point we choose the time arguments in such a way that we isolate
the long–time fluctuations of correlations. We set
\begin{equation}
    t_2 = t_1 + \tau_1,
    \qquad
    s_2 = s_1 + \tau_2,
\end{equation}
and consider the limit
\begin{equation}
    |s_1 - t_1| \to \infty,
    \qquad
    \tau_1,\tau_2 = \mathcal{O}(1),
\end{equation}
assuming that the process has reached stationarity
(\(C^\ast(t,s)=C^\ast(t-s)\)) and loses correlations at long times.
With this choice one obtains
\begin{equation}
    C^\ast(t_1,t_2) = C^\ast(\tau_1),\qquad
    C^\ast(s_1,s_2) = C^\ast(\tau_2),\qquad
    C^\ast(t_1,s_2) = C^\ast(s_1 - t_1 + \tau_2) \to 0,\qquad
    C^\ast(s_1,t_2) = C^\ast(t_1 - s_1 + \tau_1) \to 0,
    \label{eq:long_time_limits}
\end{equation}
where the last two limits vanish because the dynamics decorrelates on
large time scales.

Using \eqref{eq:wick_yyyy_again} together with
\eqref{eq:long_time_limits}, the connected four–point function simplifies to
\begin{equation}
    \big\langle
        y(t_1)y(s_1)y(t_2)y(s_2)
    \big\rangle_{\text{eff},c}
    \xrightarrow[|t_2-t_1|\to\infty]{}
    C^\ast(\tau_1)\,C^\ast(\tau_2),
\end{equation}
and therefore, in this limit,
\begin{equation}
    \mathcal{H}_{\tilde C \tilde C}
    \;\longrightarrow\;
    \mathcal{H}_{\tilde C \tilde C}(\tau_1,\tau_2)
    = -\,C^\ast(\tau_1)\,C^\ast(\tau_2),
\end{equation}
with analogous expressions for the mixed and \(CC\) blocks. Passing to Fourier space,
\begin{equation}
    \mathcal{H}_{AB}(\omega_1,\omega_2)
    = \int d\tau_1 d\tau_2\,
      e^{i\omega_1\tau_1 + i\omega_2\tau_2}
      \mathcal{H}_{AB}(\tau_1,\tau_2),
\end{equation}
one finds that the Hessian blocks are diagonal in frequency and can be
written as
\begin{align}
    \mathcal{H}_{\tilde C \tilde C}(\omega_1,\omega_2)
    &= - C^\ast(\omega_1)\,C^\ast(\omega_2)\,
    ,
    \label{eq:H_tCtC_freq}\\[2pt]
    \mathcal{H}_{C \tilde C}(\omega_1,\omega_2)
    &= \Big[-\frac{1}{\sigma^2}
            + \chi(\omega_1)\chi(\omega_2)\Big]\,
      ,
    \label{eq:H_CtC_freq}\\[2pt]
    \mathcal{H}_{C C}(\omega_1,\omega_2)
    &= 0,
    \label{eq:H_CC_freq}
\end{align}
where \(\chi(\omega)\) is the Fourier transform of the linear response function ($R^*(t,s)$) of the effective single–site process, and \(C^\ast(\omega)\) is the Fourier transform of the DMF correlator.

\paragraph{Four-point fluctuations and the observable $\Psi$}

The Gaussian theory defined by the quadratic fluctuation action
$S^{(2)}[\delta C,\delta\tilde{C}]$ governs the fluctuations of the
collective correlator $C(t,s)$ around its saddle-point value $C^\ast(t,s)$.
A convenient object to quantify these fluctuations is the four-time
observable
\begin{equation}
    \Psi(t_1,s_1,t_2,s_2)
    :=
    S\,\big\langle
        \delta C(t_1,s_1)\,\delta C(t_2,s_2)
    \big\rangle,
    \qquad
    \delta C(t,s) = C(t,s) - C^\ast(t,s),
\end{equation}
which measures the width of the distribution of two-time correlations
across different realizations of the quenched disorder and the noise. In simulations, however, one is usually interested in:
\begin{equation}
    \Psi_y(t_1,s_1,t_2,s_2)
    :=
    S\,\big\langle
        \delta C_y(t_1,s_1)\,\delta C_y(t_2,s_2)
    \big\rangle,
    \qquad
    \delta C_y(t,s) = C_y(t,s) - C^\ast_y(t,s),
\end{equation}
which can be expressed in terms of $\Psi$ as:
\begin{equation}
    \Psi_y=\frac{1}{\sigma^4}\Psi.
\end{equation}
At quadratic order, we had arrived to the path integral over $(\delta C,\delta\tilde{C})$ being Gaussian:
\begin{equation}
\begin{aligned}
    \mathcal{Z}_{\mathrm{fluc}}
    &=
    \int\mathcal{D}(\delta C)\,\mathcal{D}(\delta\tilde{C})\;
    \exp\big[-S^{(2)}[\delta C,\delta\tilde{C}]\big].
\end{aligned}
\end{equation}
For a Gaussian functional integral of the form
\begin{equation}
    S^{(2)}[\Phi]
    = \frac{1}{2}\,\Phi^\top \mathcal{H}\,\Phi,
\end{equation}
the covariance of the fluctuation fields is given by the inverse of the
Hessian $\mathcal{H}^{-1}$. In our case, packing the fields into
\begin{equation}
    \Phi :=
    \begin{pmatrix}
        \delta C \\
        \delta\tilde{C}
    \end{pmatrix},
\end{equation}
we obtain
\begin{equation}
    \big\langle
        \delta C(t_1,s_1)\,\delta C(t_2,s_2)
    \big\rangle
    = \big[\mathcal{H}^{-1}\big]_{CC}(t_1,s_1; t_2,s_2),
\end{equation}
where $[\mathcal{H}^{-1}]_{CC}$ denotes the $C$--$C$ block of the
inverse matrix.

The observable $\Psi$ that quantifies the width of the fluctuations of
the correlator is defined as
\begin{equation}
    \Psi(t_1,s_1;t_2,s_2)
    := S\,
       \big\langle
           \delta C(t_1,s_1)\,\delta C(t_2,s_2)
       \big\rangle,
\end{equation}
so that, in terms of the Hessian,
\begin{equation}
    \Psi(t_1,s_1;t_2,s_2)
    = S\,
      \big[H^{-1}\big]_{CC}(t_1,s_1; t_2,s_2)=\frac{S}{S-1}\big[\mathcal{H}^{-1}\big]_{CC}(t_1,s_1; t_2,s_2)\approx [\mathcal{H}]^{-1}_{CC},
    \label{eq:Psi_from_Hinv}
\end{equation}
where, in the last equality, we have approximated $S/(S-1)\approx 1$ in the large $S$ limit.

\paragraph{Schur–complement formula.}
To make \eqref{eq:Psi_from_Hinv} explicit, we use the general formula
for the inverse of a block matrix. Consider the abstract block form
\begin{equation}
    \mathcal{H}
    =
    \begin{pmatrix}
        A & B \\
        B^\top & D
    \end{pmatrix},
\end{equation}
where here
\begin{equation}
    A \equiv H_{CC},\qquad
    B \equiv H_{C\tilde{C}},\qquad
    D \equiv H_{\tilde{C}\tilde{C}}.
\end{equation}
Assuming that $D$ is invertible, the $C$--$C$ block of the inverse
matrix $\mathcal{H}^{-1}$ is given by the Schur complement:
\begin{equation}
    \big[\mathcal{H}^{-1}\big]_{CC}
    = \Big( A - B\,D^{-1}\,B^\top \Big)^{-1}.
    \label{eq:Schur_block}
\end{equation}
Applying this to our case, we obtain
\begin{equation}
    \big\langle
        \delta C(t_1,s_1)\,\delta C(t_2,s_2)
    \big\rangle
    =
    \Big[
        \mathcal{H}_{CC}
        - \mathcal{H}_{C\tilde{C}}\,\mathcal{H}_{\tilde{C}\tilde{C}}^{-1}\,\mathcal{H}_{\tilde{C}C}
    \Big]^{-1}\!(t_1,s_1; t_2,s_2).
    \label{eq:deltaC_cov_Schur}
\end{equation}
Therefore, the fluctuation observable $\Psi$ can be written compactly as
\begin{equation}
    \Psi(t_1,s_1;t_2,s_2)
    =
    \,
    \Big[
        \mathcal{H}_{CC}
        - \mathcal{H}_{C\tilde{C}}\,\mathcal{H}_{\tilde{C}\tilde{C}}^{-1}\,\mathcal{H}_{\tilde{C}C}
    \Big]^{-1}\!(t_1,s_1; t_2,s_2).
    \label{eq:Psi_Schur_final}
\end{equation}

\eqref{eq:Psi_Schur_final} is the Schur–complement expression
for $\Psi$: it shows that the covariance of the order-parameter
fluctuations $\delta C$ is obtained by inverting the \emph{effective}
kernel
\begin{equation}
    \mathcal{K}_{CC}
    :=
    \mathcal{H}_{CC}
    - \mathcal{H}_{C\tilde{C}}\,\mathcal{H}_{\tilde{C}\tilde{C}}^{-1}\,\mathcal{H}_{\tilde{C}C},
\end{equation}
which incorporates both the direct $CC$ block and the indirect coupling through the $\tilde{C}$ sector.

\paragraph{Fluctuations of the long-time window covariance}

Using the Schur–complement relation
\begin{equation}
    \big\langle
        \delta C(\omega)\,\delta C(-\omega)
    \big\rangle
    =
    \Big[
        \mathcal{H}_{CC}(\omega)
        - \mathcal{H}_{C\tilde C}(\omega)\,
          \mathcal{H}_{\tilde C\tilde C}^{-1}(\omega)\,
          \mathcal{H}_{\tilde C C}(\omega)
    \Big]^{-1},
\end{equation}
and the fact that here \(H_{CC}(\omega)=0\), we obtain
\begin{equation}
    \big\langle
        \delta C(\omega)\,\delta C(-\omega)
    \big\rangle
    =
    \frac{C^\ast(\omega)C^\ast(-\omega)}
        {\big[1 - \sigma^2\,\chi(\omega)\chi(-\omega)\big]^2}.
\end{equation}
The four–point observable
\(\Psi(\omega) = S\,\langle\delta C(\omega)\delta C(-\omega)\rangle\)
thus reads
\begin{equation}
    \Psi(\omega)
    =
    \bigg[
        \frac{|C^\ast(\omega)|}
             {1 - \sigma^2\,\chi(\omega)\chi(-\omega)}
    \bigg]^2.
\end{equation}
This expression relates the fluctuation spectrum of the
two–point correlator to terms of the DMF correlator \(C^\ast\) and the
linear response function \(\chi\). 
Substituting the expression for the linear response \eqref{eq:chi_omega_OU}, finally gives:
\begin{equation}
    \Psi(\omega)
    =
    \left[
        \frac{|C^\ast(\omega)|}
             {1 - \displaystyle{\frac{\sigma^2}{1+\omega^2}}}
    \right]^2.
\end{equation}
In particular, let us observe that:
\begin{equation}
    \frac{\Psi(\omega)}{|C^*(\omega)|^2}=\frac{\Psi_y(\omega)}{|C_y(\omega)|^2}= \left[
        \frac{1}
             {1 - \sigma(\omega)^2}
    \right]^2,
    \label{eq: normalized psi}
\end{equation}
with $\sigma(\omega)^2=\sigma^2/(1+\omega^2)$.

\subsection{Connection with microscopic observables}

\paragraph{Average correlations.}
Up to this point, we have derived a closed expression that links the
fluctuations of the correlations to the amplitude of the quenched
disorder, $\sigma$, within the DMF framework. We now relate these
macroscopic quantities to microscopic observables that can be directly
computed from simulations (or data).

We denote by $ C_{ij}(\omega)$ the empirical covariance matrix of
the dynamics \eqref{eq: dynamics rotated}. Concretely, $C_{ij}(\omega)$ is the frequency-dependent covariance between $y_i(t)$ and $y_j(t)$ at frequency $\omega$:
\begin{equation}
    C_{ij}(\omega)
    := \big\langle y_i(\omega)\,y_j(-\omega)\big\rangle,
\end{equation}
where $\langle\cdot\rangle$ denotes the average over both the quenched
disorder and the dynamical noise. The DMF self-covariance, averaged over sites, is defined as
\begin{equation}
    C_y(\omega)
    := \frac{1}{S}\sum_{i=1}^S C_{ii}(\omega)
     = \frac{1}{S}\sum_{i=1}^S
       \big\langle y_i(\omega)\,y_i(-\omega)\big\rangle.
\end{equation}
Thus, the DMF order parameter $C_y(\omega)$ is identified with the
site-averaged diagonal of the microscopic covariance matrix.

From the saddle-point analysis, the effective dynamics of a typical
transverse mode is given by the linear Langevin equation
\begin{equation}
    \partial_t y(t)
    = - y(t) + \eta(t) + \sqrt{D}\,\xi(t),
    \label{eq:effective_single_particle_dynamics_again}
\end{equation}
where $\xi(t)$ is a standard white noise,
$\langle \xi(t)\xi(s)\rangle = \delta(t-s)$, and $\eta(t)$ is a
zero-mean Gaussian ``mean-field'' input with covariance
\begin{equation}
    \big\langle \eta(t)\,\eta(s)\big\rangle
    = \sigma^2\,C_y(t,s).
\end{equation}
In frequency space, the power spectrum of $y(t)$ reads
\begin{equation}
    C_y(\omega)
    = \big\langle |y(\omega)|^2\big\rangle
    = |\chi(\omega)|^2\Big(C_\eta(\omega) + D\Big),
\end{equation}
where $\chi(\omega)$ is the linear response function and
$C_\eta(\omega)$ is the power spectrum of $\eta(t)$. DMF
self-consistency enforces $C_\eta(\omega) = \sigma^2 C_y(\omega)$.
Using the explicit response for the linear dynamics,
$\chi(\omega) = (1 - i\omega)^{-1}$, one obtains
\begin{equation}
    C_y(\omega)
    = \frac{D}{1 + \omega^2 - \sigma^2}.
\end{equation}
In particular, the equal-time variance $C_y(0)$ diverges as
$\sigma \to 1$ from below, indicating that the diagonal entries of the
covariance matrix grow without bound as the system approaches the edge
of chaos.

\paragraph{Second moment of the correlations.}
We now relate the second moment of the covariance matrix to the
macroscopic four-point observable obtained in the beyond-DMF analysis.
To this end, we introduce the empirical scalar order parameter for a given realization of the quenched disorder:
\begin{equation}
    \widehat{\mathcal C}(\omega)
    := \frac{1}{S}\sum_{i=1}^S
       y_i(\omega)\,y_i(-\omega),
\end{equation}
so that its average is precisely the DMF self-covariance,
\begin{equation}
    \big\langle \widehat{\mathcal C}(\omega)\big\rangle
    = C_y(\omega).
\end{equation}
Fluctuations of this order parameter are described by
\begin{equation}
    \delta\widehat{\mathcal C}(\omega)
    := \widehat{\mathcal C}(\omega) - C_y(\omega).
\end{equation}
The four-point object computed in the field-theoretic expansion is the precisely the variance of $\widehat{\mathcal C}(\omega)$,
\begin{equation}
    \big\langle
        \delta\widehat{\mathcal C}(\omega)\,
        \delta\widehat{\mathcal C}(-\omega)
    \big\rangle
    =
    \big\langle
        \widehat{\mathcal C}(\omega)\,
        \widehat{\mathcal C}(-\omega)
    \big\rangle
    - \big|C_y(\omega)\big|^2.
    \label{eq:deltaC_variance_def}
\end{equation}
Using the definition of $\widehat{\mathcal C}(\omega)$, the first term
can be written as
\begin{equation}
\begin{aligned}
    \big\langle
        \widehat{\mathcal C}(\omega)\,
        \widehat{\mathcal C}(-\omega)
    \big\rangle
    &=
    \bigg\langle
        \frac{1}{S}\sum_{i=1}^S y_i(\omega)\,y_i(-\omega)\;
        \frac{1}{S}\sum_{j=1}^S y_j(-\omega)\,y_j(\omega)
    \bigg\rangle\\[4pt]
    &=
    \frac{1}{S^2}\sum_{i,j=1}^S
    \big\langle
        y_i(\omega)\,y_i(-\omega)\,
        y_j(-\omega)\,y_j(\omega)
    \big\rangle.
\end{aligned}
\end{equation}
Since, to second order in $S$, dynamics are Gaussian, Wick's theorem gives, for each pair
$(i,j)$,
\begin{equation}
\begin{aligned}
    \big\langle
        y_i(\omega)\,y_i(-\omega)\,
        y_j(-\omega)\,y_j(\omega)
    \big\rangle
    &=
    \big\langle y_i(\omega)y_i(-\omega)\big\rangle
    \big\langle y_j(-\omega)y_j(\omega)\big\rangle\\
    &\quad+
    \big\langle y_i(\omega)y_j(-\omega)\big\rangle
    \big\langle y_i(-\omega)y_j(\omega)\big\rangle\\
    &\quad+
    \big\langle y_i(\omega)y_j(\omega)\big\rangle
    \big\langle y_i(-\omega)y_j(-\omega)\big\rangle.
\end{aligned}
\end{equation}
In a stationary real process, Fourier components of the same sign are
uncorrelated for $\omega\neq 0$, so that
$\langle y_i(\omega)y_j(\omega)\rangle = 0$ and the last term vanishes.
In terms of the microscopic covariances $C_{ij}(\omega)$, we obtain
\begin{equation}
    \big\langle
        y_i(\omega)\,y_i(-\omega)\,
        y_j(-\omega)\,y_j(\omega)
    \big\rangle
    =
    C_{ii}(\omega)\,C_{jj}(\omega)
    + C_{ij}(\omega)\,C_{ij}(-\omega).
\end{equation}
Substituting this into the expression for
$\langle \widehat{\mathcal C}(\omega)\widehat{\mathcal C}(-\omega)\rangle$
and using
\begin{equation}
    \big|C_y(\omega)\big|^2
    = \left(\frac{1}{S}\sum_i C_{ii}(\omega)\right)^2
    = \frac{1}{S^2}\sum_{i,j} C_{ii}(\omega)\,C_{jj}(\omega),
\end{equation}
\eqref{eq:deltaC_variance_def} simplifies to
\begin{equation}
\begin{aligned}
    \big\langle
        \delta\widehat{\mathcal C}(\omega)\,
        \delta\widehat{\mathcal C}(-\omega)
    \big\rangle
    &=
    \frac{1}{S^2}\sum_{i,j}
    \Big[
        C_{ii}(\omega)\,C_{jj}(\omega)
        + C_{ij}(\omega)\,C_{ij}(-\omega)
    \Big]
    - \frac{1}{S^2}\sum_{i,j} C_{ii}(\omega)\,C_{jj}(\omega)\\[4pt]
    &=
    \frac{1}{S^2}\sum_{i,j} C_{ij}(\omega)\,C_{ij}(-\omega).
\end{aligned}
\end{equation}
Therefore, the macroscopic four-point fluctuation of the DMF order
parameter is directly related to the microscopic second moment of the
covariance matrix:
\begin{equation}
    \Psi_y(\omega)
    := S\,\big\langle
         \delta\widehat{\mathcal C}(\omega)\,
         \delta\widehat{\mathcal C}(-\omega)
       \big\rangle
    = \frac{1}{S}\sum_{i,j=1}^S
      C_{ij}(\omega)\,C_{ij}(-\omega).
    \label{eq:Psi_micro_link_clean}
\end{equation}
\paragraph{Normalized width and inference of the effective coupling.}

The relation \eqref{eq:Psi_micro_link_clean} provides the formal bridge
between the DMF four-point function and the microscopic second moment
of the covariance matrix. We now introduce the
normalized width of the covariance distribution, which is the quantity that we use in practice to estimate the variance of interactions, $\sigma^2$.

We start by defining the second moment of the off-diagonal covariances as:
\begin{equation}
    M_2(\omega)
    := \frac{1}{S(S-1)}
       \sum_{i\neq j} \big|C_{ij}(\omega)\big|^2.
\end{equation}
The (dimensionless) normalized width of the covariance distribution is
then defined by
\begin{equation}
    \Delta^2(\omega)
    := \frac{M_2(\omega)}{C_y(\omega)^2}.
    \label{eq:Delta_def_S}
\end{equation}
To relate $\Delta^2(\omega)$ to $\Psi_y(\omega)$, we decompose the
double sum into diagonal and off-diagonal contributions:
\begin{equation}
\begin{aligned}
    \sum_{i,j=1}^S C_{ij}(\omega)\,C_{ij}(-\omega)
    &=
    \sum_{i=1}^S C_{ii}(\omega)\,C_{ii}(-\omega)
    + \sum_{i\neq j} C_{ij}(\omega)\,C_{ij}(-\omega).
\end{aligned}
\end{equation}
Assuming exchangeability of units, the diagonal elements are all equal,
$C_{ii}(\omega)\approx C_y(\omega)$, so that
\begin{equation}
    \sum_{i=1}^S C_{ii}(\omega)\,C_{ii}(-\omega)
    \approx S\,|C_y(\omega)|^2,
\end{equation}
and the off-diagonal part can be expressed in terms of $M_2(\omega)$ as
\begin{equation}
    \sum_{i\neq j} C_{ij}(\omega)\,C_{ij}(-\omega)
    = S(S-1)\,M_2(\omega).
\end{equation}
Inserting these into the definition of $\Psi_y(\omega)$ yields
\begin{equation}
\begin{aligned}
    \Psi_y(\omega)
    &= \frac{1}{S}
       \left[
         S\,|C_y(\omega)|^2
         + S(S-1)\,M_2(\omega)
       \right]\\[4pt]
    &= |C_y(\omega)|^2 + (S-1)\,M_2(\omega).
\end{aligned}
\end{equation}
Solving for $M_2(\omega)$ and using the definition
\eqref{eq:Delta_def_S}, we obtain
\begin{equation}
    M_2(\omega)
    = \frac{\Psi_y(\omega) - |C_y(\omega)|^2}{S-1},
    \qquad
    \Delta^2(\omega)
    = \frac{\Psi_y(\omega)/|C_y(\omega)|^2 - 1}{S-1}.
    \label{eq:Delta_Psi_relation_S}
\end{equation}
In the large-$S$ limit, this simplifies to
\begin{equation}
    S\,\Delta^2(\omega)
    \simeq
    \frac{\Psi_y(\omega)}{|C_y(\omega)|^2} - 1.
    \label{eq:SDelta_Psi_relation_S}
\end{equation}

On the other hand, plugging this expression into \eqref{eq: normalized psi} gives:
\begin{equation}
    \frac{\Psi_y(\omega)}{|C_y(\omega)|^2}
    =
    \frac{1}{\bigl(1 - \sigma_{\mathrm{eff}}(\omega)^2\bigr)^2},
\end{equation}
where $\sigma_{\mathrm{eff}}(\omega)$ is the effective variance of the interactions at frequency $\omega$, inferred from the data. Inserting this into \eqref{eq:SDelta_Psi_relation_S}, we arrive at
\begin{equation}
    S\,\Delta^2(\omega)
    =
    \frac{1}{\bigl(1 - \sigma_{\mathrm{eff}}(\omega)^2\bigr)^2}
    - 1.
    \label{eq:SDelta_g_relation_S}
\end{equation}
Solving for $\sigma_{\mathrm{eff}}(\omega)$ gives
\begin{equation}
    \sigma_{\mathrm{eff}}(\omega)
    =
    \sqrt{
        1 - \sqrt{
            \frac{1}{1 + S\,\Delta^2(\omega)}
        }
    }.
    \label{eq:g_from_Delta_S}
\end{equation}
which allows us to estimate the effective variance of interactions $\sigma_\text{eff}(\omega)$ just by means of the moments of the histogram of covariances.

\subsection{Back to the original coordinates.}
In the previous sections, the beyond-DMF four-point observable was
related to the microscopic covariance matrix of the transverse
variables $y_i$ as
\begin{equation}
    \Psi_y(\omega)
    = \frac{1}{S}\sum_{i,j=1}^S
      C^{(y)}_{ij}(\omega)\,C^{(y)}_{ij}(-\omega),
    \label{eq:Psi_y_def}
\end{equation}
where
\begin{equation}
    C^{(y)}_{ij}(\omega)
    := \big\langle y_i(\omega)\,y_j(-\omega)\big\rangle.
\end{equation}
We now show that the same quantity, computed in the original coordinates $\mathbf{x}(t)$, is identical. This justifies computing the normalized width directly from the covariance of $\mathbf{x}$. Indeed, let $U$ be the orthogonal matrix defining the change of basis,
\begin{equation}
    \mathbf{X}(t) = U^\top \mathbf{x}(t),
    \qquad
    \mathbf{x}(t) = U\,\mathbf{X}(t),
\end{equation}
with $\mathbf{X}(t) = (X_0(t),y_1(t),\dots,y_{S-1}(t))^\top$. Denote by
$C^{(x)}(\omega)$ and $C^{(y)}(\omega)$ the covariance matrices in the
original and rotated bases, respectively,
\begin{equation}
    C^{(x)}_{ij}(\omega)
    := \big\langle x_i(\omega)\,x_j(-\omega)\big\rangle,
    \qquad
    C^{(y)}_{ab}(\omega)
    := \big\langle X_a(\omega)\,X_b(-\omega)\big\rangle.
\end{equation}
By orthogonality of $U$, these are related by
\begin{equation}
    C^{(x)}(\omega)
    = U\,C^{(y)}(\omega)\,U^\top.
    \label{eq:Cx_Cy_relation}
\end{equation}

We define the four-point observable in the original basis as
\begin{equation}
    \Psi_x(\omega)
    := \frac{1}{S}\sum_{i,j=1}^S
       C^{(x)}_{ij}(\omega)\,C^{(x)}_{ij}(-\omega).
    \label{eq:Psi_x_def}
\end{equation}
For a real stationary process one has
$C^{(x)}_{ij}(-\omega) = C^{(x)}_{ij}(\omega)^\dagger$ (where $\dagger$ denotes the complex conjugate), so that
\begin{equation}
    \sum_{i,j} C^{(x)}_{ij}(\omega)\,C^{(x)}_{ij}(-\omega)
    = \sum_{i,j} \big|C^{(x)}_{ij}(\omega)\big|^2
    = \big\|C^{(x)}(\omega)\big\|_{\mathrm{F}}^2,
\end{equation}
where $\|\cdot\|_{\mathrm{F}}$ denotes the Frobenius norm,
$\|A\|_{\mathrm{F}}^2 := \sum_{ij} |A_{ij}|^2 =
\mathrm{Tr}(A A^\dagger)$. Similarly,
\begin{equation}
    \sum_{i,j} C^{(y)}_{ij}(\omega)\,C^{(y)}_{ij}(-\omega)
    = \big\|C^{(y)}(\omega)\big\|_{\mathrm{F}}^2.
\end{equation}
Thus,
\begin{equation}
    \Psi_x(\omega)
    = \frac{1}{S}\big\|C^{(x)}(\omega)\big\|_{\mathrm{F}}^2,
    \qquad
    \Psi_y(\omega)
    = \frac{1}{S}\big\|C^{(y)}(\omega)\big\|_{\mathrm{F}}^2.
\end{equation}

Using the relation \eqref{eq:Cx_Cy_relation} and the invariance of the
Frobenius norm under orthogonal similarity transformations, we obtain
\begin{equation}
    \big\|C^{(x)}(\omega)\big\|_{\mathrm{F}}^2
    = \big\|C^{(y)}(\omega)\big\|_{\mathrm{F}}^2,
\end{equation}
and hence
\begin{equation}
    \Psi_x(\omega)
    = \Psi_y(\omega).
\end{equation}

\subsection{From long- to short-time window correlations}
\label{subsec: g DMFT}
Throughout this whole section, we derived the relation between the normalized width of covariances and the diversity of interactions, $\sigma$. At $\omega=0$, this corresponds to studying the dynamics at very long times. However, in most cases, the empirically available object is the short-time window covariance matrix. In order to derive a similar expression for the short-time window covariance, we have to come back from the frequency domain to the time domain, exploiting the fact that, for a stationary process,
\begin{equation}
    C_{ij}(\tau=0)
    = \langle y_i(t) y_j(t+0)\rangle
    = \int_{\mathbb{R}} \frac{d\omega}{2\pi}\, C_{ij}(\omega) e^{-i\omega\cdot 0}
    = \int_{\mathbb{R}} \frac{d\omega}{2\pi}\, C_{ij}(\omega).
\end{equation}
For example, the average self-covariance (diagonal terms in the short-time window covariance matrix) is given by
\begin{equation}
    C_y^\text{short}
    = \int_{\mathbb{R}} \frac{D}{1+\omega^2-\sigma^2}\,\frac{d\omega}{2\pi}
    = \frac{D}{2\sqrt{1-\sigma^2}},
\end{equation}
which describes the divergence of the self-covariance terms when the system approaches the edge of chaos.

On the other hand, in order to calculate the fluctuations in the correlations, we must Fourier-transform back the expression for the fourth-point moment in frequency space,
\begin{equation}
    \Psi(\omega_1,\omega_2)
    = \frac{C(\omega_1)C(\omega_2)}{|1-\sigma^2\chi(\omega_1)\chi(\omega_2)|^2},
\end{equation}
where $C(\omega)=D/(1+\omega^2-\sigma^2)$ and $\chi(\omega)=1/(1-i\omega)$. The short-time connected four-point function is then
\begin{equation}
    \Psi^\text{short}(\tau_1=0, \tau_2=0)
    = \big\langle \delta C(\tau_1=0)\,\delta C(\tau_2=0)\big\rangle
    = \int \frac{d\omega_1}{2\pi}\,\frac{d\omega_2}{2\pi}\,
    \frac{C(\omega_1)C(\omega_2)}{|1-\sigma^2\chi(\omega_1)\chi(\omega_2)|^2}.
\end{equation}

Directly working with the modulus in the complex plane is not convenient. Instead, we introduce the analytic function $\tilde\chi(\omega)=1/(1+i\omega)$, which coincides with $\chi^*(\omega)$ on the real axis, and rewrite
\begin{equation}
    |1-\sigma^2\chi(\omega_1)\chi(\omega_2)|^2
    = \bigl(1-\sigma^2\chi(\omega_1)\chi(\omega_2)\bigr)
      \bigl(1-\sigma^2\tilde\chi(\omega_1)\tilde\chi(\omega_2)\bigr).
\end{equation}
A straightforward algebra then shows that, for real $\omega_1,\omega_2$,
\begin{equation}
    |1-\sigma^2\chi(\omega_1)\chi(\omega_2)|^2
    = \frac{(\lambda-\omega_1\omega_2)^2+(\omega_1+\omega_2)^2}
           {(1+\omega_1^2)(1+\omega_2^2)},
    \qquad \lambda := 1-\sigma^2.
\end{equation}
Plugging this back into $\Psi(\omega_1,\omega_2)$ yields the fully rational form
\begin{equation}
    \Psi(\omega_1,\omega_2)
    = \frac{D^2}{(\omega_1^2+\lambda)(\omega_2^2+\lambda)}\,
      \frac{(1+\omega_1^2)(1+\omega_2^2)}
           {(\lambda-\omega_1\omega_2)^2+(\omega_1+\omega_2)^2}.
\end{equation}
Therefore
\begin{equation}
    \Psi^\text{short}(0,0)
    = \frac{D^2}{(2\pi)^2}
      \int_{\mathbb{R}} d\omega_2\,\frac{1+\omega_2^2}{\omega_2^2+\lambda}
      \int_{\mathbb{R}} d\omega_1\,
      \frac{1+\omega_1^2}
           {(\omega_1^2+\lambda)\,\big[(\lambda-\omega_1\omega_2)^2+(\omega_1+\omega_2)^2\big]}.
\end{equation}
We now perform the $\omega_1$ integral by residues. For fixed real $\omega_2$, the integrand has poles at
\begin{equation}
    \omega_1 = \pm i\sqrt{\lambda},\qquad
    \omega_1 = \frac{\lambda+i\omega_2}{\omega_2-i},\qquad
    \omega_1 = \frac{\lambda-i\omega_2}{\omega_2+i}.
\end{equation}
Closing the contour in the upper half-plane, only two of these poles contribute:
\begin{equation}
    \omega_1 = i\sqrt{\lambda},\qquad
    \omega_1 = \omega_+ := \frac{\lambda+i\omega_2}{\omega_2-i}.
\end{equation}
Evaluating the residues at these poles and summing them, one obtains the inner integral in closed form,
\begin{equation}
    \int_{\mathbb{R}} d\omega_1\,
    \frac{1+\omega_1^2}
         {(\omega_1^2+\lambda)\,\big[(\lambda-\omega_1\omega_2)^2+(\omega_1+\omega_2)^2\big]}
    = \pi\,\frac{\lambda+1-\sqrt{\lambda} + \dfrac{\omega_2^2}{\sqrt{\lambda}}}
              {(\lambda+\omega_2^2)^2}
    \equiv J(\omega_2).
\end{equation}
Thus the short-time four-point amplitude reduces to a single real integral:
\begin{equation}
    \Psi^\text{short}(0,0)
    = \frac{D^2}{4\pi}
      \int_{\mathbb{R}} d\omega_2\,
      \frac{(1+\omega_2^2)\,\bigl(\lambda+1-\sqrt{\lambda} + \dfrac{\omega_2^2}{\sqrt{\lambda}}\bigr)}
           {(\lambda+\omega_2^2)^3}.
\end{equation}
This last integral can be evaluated again by residues or by using standard one-dimensional integrals of the form $\int d\omega/(\omega^2+\lambda)^n$. After a straightforward but somewhat lengthy algebra one finds
\begin{equation}
    \Psi^\text{short}(0,0)
    = \frac{D^2}{32}\left(
        \frac{2}{\lambda}
      - \frac{2}{\lambda^2}
      + \frac{1}{\sqrt{\lambda}}
      + \frac{4}{\lambda^{3/2}}
      + \frac{3}{\lambda^{5/2}}
    \right).
\end{equation}
Dividing by the squared short-time self-covariance,
\begin{equation}
    \big(C_y^\text{short}\big)^2
    = \left(\frac{D}{2\sqrt{\lambda}}\right)^2
    = \frac{D^2}{4\lambda},
\end{equation}
we obtain
\begin{equation}
    \frac{\Psi^\text{short}}{\big(C_y^\text{short}\big)^2}
    = \frac{1}{8}\left(
        2 - \frac{2}{\lambda}
      + \sqrt{\lambda}
      + \frac{4}{\sqrt{\lambda}}
      + \frac{3}{\lambda^{3/2}}
    \right).
\end{equation}
We now define the normalized variance of short-time correlations as
\begin{equation}
\label{eq: g estimation}
    S\Delta_\text{short}^2
    := \frac{\Psi^\text{short}}{\big(C_y^\text{short}\big)^2} - 1
    = \frac{1}{8}\left(
        -6 - \frac{2}{\lambda}
        + \sqrt{\lambda}
        + \frac{4}{\sqrt{\lambda}}
        + \frac{3}{\lambda^{3/2}}
    \right).
\end{equation}
In terms of $x=\sqrt{\lambda}$, this relation can be written as a quartic equation,
\begin{equation}
    8 S\Delta_\text{short}^2 x^3
    = -6x^3 - 2x + x^4 + 4x^2 + 3,
\end{equation}
or equivalently
\begin{equation}
    x^4 - (8S\Delta_\text{short}^2 + 6)x^3 + 4x^2 - 2x + 3 = 0,
    \qquad x=\sqrt{\lambda},\ \ \lambda=1-\sigma^2.
\end{equation}
This algebraic equation implicitly relates the interaction variance $\sigma^2$ to the (normalized) fluctuations of short-time window correlations. Since we are interested in the regime close to the edge of chaos, it is convenient to consider the expansion for $\lambda\simeq 1$ (small $\sigma$). Writing $\lambda = 1-\varepsilon$ with $\varepsilon\ll 1$, we find
\begin{equation}
    S\Delta_\text{short}^2
    = \frac{\varepsilon}{2} + \frac{5}{8}\varepsilon^2 + \mathcal{O}(\varepsilon^3),
\end{equation}
which can be inverted to obtain
\begin{equation}
    \varepsilon \equiv 1-\lambda
    \simeq 2 S\Delta_\text{short}^2 - 5\left(S\Delta_\text{short}^2\right)^2 + \mathcal{O}\big((S\Delta_\text{short}^2)^3\big).
\end{equation}
Using $\lambda = 1-\sigma^2$, this finally yields the near-critical estimate
\begin{equation}
    \sigma^2
    \simeq 2 S\Delta_\text{short}^2
          - 5\left(S\Delta_\text{short}^2\right)^2
          + \mathcal{O}\big((S\Delta_\text{short}^2)^3\big),
\end{equation}
which can be used as a practical estimator relating the fluctuations of short-time window correlations to the diversity of interactions in the large--$S$ limit. 

All of the calculations shown here are for a Gaussian matrix with connectivity $C=1$. In the main text, we discuss the general case in which the connectivity $C$ is fixed to a given value. In that case, the total variance of the interactions, let us denote it as $g$, is nothing but $\sigma\cdot \sqrt{C}$ and it can be calculated by solving \eqref{eq: g estimation}, where $\lambda = 1-g^2$ in that case.

\begin{figure}[!t]
\centering
\includegraphics[width=1.0\textwidth]{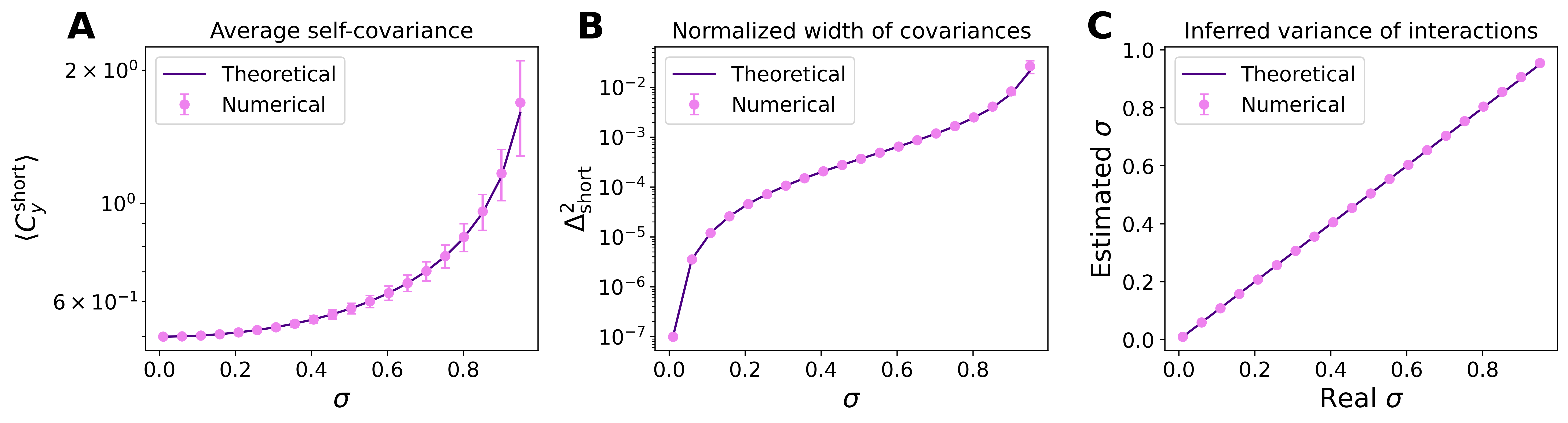}
\caption{\textbf{Predictions from DMFT match numerical experiments from random networks of $S=500$ species, averaged over $M=10^3$ independent realizations}. \textbf{A}. The average self covariance scales with the $\sigma$, the fluctuations of the quenched disorder, as $(1-\sigma^2)^{-1/2}$, when the system approaches the edge of instability. Numerical simulations, carried out by solving the Lyapunov equation numerically (in order to avoid possible finite-sampling effects) match theory perfectly. \textbf{B}. The normalized width of correlations, $\Delta_\text{short}^2$ diverges as $(1-\sigma^2)^{-3/2}$ as $\sigma$ approaches $1$. \textbf{C}. Using DMFT, $\sigma$ can be inferred just by measuring the normalized width of covariances. As observed, the estimator matches perfectly the real value of $\sigma$.}
\label{fig: SI_3}
\end{figure}



\section{The emergence of chaos}

The idea that irregular fluctuations in ecological community abundances could be driven by deterministic chaotic dynamics has been debated—and often contested—in ecology for decades \cite{may_simple_1976, bjornstad_nonlinearity_2015}. Nonetheless, controlled experiments have reported clear signatures of chaos, including aperiodic trajectories and strange attractors, in insect populations \cite{costantino_experimentally_1995, costantino_chaotic_1997} and in simplified marine communities (barnacles, algae, and mussels) monitored in a marine reserve \cite{beninca_species_2015}. More recently, accumulating empirical and theoretical evidence suggests that chaotic dynamics may be more common in natural ecosystems than previously assumed \cite{Rogers}.

On the other hand, a growing body of theoretical and numerical work indicates that communities with random interactions can exhibit chaotic dynamics in generalized Lotka–Volterra models \cite{Bunin2024, Roy-chaos, DeMonte-chaos}, as well as in more complex consumer–resource models \cite{Mehta-PRL2024}.

In the main text, we argue that microbial communities operate in a fixed-point regime, close to the transition to chaos. This working hypothesis is attractive because it yields well-defined long-time mean abundances and maintains marginal statistics compatible with Grilli’s laws \cite{Grilli2020}. By contrast, in the chaotic regime the DMFT description contains a nonvanishing self-consistent noise term, which drives persistent fluctuations of species abundances. In Section \ref{Sec: Grilli DMFT}, we show that these fluctuations generically spoil the marginal patterns predicted by Grilli’s laws. This motivates focusing on the fixed-point side of the transition.

This section provides a more detailed discussion of chaotic dynamics in the interacting stochastic logistic model introduced in the main text. We first analyze the deterministic (noise-free) limit, and then extend the discussion to the fully stochastic setting with environmental noise.

\subsection{Chaos in the deterministic limit}
\paragraph{Chaos is fueled by immigration, and impossible with extinctions.} Whether a simple model with random interactions can sustain chaos is an interesting question in its own right, and it has motivated a substantial body of work in recent years. In generalized Lotka–Volterra models, chaotic dynamics typically entail large abundance fluctuations; in ecological settings, such fluctuations often drive some species to extinction \cite{DeMonte-chaos, Bunin2024}. In a purely deterministic ODE formulation, however, extinction is not an absorbing event: species abundances remain strictly positive and can only approach zero asymptotically. This idealization is not biologically realistic: real communities are composed of a finite number of individuals, so sufficiently low abundances effectively imply extinction.

The only way to circumvent this issue within a deterministic ODE description is to modify the dynamics near vanishing abundances. A common choice is to include an infinitesimally small (but nonzero) immigration rate $\lambda$, which prevents abundances from collapsing to zero and thereby avoids effective extinctions \cite{Bunin2024}. \footnote{A more rigorous alternative is to include demographic noise, which can drive populations to extinction and trap them at zero abundance through a genuine absorbing state \cite{Nature, Dornic}.} From a modeling perspective, this corresponds to allowing individuals to enter the community from an external pool, i.e., to treating the system as open rather than isolated and thus able to exchange biomass with its surroundings \cite{Fisher-chaos,Fisher2023} The resulting dynamics can be written as
\begin{equation}
    \label{eq: GLV migration}
    \dot{x}_i(t)
    = \frac{x_i(t)}{\tau}\left[ 1 + \sum_{j=1}^S A_{ij}\,x_j(t) \right]+\lambda.
\end{equation}
We can therefore treat extinctions in three different ways:
\begin{itemize}
    \item \textbf{Route 1:} Ignore extinctions altogether. Species can reach arbitrarily low abundances, which is mathematically consistent but biologically unrealistic.
    \item \textbf{Route 2:} Impose an extinction threshold $T_e$ and remove species once $x_i(t) < T_e$. This approximates demographic discreteness (finite numbers of individuals).
    \item \textbf{Route 3:} Add a small immigration rate $\lambda$ to prevent long-time collapse to zero, mimicking exchange with a surrounding environment.
\end{itemize}
Using these different approaches, we quantify sensitivity to initial conditions by computing the largest Lyapunov exponent (LLE), which measures the exponential divergence of nearby trajectories and is a standard diagnostic of chaos in deterministic dynamical systems \cite{eckmann_ergodic_1985}.
 
\begin{figure}[t]
    \centering
\includegraphics[width=\linewidth]{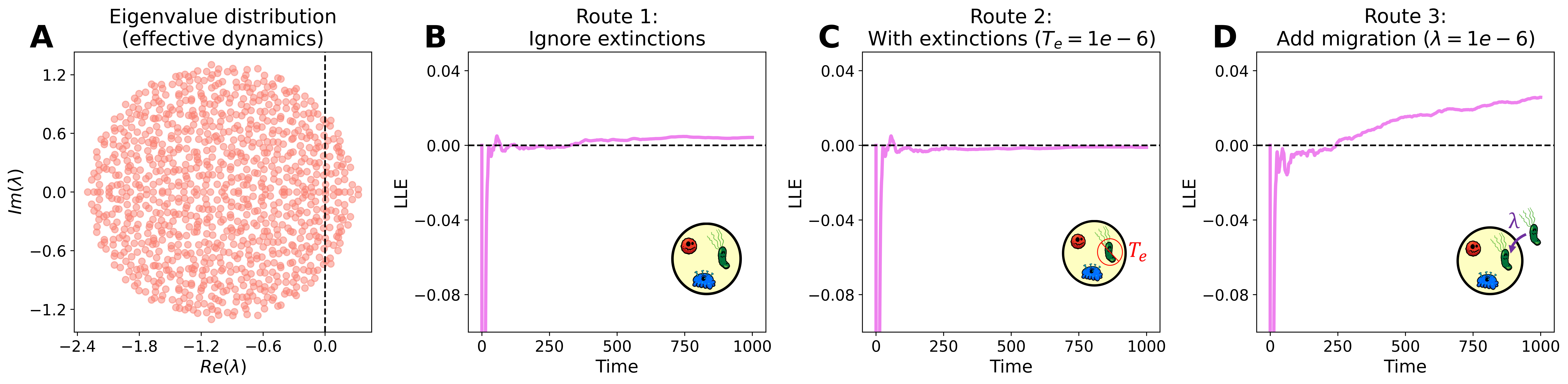}
    \caption{\textbf{Chaos in the generalized Lotka-Volterra model under different treatments of extinction in a system with $S=10^3$, connectivity $C=1.0$, and interaction strength $\sigma=1.3$}.
\textbf{(A)} Eigenvalue distribution of the linearized effective dynamics (dashed line marks $\mathrm{Re}(\lambda)=0$).
\textbf{(B--D)} Time evolution of the largest Lyapunov exponent (LLE) for three protocols:
\textbf{(B)} Route 1 (no explicit extinction handling),
\textbf{(C)} Route 2 (remove species once $x_i<T_e$ with $T_e=10^{-6}$),
and \textbf{(D)} Route 3 (add migration $\lambda=10^{-6}$).
Routes 1 and 3 exhibit sustained positive LLE, indicating persistent chaos, whereas Route 2 displays only transient positivity before extinctions reduce the effective dimensionality and the dynamics relax to a stable state. Migration replenishes depleted species, effectively fueling the fluctuations that sustain chaotic dynamics.
}
    \label{fig: SI_chaos_1}
\end{figure}

 In Figure \ref{fig: SI_chaos_1}, we present numerical estimates of the Lyapunov exponent \cite{benettin_lyapunov_1980, pikovskij_lyapunov_2016} obtained under the three extinction-handling protocols, for a system with $S=10^3$, connectivity $C=1.0$, and interaction strength $\sigma=1.3$. Panel A shows the eigenvalue spectrum of the linearized dynamics, revealing that many modes are linearly unstable. Panels B--D report the time-dependent Lyapunov exponent for the three routes (ignoring extinctions, imposing an extinction threshold, and adding immigration). We observe that Routes 1 and 3 yield sustained chaotic dynamics, whereas Route 2 does not. Under Route 2, the Lyapunov exponent can transiently become positive: the dynamics initially display chaos, but as extinctions accumulate, the effective dimensionality of the system is reduced. The system then slowly relaxes toward a lower-dimensional attractor whose dynamics become stable again. By contrast, immigration enables chaos to persist by continually replenishing depleted species, effectively fueling the fluctuations that sustain the chaotic state. 

From here on, we focus on the study of the system with a small immigration rate, that will be set to $\lambda=10^{-6}$, unless otherwise stated. Figure 2
in the main text is reproduced  following this choice.

\begin{figure}[t]
    \centering
    \includegraphics[width=\linewidth]{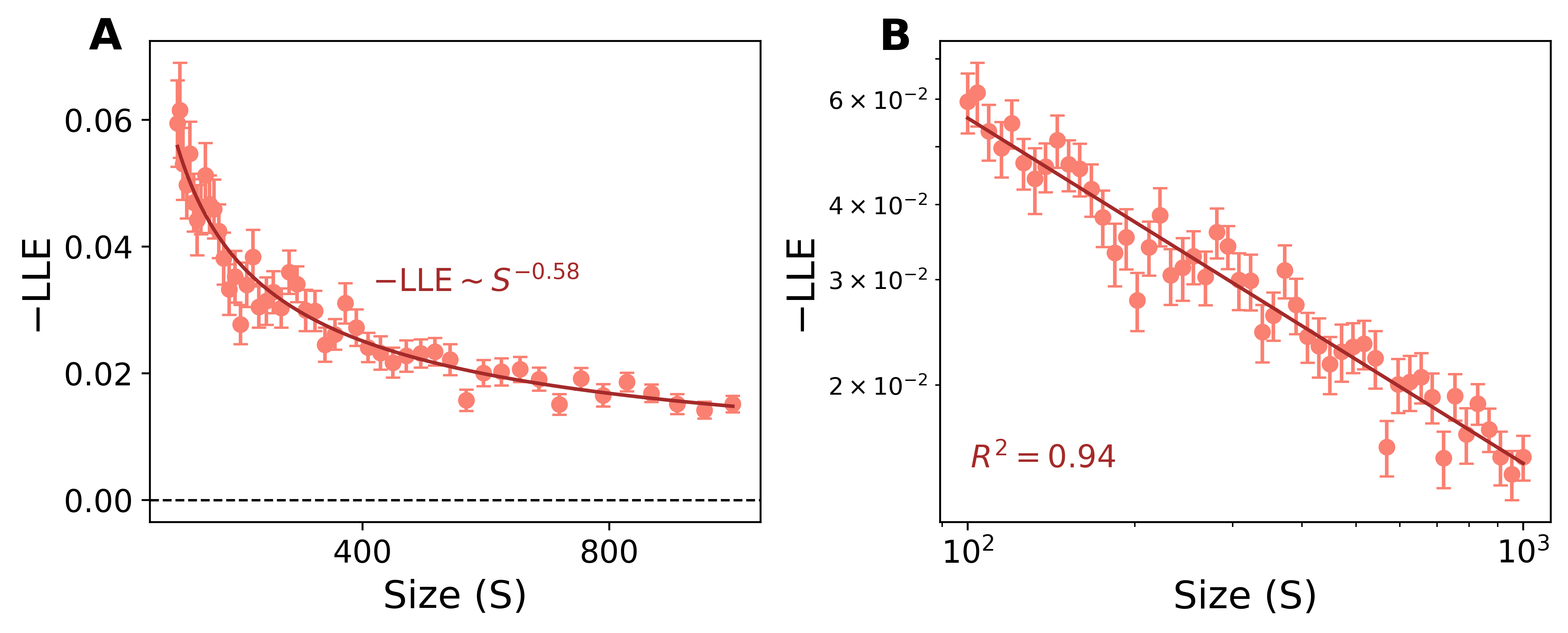}
    \caption{\textbf{Finite-size scaling of the largest Lyapunov exponent at the DMFT-predicted boundary.}
    Largest Lyapunov exponent (LLE) measured at $(\sigma,C)=(1,1)$ as a function of system size $S$ (points represent mean values, while error bars stand for standard deviations over realizations -normalized by the square root of the number of measurements, which was $M=50$). \textbf{A}: linear axes; \textbf{B}: log--log representation. The solid line shows a power-law fit, $-\mathrm{LLE}\sim S^{-\alpha}$, with $R^2$ reported in the log--log panel. The observed decay of the (negative) LLE towards zero with increasing $S$ indicates that the system approaches marginal stability as $S\to\infty$, consistent with the thermodynamic-limit transition occurring at $\sigma\sqrt{C}=1$.}
    \label{fig:SI_chaos_2}
\end{figure}

\paragraph{The boundary of chaos in the thermodynamic limit.}
The DMFT description of the generalized Lotka--Volterra model used in this work suggests that the transition to chaos occurs at the marginal stability condition
\begin{equation}
\sigma \sqrt{C}=1,
\end{equation}
at least in the thermodynamic limit ($S\to\infty$). For finite communities, finite-size effects can shift the apparent transition, effectively delaying the onset of chaotic dynamics. We test this prediction by measuring the largest Lyapunov exponent (LLE) at the putative critical point $(\sigma,C)=(1,1)$ for increasing system size. As shown in Fig.~\ref{fig:SI_chaos_2}, the (negative) LLE decreases in magnitude with $S$ and is consistent with approaching zero as $S$ grows, indicating asymptotic marginal stability. This supports the DMFT conclusion that, in the thermodynamic limit, the onset of chaos is located precisely at $\sigma\sqrt{C}=1$.

\subsection{Noisy chaos}
Let us now study chaos in the presence of stochastic fluctuations,
which we model as a multiplicative noise term. Explicitly, we consider the
ISLM analyzed in the main text:
\begin{equation}
    \label{eq: GLV migration and noise}
    \dot{x}_i(t)
    = \frac{x_i(t)}{\tau}\left[ 1 + \sum_{j=1}^S A_{ij}\,x_j(t) \right]
      + \lambda
      + \sigma_0 x_i(t)\,\xi_i(t).
\end{equation}
The interplay between noise and chaos has been studied extensively \cite{bao_competitive_2011, fiasconaro_noise_2004}. Depending on the context, noise can be stabilizing—for instance by suppressing chaotic dynamics \cite{schuecker_optimal_2018}—or destabilizing, in the sense that it can induce chaotic-like behavior in systems that are non-chaotic in the deterministic limit \cite{scheuring_only_2007, bassols-cornudella_noise-induced_2023}.

In the context of dynamical systems, the theory of random dynamical systems
(RDS) \cite{arnold_random_1998} provides a natural generalization of chaos in
the presence of noise. The stochastic dynamics generated by \eqref{eq: GLV migration and noise} defines a random flow
$\Phi(t,\omega,x)$ driven by the noise realization $\omega$ (the collection of
paths $\{\xi_i(t)\}_i$). For each fixed realization $\omega$, the long-time
behaviour is captured by a random attractor $A(\omega)$ and an associated
sample Lyapunov spectrum. In particular, the largest Lyapunov exponent (LLE)
$\lambda_1(\omega)$ quantifies the exponential divergence of nearby
trajectories for that noise realization. Following standard practice \cite{lajoie_chaos_2013}, we characterize chaotic behaviour under noisy dynamics by the (disorder- and noise-averaged) LLE, i.e. by the expectation of $\lambda_1(\omega)$ over realizations of the noise $\xi_i(t)$ and the quenched disorder $A_{ij}$.

\begin{figure}[t]
    \centering
    \includegraphics[width=0.9\textwidth]{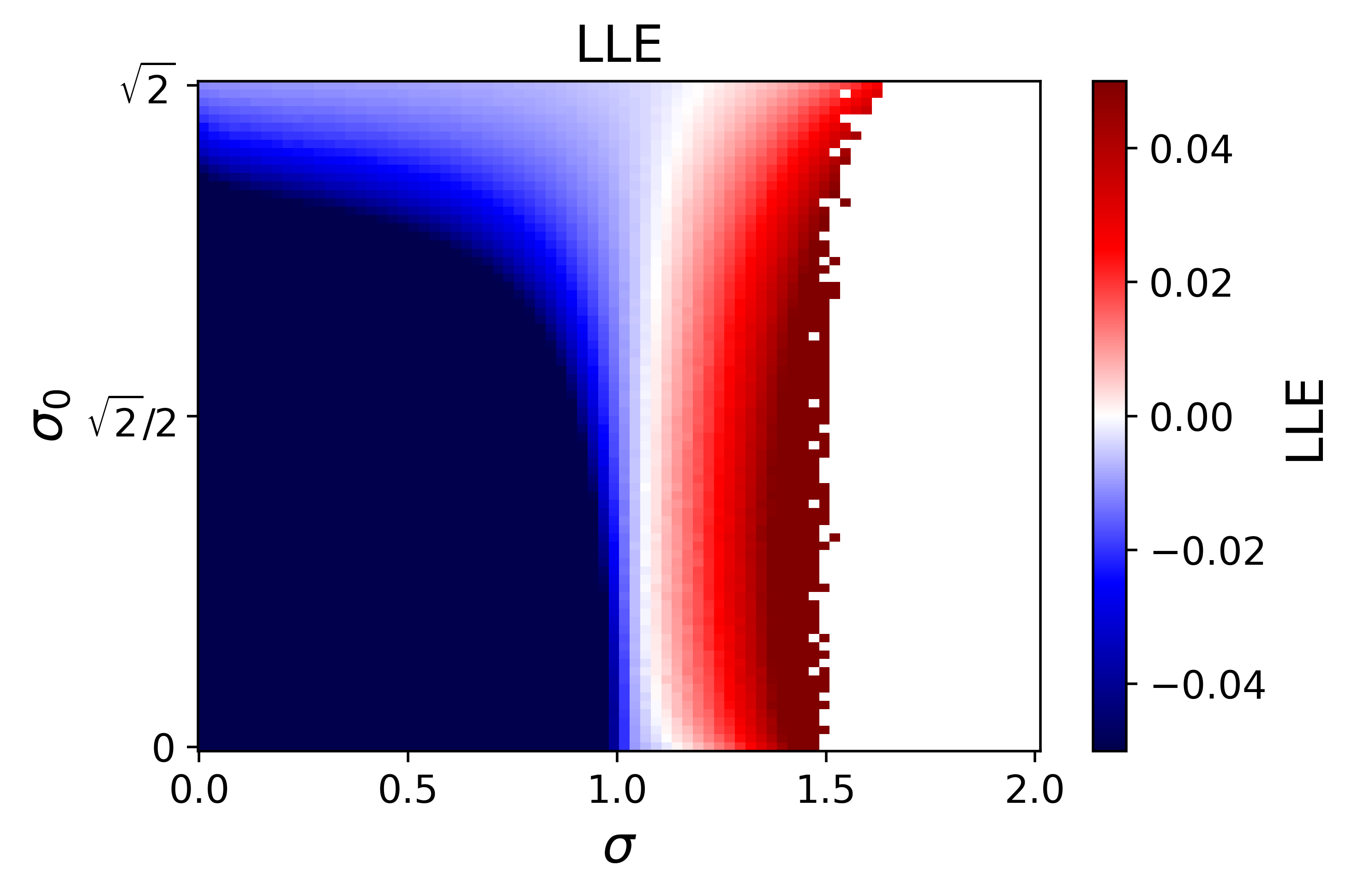}
    \caption{
        \textbf{Largest Lyapunov exponent (LLE) as a function of the interaction strength
        $\sigma$ and noise amplitude $\sigma_0$ for the noisy ISLM in
        \eqref{eq: GLV migration and noise}}. The color map shows the
        disorder- and noise-averaged LLE over $M = 50$ realizations of the
        interaction matrix and noise, for a system of size $S = 10^3$ as a function of $\sigma$ and $\sigma_0$. The white
        contour marks the locus where the LLE crosses zero, separating the
        chaotic regime (LLE $> 0$) from the stable or marginally stable regime
        (LLE $\le 0$).
    }
    \label{fig:SI_chaos_3}
\end{figure}

To quantify how noise modulates chaos, we computed the largest Lyapunov
exponent (LLE) across the $(\sigma,\sigma_0)$ plane for large systems. As we proved in section \ref{sec: The ISLM}, there is a natural bound for how big the fluctuations can be, i.e., $\sigma_0<\sqrt{2/\tau}$ which, with this choice of parameters ($\tau=1$) translates to $\sigma_0<\sqrt{2}$. Figure \ref{fig:SI_chaos_3} shows the resulting phase diagram for $S=10^3$
species, averaged over $M=50$ realizations of the interaction matrix and the
noisy drive. The white contour indicates the region where the LLE vanishes and
thus separates a regime with robust noisy chaos (LLE $>0$) from a regime where
the dynamics are stable or marginal (LLE $\le 0$). This confirms that the chaotic
phase observed in the deterministic limit persists under stochastic
fluctuations over a broad range of noise amplitudes.

\section{Empirical analysis of gut microbiome cohorts}

In the main text, we analyzed gut microbiome data across several disease cohorts and healthy controls (inflammatory bowel disease -IBD-, colorectal cancer -CRC-, \textit{Clostridioides difficile} infection -CDI-,
irritable bowel syndrome -IBS-, and their corresponding healthy controls), and compared their inferred distances to criticality (Figure 5). Because the IBD dataset contains the largest number of samples, it provides particularly reliable estimates of the distance to criticality and serves as a natural benchmark (see Table \ref{tab:disease_samples_taxa}). In this section, we describe in detail the datasets and the subsampling-and-inference pipeline used to obtain the main-text results, and we systematically examine how these results depend on methodological choices. In particular, we report the specific subsampling parameters $(r,p)$ used for each cohort, and we test the robustness of the inferred distances to criticality against variations in these parameters and in the number of samples.

\subsection{Description of the data}

For the disease analyses, we used the following datasets:
\begin{itemize}
    \item IBD: Metagenomic data were obtained from the Inflammatory Bowel Disease Multi’omics Database (https://ibdmdb.org/) and \cite{LloydPrice}.
    \item IBS: Metagenomic data were obtained from \cite{Mars}. 
    \item CDI: Metagenomic data were obtained from \cite{Ferretti}.
    \item CRC: Metagenomic data were obtained from \cite{Yachida}. 
    \item Independent Healthy cohorts: Two independent healthy-control cohorts based on 16S rRNA sequencing were obtained from \cite{Park}.
\end{itemize}
All these datasets were used in the pre-processed form curated by their original authors. We refer the reader to the corresponding publications for detailed descriptions of pre-processing pipelines. For our analyses, the only additional step was normalizing count data to obtain relative abundances for each sample. The resulting total number of samples and species is shown at Table \ref{tab:disease_samples_taxa}.

\begin{table}[h!]
    \centering
    \begin{tabular}{llll}
        \hline
        Disease & Group         & \# Samples & \# Taxa \\
        \hline
        IBD  & CD             & 748 & 329  \\
             & UC             & 454 & 378  \\
             & Healthy         & 428 & 284  \\
        \hline
        CRC  & Healthy        & 251 & 517  \\
             & Stage\_0       & 73  & 421  \\
             & Stage\_I\_II   & 111 & 479  \\
             & Stage\_III\_IV & 74  & 436  \\
        \hline
        CDI  & Healthy        & 186 & 639 \\
             & Unhealthy      & 775 & 234 \\
        \hline
        IBS  & Healthy        & 152 & 1525 \\
             & IBS-C          & 148 & 1679 \\
             & IBS-D          & 176 & 1601 \\
        \hline
        Healthy  & Healthy 1    & 642 & 292 (Genus)  \\
        Healthy  & Healthy 2    & 954 & 367 (Genus)  \\
    \end{tabular}
    \caption{Number of samples and taxa, generally species unless otherwise stated, per disease group.}
    \label{tab:disease_samples_taxa}
\end{table}

\subsection{Subsampling pipeline based on the species--samples ratios \(r\) and \(p\)}

For each disease cohort and diagnostic group (Table~\ref{tab:disease_samples_taxa}) we start from an abundance matrix
\[
    X \in \mathbb{R}^{S \times T},
\]
where rows index taxa (species) and columns index samples. We then apply the following subsampling pipeline to estimate the distance to criticality from correlations.

\paragraph{Definition of the control parameters \(p\) and \(r\).}
We control two dimensionless ratios:
\begin{enumerate}
    \item A \emph{subset fraction} \(p \in (0,1]\), which fixes the number of samples used in each subsample,
    \begin{equation}
        T_2 = \lfloor p\,T \rfloor,
    \end{equation}
    where $\lfloor x \rfloor$ denotes the floor function of $x$.
    \item A \emph{species--samples ratio} \(r>0\), defined at the subsample level as
    \begin{equation}
        r \equiv \frac{S_{\mathrm{eff}}}{T_2},
    \end{equation}
    where \(S_{\mathrm{eff}}\) is the number of taxa that remain in a given subsample after all filters (see below). In practice \(r\) is enforced as an upper bound on \(S_{\mathrm{eff}}/T_2\) (we cannot create taxa if too few are present).
\end{enumerate}

\paragraph{Per-subsample procedure.}
Given \(X\), and fixed \((r,p)\), we generate \(M\) subsamples, indexed by \(m = 1,\dots,M\). For each subsample we:

\begin{enumerate}
    \item \textbf{Sample columns (samples).} We choose a set of columns \(J_m \subset \{1,\dots,T\}\) of size
    \begin{equation}
        |J_m| = T_2 = \lfloor p\,T \rfloor,
    \end{equation}
    uniformly at random without replacement, and form the restricted abundance matrix
    \begin{equation}
        X^{(m)}_{\mathrm{full}} = X[:,J_m] \in \mathbb{R}^{S \times T_2},
    \end{equation}
    where we used $X[:,J_m]$ to indicate the submatrix formed by taking all the rows but only the columns in $J_m$.

    \item \textbf{Presence filter within the subsample.}
    We define presence of taxon \(i\) in sample \(t\) within this subsample by
    \begin{equation}
        \mathrm{present}_{i t} =
        \begin{cases}
            1, & X^{(m)}_{\mathrm{full}}(i,t) \ge \tau,\\
            0, & \text{otherwise},
        \end{cases}
    \end{equation}
    with a small threshold \(\tau\) (we used \(\tau = 10^{-5}\)). For each taxon we compute the number of presences,
    \begin{equation}
        k_i^{(m)} = \sum_{t=1}^{T_2} \mathrm{present}_{i t}.
    \end{equation}
    We retain only taxa with at least \(\texttt{min\_present}\) presences (we used \(\texttt{min\_present}=3\)):
    \begin{equation}
        \mathcal{I}_m = \bigl\{ i \in \{1,\dots,S\} : k_i^{(m)} \ge \texttt{min\_present} \bigr\}.
    \end{equation}
    If \(|\mathcal{I}_m| < \texttt{min\_rows}\) (with \(\texttt{min\_rows}=3\)), this subsample is discarded. Otherwise we define the restricted matrix
    \begin{equation}
        X^{(m)} = X^{(m)}_{\mathrm{full}}[\mathcal{I}_m,:] \in \mathbb{R}^{S_{\text{raw}} \times T_2},
    \end{equation}
    where \(S_{\text{raw}} = |\mathcal{I}_m|\).

    \item \textbf{Enforcing the species--samples ratio.}
    The target ratio is \(r = S_{\mathrm{eff}}/T_2\); internally we enforce a lower bound on the samples-per-species ratio
    \begin{equation}
        \rho_{\mathrm{min}} \equiv \frac{T_2}{S_{\mathrm{eff}}} = \frac{1}{r}.
    \end{equation}
    Given \(T_2\) and the current number of taxa \(S_{\text{raw}}\), we compute a maximal allowed number of taxa
    \begin{equation}
        S_{\max} = \left\lfloor \frac{T_2}{\rho_{\mathrm{min}}} \right\rfloor
                  = \left\lfloor r\,T_2 \right\rfloor.
    \end{equation}
    If \(S_{\text{raw}} \le S_{\max}\) we keep all taxa. If \(S_{\text{raw}} > S_{\max}\), we rank taxa according to (i) their occupancy within the subsample,
    \begin{equation}
        \mathrm{occ}_i^{(m)} = \frac{k_i^{(m)}}{T_2},
    \end{equation}
    and (ii) their total abundance,
    \begin{equation}
        A_i^{(m)} = \sum_{t=1}^{T_2} X^{(m)}(i,t),
    \end{equation}
    and retain the \(S_{\max}\) taxa with highest \(\mathrm{occ}_i^{(m)}\) (breaking ties by larger \(A_i^{(m)}\)). This yields a final matrix
    \begin{equation}
        \tilde{X}^{(m)} \in \mathbb{R}^{S_{\mathrm{eff}}^{(m)} \times T_2},
    \end{equation}
    with
    \begin{equation}
        S_{\mathrm{eff}}^{(m)} \le S_{\max}
        \quad\Rightarrow\quad
        \frac{S_{\mathrm{eff}}^{(m)}}{T_2} \le r.
    \end{equation}
    The \emph{effective} ratio in subsample \(m\) is then
    \begin{equation}
        r_{\mathrm{eff}}^{(m)} = \frac{S_{\mathrm{eff}}^{(m)}}{T_2},
    \end{equation}
    and we report its average \(\langle r_{\mathrm{eff}}\rangle\) across subsamples.

    \item \textbf{Standardization and removal of constant taxa.}
    For each taxon \(i\) in \(\tilde{X}^{(m)}\) we compute its mean and standard deviation across the \(T_2\) samples,
    \begin{equation}
        \mu_i^{(m)} = \frac{1}{T_2} \sum_{t=1}^{T_2} \tilde{X}^{(m)}(i,t), 
        \qquad
        \sigma_i^{(m)} = \sqrt{\frac{1}{T_2-1} \sum_{t=1}^{T_2}
            \bigl(\tilde{X}^{(m)}(i,t) - \mu_i^{(m)}\bigr)^2}.
    \end{equation}
    Taxa with vanishing variance (\(\sigma_i^{(m)} \le \varepsilon\), with a small numerical threshold \(\varepsilon\)) are discarded. For the remaining taxa we construct \(z\)-scored abundances
    \begin{equation}
        Z^{(m)}(i,t) =
        \frac{ \tilde{X}^{(m)}(i,t) - \mu_i^{(m)} }{\sigma_i^{(m)}}.
    \end{equation}

    \item \textbf{Correlation matrix for the subsample.}
    From the standardized matrix \(Z^{(m)}\) we compute the empirical Pearson correlation matrix
    \begin{equation}
        C^{(m)} = \frac{1}{T_2 - 1} Z^{(m)} Z^{(m)\top},
    \end{equation}
    and explicitly set the diagonal to one, \(C^{(m)}_{ii} = 1\).

    \item \textbf{Distance-to-criticality estimator.}
    For each subsample, we feed \(C^{(m)}\) into the dynamical mean-field theory (DMFT)--based estimator of the distance to criticality, \(\hat{g}^{(m)}\), derived in section \ref{subsec: g DMFT}. 
\end{enumerate}

\paragraph{Aggregation across subsamples and cohorts.}
For each diagnostic group and choice of \((r,p)\) we thus obtain a collection of finite estimates
\(\{\hat{g}^{(m)}\}_{m \in \mathcal{M}}\), where \(\mathcal{M}\) is the set of successful subsamples. We characterize the distribution of \(\hat{g}\) by its empirical histogram \(\rho(\hat{g})\), as well as its mean and spread, and compare these distributions across disease groups and across randomized controls (obtained by applying the same pipeline to suitably shuffled abundance matrices).

\paragraph{Choice of the control parameters \(p\) and \(r\).}
Our subsampling scheme is controlled by two dimensionless parameters: the subset fraction \(p\) and the species--samples ratio \(r\). The parameter \(p \in (0,1]\) fixes the number of samples used in each subsample, meaning that a smaller value of \(p\) reduces \(T_2\), so each subsample is built from fewer samples. This has two main consequences: (i) individual correlation matrices are more weakly sampled (noisier estimates), and (ii) the variability between different subsamples increases, because each subsample sees a more distinct subset of the cohort.

The parameter \(r > 0\) controls the species-to-samples ratio of each subsample. For fixed \(T_2\), smaller values of \(r\) correspond to smaller \(S_{\mathrm{eff}}\), i.e.\ fewer rows in the abundance matrix. This improves the sampling quality of the correlation matrix (each entry is supported by relatively more samples per degree of freedom), but at the cost of reducing the dimensionality of the system and potentially discarding biologically relevant taxa. 

In practice, \(p\) and \(r\) cannot both be chosen arbitrarily small: too small \(p\) yields very few samples per subsample, and too small \(r\) then forces \(S_{\mathrm{eff}}\) to become very small as well, eventually leaving an insufficient number of species to define a meaningful correlation structure. Our choices of \(p\) and \(r\) therefore reflect a compromise between (i) having many, diverse subsamples (small \(p\)), (ii) having well-sampled correlation matrices (small \(r\)), and (iii) retaining a sufficiently large effective system size \(S_{\mathrm{eff}}\).

\subsection{Assessing group separation via the area under the ROC curve}
To quantify how well the inferred distances to criticality $\hat{g}$ separate two groups
of samples (e.g.\ Healthy vs.\ Diseased), we use the area under the receiver operating
characteristic (ROC) curve, AUC \cite{mannwhitney1947, hanley1982, fawcett2006}. Let $X = \{\hat{g}_i^{(A)}\}_{i=1}^{n_A}$ and
$Y = \{\hat{g}_j^{(B)}\}_{j=1}^{n_B}$ denote the scores for two groups $A$ and $B$. The AUC
associated with using $\hat{g}$ as a scalar classifier of group membership is defined as
the probability that a randomly chosen sample from group $A$ has a larger score than a
randomly chosen sample from group $B$:
\begin{equation}
    \mathrm{AUC} \;\equiv\; \mathbb{P}\!\left(\hat{g}^{(A)} > \hat{g}^{(B)}\right)
    \;=\;
    \frac{1}{n_A n_B}
    \sum_{i=1}^{n_A} \sum_{j=1}^{n_B}
    \mathbb{I}\!\Bigl[\hat{g}_i^{(A)} > \hat{g}_j^{(B)}\Bigr],
\end{equation}
where $\mathbb{I}[\cdot]$ is the indicator function. An AUC of $0.5$ corresponds to chance
level (no separation between the two distributions of $\hat{g}$), while $\mathrm{AUC} \to 1$
indicates that group $A$ almost always has larger $\hat{g}$ than group $B$
(near-perfect separation in the chosen direction). Values $\mathrm{AUC} < 0.5$ simply mean
that $B$ tends to have larger $\hat{g}$ than $A$, i.e.\ the direction of separation is reversed.

In practice, we estimate the AUC using the Mann--Whitney $U$ statistic, which is known to be
equivalent to the empirical probability $\mathbb{P}(\hat{g}^{(A)} > \hat{g}^{(B)})$. Writing
$U$ for the Mann--Whitney statistic comparing $X$ and $Y$, with sample sizes
$n_A = |X|$ and $n_B = |Y|$, the empirical AUC is
\begin{equation}
    \widehat{\mathrm{AUC}}
    \;=\;
    \frac{U}{n_A n_B}.
\end{equation}

This representation naturally yields a hypothesis test for separation between the two groups.
Under the null hypothesis that both groups have the same underlying distribution of distances
to criticality, the scores are exchangeable and the AUC equals $0.5$:
\begin{equation}
    H_0: \quad F_A(\hat{g}) = F_B(\hat{g})
    \qquad\Longleftrightarrow\qquad
    \mathbb{P}\!\left(\hat{g}^{(A)} > \hat{g}^{(B)}\right) = \tfrac{1}{2},
\end{equation}
where $F_A$ and $F_B$ are the cumulative distribution functions of $\hat{g}$ in groups $A$
and $B$. The alternative hypothesis of separation (in either direction) is
\begin{equation}
    H_1: \quad F_A(\hat{g}) \neq F_B(\hat{g})
    \qquad\Longleftrightarrow\qquad
    \mathbb{P}\!\left(\hat{g}^{(A)} > \hat{g}^{(B)}\right) \neq \tfrac{1}{2},
\end{equation}
which corresponds to $\mathrm{AUC} \neq 0.5$. The $p$-value associated with this contrast is obtained from the null distribution of the Mann--Whitney statistic $U$ (or, equivalently, from the asymptotic normal approximation
for large sample sizes). In our analysis, we report both the estimated AUC and its associated
$p$-value. Large AUC values close to $1$ together with very small $p$-values indicate that the inferred distances to criticality provide a strong and statistically significant separation
between the two groups under comparison. To decide whether two groups are effectively separated by our methodology, we adopted the
following criterion. Given two groups $A$ and $B$ with AUC
\[
\mathrm{AUC} = \mathbb{P}\bigl(\hat{g}^{(A)} > \hat{g}^{(B)}\bigr),
\]
we define
\[
\mathrm{AUC}^\star = \max\bigl(\mathrm{AUC},\,1-\mathrm{AUC}\bigr),
\]
so that $\mathrm{AUC}^\star$ measures the distance from chance level ($0.5$) irrespective of
which group has larger distances to criticality. We call two groups \emph{separable} if
$\mathrm{AUC}^\star \geq 0.8$, i.e.\ if at least $80\%$ of cross-group pairs are ordered
consistently in one direction. As we will see, all reported Mann--Whitney $p$-values are extremely small,
so the main focus is on this criterion rather than on formal significance.

\subsection{Values of $p$ and $r$ used for Figure 5 in the main text}

In order to produce the figure shown in the main text (Figure 5), we used the values of $p$ and $r$ listed in Table~\ref{tab:all_cohorts_params}. These parameters were chosen so as to balance three requirements: (i) obtain an effective number of taxa per subsample $S_\text{eff}$ of order $10^2$ whenever the cohort size allows it (and at least a few tens for smaller cohorts), thereby justifying the use of the DMFT estimator of $g$, which performs better at large $S$; (ii) keep the species--samples ratio $r = S_\text{eff}/T_2$ in a moderate range ($r \sim 0.4$--$0.8$), ensuring that each retained taxon is reasonably well sampled; and (iii) use subset fractions $p = T_2/T$ that are not too large (typically $p \lesssim 0.7$, with $p = 0.5$ for IBS) so that many distinct subsamples can be drawn without repeatedly reusing the same samples. In the following subsections we explore how our conclusions are affected by systematic changes of $r$ and $p$.

\begin{table}[h!]
    \centering
    \begin{tabular}{llcccc}
        \hline
        Disease & Group   & $S_{\mathrm{eff}}$ & $T_2$ 
                & $r = S_{\mathrm{eff}}/T_2$ 
                & $p = T_2/T$ \\
        \hline
        IBD     & Healthy & 89  & 200 & $89/200 \approx 0.445$ & $200/428 \approx 0.467$ \\
                & CD      & 80  & 200 & $80/200 \approx 0.400$ & $200/748 \approx 0.267$ \\
                & UC      & 79  & 200 & $79/200 \approx 0.395$ & $200/454 \approx 0.441$ \\
        \hline
        CRC     & Healthy     & 123 & 176 & $123/176 \approx 0.699$ & $176/251 \approx 0.701$ \\
                & Stage 0     & 36  & 51  & $36/51 \approx 0.706$   & $51/73 \approx 0.699$ \\
                & Stage I--II & 55  & 78  & $55/78 \approx 0.705$   & $78/111 \approx 0.703$ \\
                & Stage III--IV & 36 & 52 & $36/52 \approx 0.692$   & $52/74 \approx 0.703$ \\
        \hline
        CDI     & Healthy     & 91  & 130 & $91/130 \approx 0.700$  & $130/186 \approx 0.699$ \\
                & Unhealthy   & 90  & 129 & $90/129 \approx 0.698$  & $129/775 \approx 0.166$ \\
        \hline
        IBS     & Healthy     & 61  & 76  & $61/76 \approx 0.803$   & $76/152 \approx 0.500$ \\
                & IBS-C       & 59  & 74  & $59/74 \approx 0.797$   & $74/148 \approx 0.500$ \\
                & IBS-D       & 70  & 88  & $70/88 \approx 0.795$   & $88/176 \approx 0.500$ \\
        \hline
    \end{tabular}
    \caption{Subsampling parameters used for each cohort. 
    Here $S_{\mathrm{eff}}$ is the typical effective number of taxa retained per subsample, 
    $T_2$ is the number of samples per subsample, 
    $r = S_{\mathrm{eff}}/T_2$ is the species--samples ratio, 
    and $p = T_2/T$ is the subset fraction of samples used in each subsample.}
    \label{tab:all_cohorts_params}
\end{table}

\subsection{Group separations for Figure 5 in the main text}

The corresponding group separations in terms of the inferred distance to criticality are
summarized in Table \ref{tab:auc_maintext_params}, which reports the pairwise AUCs between
diagnostic subgroups for the same $(r,p)$ choices used in Figure 5 (see \ref{tab:all_cohorts_params}), together with an
indicator of whether each comparison satisfies our separability criterion
$\max(\mathrm{AUC}, 1-\mathrm{AUC}) \ge 0.8$. For all cohorts, the contrast between healthy
and pooled non-healthy samples yields AUCs very close to $1$ and is classified as
``Separable'', indicating almost perfect discrimination based on $\hat{g}$. Within cohorts,
several subgroup contrasts also show strong separation (e.g. healthy vs. individual tumor
stages in CRC, or healthy vs. IBS-C/IBS-D in IBS), while a few comparisons display AUC
values well below $0.5$ (such as CD vs.\ UC or IBS-C vs.\ IBS-D), which are likewise marked
as separable but correspond to a reversed ordering of the scores. Only in the comparison between Stage 0 and Stage I-II, we get a score that yields non separability between groups. Together, these results
confirm that the qualitative patterns visible in Figure 5 reflect robust and statistically
significant differences in distance to criticality between diagnostic groups.

\begin{table}[h!]
    \centering
    \begin{tabular}{llcc}
        \hline
        Disease & Pair & AUC & Separable? \\
        \hline
        IBD & Healthy vs CD                  & 1.000 & Yes \\
            & Healthy vs UC                  & 1.000 & Yes \\
            & CD vs UC                       & 0.101 & Yes \\
            & Healthy vs pooled non-Healthy  & 1.000 & Yes \\
        \hline
        CRC & Healthy vs Stage 0             & 0.999 & Yes \\
            & Healthy vs Stage I--II         & 1.000 & Yes \\
            & Healthy vs Stage III--IV       & 1.000 & Yes \\
            & Stage 0 vs Stage I--II         & 0.696 & No  \\
            & Stage 0 vs Stage III--IV       & 0.952 & Yes \\
            & Stage I--II vs Stage III--IV   & 0.949 & Yes \\
            & Healthy vs pooled non-Healthy  & 1.000 & Yes \\
        \hline
        CDI & Healthy vs Unhealthy           & 1.000 & Yes \\
            & Healthy vs pooled non-Healthy  & 1.000 & Yes \\
        \hline
        IBS & Healthy vs IBS-C               & 0.995 & Yes \\
            & Healthy vs IBS-D               & 0.921 & Yes \\
            & IBS-C vs IBS-D                 & 0.102 & Yes \\
            & Healthy vs pooled non-Healthy  & 0.958 & Yes \\
        \hline
    \end{tabular}
    \caption{
        Pairwise areas under the ROC curve (AUC) between diagnostic groups within each
        disease cohort, computed from the distributions of inferred distance to criticality
        $\hat{g}$ shown in the main-text Figure~5 (using the subsampling parameters in
        Table~\ref{tab:all_cohorts_params}). ``Pooled non-Healthy'' denotes the union of all
        non-healthy subgroups within a cohort (e.g.\ CD and UC for IBD, all tumor stages for
        CRC). The ``Separable?'' column indicates whether the comparison meets the separability
        criterion $\max(\mathrm{AUC}, 1-\mathrm{AUC}) \ge 0.8$, so that very small AUC values
        (e.g.\ CD vs UC or IBS-C vs IBS-D) correspond to strong separation with reversed ordering
        of the scores. In all cases, the corresponding Mann--Whitney $U$-test $p$-values are
        numerically extremely small (machine underflow to $p \approx 0$).
    }
    \label{tab:auc_maintext_params}
\end{table}

\subsection{Robustness of the distance-to-criticality estimates under varying sampling conditions}

In this section, we revisit the same disease cohorts analyzed in the main text and systematically vary the sampling
parameters used in our pipeline. For each group, we estimate the distance to
criticality from abundance tables by repeatedly subsampling taxa and samples
according to the $(r,p)$-scheme introduced in the previous section: $p$ controls the
fraction of samples used in each subsample, while $r$ sets the effective
species--samples ratio within that subsample. By exploring a range of $(r,p)$
values around the choices used to generate Figure 5, we assess whether the
qualitative separation between healthy and diseased microbiomes, and the
relative ordering of cohorts, are robust to changes in sampling depth and
effective dimensionality.

\paragraph{Results for $p=0.5$}

We first investigated the effect of varying the effective species--samples ratio $r$ at fixed
subset fraction $p = 0.5$. For each disease cohort (IBD, CRC, CDI and IBS) and each target
ratio $r_{\text{target}} \in \{0.2, 0.7, 0.8\}$, we drew repeated subsamples according to the
$(r,p)$-scheme described above and estimated the distance to criticality $\hat{g}$ for each
subsample. We then quantified group separation using the AUC between the empirical
distributions of $\hat{g}$, as detailed in the previous section.

Table~\ref{tab:auc_summary_rp} summarizes the AUC values for the comparison between healthy
subjects and pooled non-healthy samples (all disease stages or subtypes combined) across
different $r_{\text{target}}$ at fixed $p = 0.5$. For IBD, CRC and CDI, the AUC between
healthy and pooled non-healthy remains very high (typically $\mathrm{AUC} \gtrsim 0.96$)
for all three values of $r_{\text{target}}$, indicating robust separation between healthy
and diseased microbiomes that does not depend sensitively on the precise choice of
species--samples ratio. For IBS, the separation is strong for $r_{\text{target}} = 0.7$ and
$r_{\text{target}} = 0.8$ ($\mathrm{AUC} \geq 0.85$), but weaker for
$r_{\text{target}} = 0.2$ ($\mathrm{AUC} \approx 0.59$), suggesting that in this cohort a very
aggressive reduction of the effective number of taxa degrades the ability to distinguish
healthy from symptomatic subjects.

Tables~\ref{tab:IBD_pairwise_auc}--\ref{tab:IBS_pairwise_auc} report the pairwise AUCs between
different disease subgroups within each cohort. For IBD, healthy vs.\ CD and healthy vs.\ UC
are always strongly separable for all $r_{\text{target}}$, while CD vs.\ UC displays strong
separation only at $r_{\text{target}} = 0.7$ and becomes less separated as $r$ deviates from
this value. In CRC, the healthy group is well separated from all tumor stages across all
$r_{\text{target}}$, and some stage--stage comparisons (e.g.\ Stage~I--II vs.\ Stage~III--IV)
also reach $\mathrm{AUC}^\star \geq 0.8$ for certain values of $r_{\text{target}}$. In IBS,
healthy vs.\ IBS-C and healthy vs.\ IBS-D are separable for the larger $r_{\text{target}}$,
and the IBS-C vs.\ IBS-D comparison yields $\mathrm{AUC} \ll 0.5$ for $r_{\text{target}} = 0.7$
and $0.8$, indicating strong separation but with reversed ordering of the scores.

Figure \ref{fig:SI4_p05_resh} illustrates these results. Each row
corresponds to a different target species--samples ratio $r_{\text{target}}$ (with $p = 0.5$
fixed), and each column to one disease cohort (IBD, CRC, CDI and IBS). In the main panels we
show the empirical distributions $\rho(\hat{g})$ for the different diagnostic groups within
each cohort, estimated from the original abundance tables. We observe a clear and systematic
shift between healthy and diseased groups across a broad range of $r_{\text{target}}$, in
line with the AUC values reported in Table~\ref{tab:auc_summary_rp}. The insets display the
same analysis applied to reshuffled abundance tables, in which the temporal/sample structure
is destroyed by random permutation of the entries. In this null model, the distributions
$\rho(\hat{g})$ for the different groups largely collapse onto each other, and the strong
between-group separations observed in the original data are no longer present. This confirms
that the separation in distance to criticality is not a trivial consequence of sample size or
marginal abundance distributions, but instead reflects genuine structure in the empirical
microbiome data.

Table~\ref{tab:rp_nominal_counts} summarizes, for each disease group, the nominal number of
samples $T_2$ and effective taxa $S_{\mathrm{eff}}$ used in each subsample when fixing
$p = 0.5$ and $r_{\text{target}} \in \{0.2, 0.7, 0.8\}$. These values are computed as
$T_2 = \lfloor p T\rfloor$ and $S_{\mathrm{eff}} \approx \lfloor r_{\text{target}} T_2\rfloor$
from the total number of samples $T$ in each group (Table~\ref{tab:disease_samples_taxa}),
and thus represent upper bounds: the presence and variance filters in our pipeline can only
reduce $S_{\mathrm{eff}}$ further.

For the IBD, CDI and IBS cohorts, the subsamples retain a reasonably large number of samples
and taxa over most parameter choices. Typically, $T_2 \gtrsim 70$ and $S_{\mathrm{eff}}$
lies in the tens to a few hundred for $r_{\text{target}} = 0.7$ and $0.8$. This is consistent
with the robust separability we observe in the corresponding AUCs: healthy vs.\ pooled
non-healthy comparisons for IBD, CRC and CDI yield $\mathrm{AUC}^\star \gtrsim 0.96$ for all
$r_{\text{target}}$, and IBS shows strong separation for $r_{\text{target}} = 0.7$ and
$0.8$ (see Table~\ref{tab:auc_summary_rp}). In contrast, for IBS at $r_{\text{target}} = 0.2$,
the effective number of taxa is reduced to $\sim 15$–$18$ per subsample, and the corresponding
healthy vs.\ pooled IBS comparison shows only weak separation
($\mathrm{AUC} \approx 0.59$), indicating that an overly aggressive reduction of
$S_{\mathrm{eff}}$ can degrade performance.

The CRC cohort is the most data-limited at the level of individual disease stages. While
the healthy CRC group retains a moderate sampling depth ($T_2 = 126$ and
$S_{\mathrm{eff}} \approx 88$–$101$ depending on $r_{\text{target}}$), the stage-specific
groups have much smaller subsamples: for $p = 0.5$ one typically has
$T_2 \approx 36$--$56$ and $S_{\mathrm{eff}} \approx 7$--$45$ for Stage~0, Stage~I--II and
Stage~III--IV (Table~\ref{tab:rp_nominal_counts}). After applying occupancy and variance
filters, the actual $S_{\mathrm{eff}}$ used in the correlation estimates may be even lower.
This helps explain why, although healthy vs.\ pooled non-healthy CRC exhibits very high AUCs
($\mathrm{AUC}^\star \gtrsim 0.96$ for all $r_{\text{target}}$), some pairwise comparisons
between individual stages show more variable behavior and should be interpreted with greater
caution. In particular, the strong separations observed between certain stage pairs
(e.g.\ Stage~I--II vs.\ Stage~III--IV at $r_{\text{target}} = 0.7$ or $0.8$) are based on
subsamples with relatively few taxa and samples, and thus provide suggestive but less
statistically robust evidence than the healthy vs.\ pooled CRC contrasts.

\begin{table}[h!]
    \centering
    \begin{tabular}{llrrrrr}
        \hline
        Disease & Group &
        $T$ &
        $T_2$ ($p=0.5$) &
        $S_{\mathrm{eff}}(r=0.2)$ &
        $S_{\mathrm{eff}}(r=0.7)$ &
        $S_{\mathrm{eff}}(r=0.8)$ \\
        \hline
        IBD & Healthy (nonIBD) & 428 & 214 &  43 & 150 & 171 \\
            & CD               & 748 & 374 &  75 & 262 & 299 \\
            & UC               & 454 & 227 &  45 & 159 & 182 \\
        \hline
        CRC & Healthy          & 251 & 126 &  25 &  88 & 101 \\
            & Stage~0          &  73 &  36 &   7 &  25 &  29 \\
            & Stage~I--II      & 111 &  56 &  11 &  39 &  45 \\
            & Stage~III--IV    &  74 &  37 &   7 &  26 &  30 \\
        \hline
        CDI & Healthy          & 186 &  93 &  19 &  65 &  74 \\
            & Unhealthy        & 234 & 117 &  23 &  82 &  94 \\
        \hline
        IBS & Healthy          & 152 &  76 &  15 &  53 &  61 \\
            & IBS-C            & 148 &  74 &  15 &  52 &  59 \\
            & IBS-D            & 176 &  88 &  18 &  62 &  70 \\
        \hline
    \end{tabular}
    \caption{
        Nominal number of samples $T_2$ and effective taxa $S_{\mathrm{eff}}$ per subsample
        for $p = 0.5$ and $r_{\text{target}} \in \{0.2, 0.7, 0.8\}$, computed as
        $T_2 = \mathrm{round}(pT)$ and $S_{\mathrm{eff}} \approx \mathrm{round}(r_{\text{target}} T_2)$
        from the total number of samples $T$ in each group (see
        Table~\ref{tab:disease_samples_taxa}). In the actual pipeline, additional
        presence and variance filters can only reduce $S_{\mathrm{eff}}$, so these
        values should be interpreted as upper bounds.
    }
    \label{tab:rp_nominal_counts}
\end{table}

\begin{figure}[h!]
    \centering
    \includegraphics[width=\linewidth]{FIGURE_SI_4_p=0.5_resh}
    \caption{
        \textbf{Distributions of inferred distance to criticality for different disease cohorts at
        fixed subset fraction $p = 0.5$ and varying target species--samples ratio
        $r_{\text{target}} \in \{0.2, 0.7, 0.8\}$}. Rows correspond to different values of
        $r_{\text{target}}$, and columns to IBD, CRC, CDI and IBS, respectively. Within each
        panel, we show $\rho(\hat{g})$ for healthy and diseased subgroups estimated from the
        original abundance tables. Insets display the corresponding distributions obtained
        after reshuffling the abundance data (random permutation of entries), which largely
        collapses the group-specific curves and eliminates the strong separations observed
        in the original data. Together with the AUC values reported in
        Tables~\ref{tab:auc_summary_rp}--\ref{tab:IBS_pairwise_auc}, this figure demonstrates
        that the separation in distance to criticality between healthy and diseased microbiomes
        is robust to changes in $(r,p)$ and is not reproduced by simple reshuffling controls.}
    \label{fig:SI4_p05_resh}
\end{figure}

\begin{table}[h!]
    \centering
    \begin{tabular}{lcccc}
        \hline
        $r_{\text{target}}$ & IBD & CRC & CDI & IBS \\
        \hline
        0.7 & 
        AUC $= 1.000$ (Yes) & 
        AUC $= 0.998$ (Yes) & 
        AUC $= 1.000$ (Yes) & 
        AUC $= 0.852$ (Yes) \\
        0.2 & 
        AUC $= 0.998$ (Yes) & 
        AUC $= 0.964$ (Yes) & 
        AUC $= 1.000$ (Yes) & 
        AUC $= 0.591$ (No) \\
        0.8 & 
        AUC $= 1.000$ (Yes) & 
        AUC $= 0.997$ (Yes) & 
        AUC $= 1.000$ (Yes) & 
        AUC $= 0.957$ (Yes) \\
        \hline
    \end{tabular}
    \caption{
        Area under the ROC curve (AUC) for the comparison between healthy and
        pooled non-healthy samples, for different target species--samples ratios
        $r_{\text{target}}$ (with $p = 0.5$ fixed). ``Yes'' indicates successful
        separation, defined as $\mathrm{AUC} \ge 0.8$. In all cases the
        corresponding Mann--Whitney $p$-values are $p \ll 10^{-5}$.
    }
    \label{tab:auc_summary_rp}
\end{table}

\begin{table}[h!]
  \centering
  \begin{tabular}{lccc}
    \hline
    Pair & $r_{\text{target}}=0.7$ & $r_{\text{target}}=0.2$ & $r_{\text{target}}=0.8$ \\
    \hline
    Healthy vs CD &
      $0.999$ (Yes) &
      $0.999$ (Yes) &
      $1.000$ (Yes) \\
    Healthy vs UC &
      $1.000$ (Yes) &
      $0.998$ (Yes) &
      $1.000$ (Yes) \\
    CD vs UC &
      $0.838$ (Yes) &
      $0.734$ (No)  &
      $0.634$ (No)  \\
    \hline
  \end{tabular}
  \caption{Pairwise AUCs between IBD subgroups for different
  target ratios $r_{\text{target}}$ (with $p=0.5$). ``Yes'' indicates
  strong separation, defined as $\max(\mathrm{AUC},1-\mathrm{AUC}) \ge 0.8$.}
  \label{tab:IBD_pairwise_auc}
\end{table}

\begin{table}[h!]
  \centering
  \begin{tabular}{lccc}
    \hline
    Pair & $r_{\text{target}}=0.7$ & $r_{\text{target}}=0.2$ & $r_{\text{target}}=0.8$ \\
    \hline
    Healthy vs Stage 0 &
      $0.994$ (Yes) &
      $0.919$ (Yes) &
      $0.990$ (Yes) \\
    Healthy vs Stage I--II &
      $1.000$ (Yes) &
      $0.988$ (Yes) &
      $1.000$ (Yes) \\
    Healthy vs Stage III--IV &
      $1.000$ (Yes) &
      $0.986$ (Yes) &
      $1.000$ (Yes) \\
    Stage 0 vs Stage I--II &
      $0.530$ (No) &
      $0.577$ (No) &
      $0.595$ (No) \\
    Stage 0 vs Stage III--IV &
      $0.762$ (No) &
      $0.638$ (No) &
      $0.818$ (Yes) \\
    Stage I--II vs Stage III--IV &
      $0.812$ (Yes) &
      $0.581$ (No) &
      $0.835$ (Yes) \\
    \hline
  \end{tabular}
  \caption{Pairwise AUCs between CRC subgroups for different
  target ratios $r_{\text{target}}$ (with $p=0.5$). ``Yes'' indicates
  strong separation, defined as $\max(\mathrm{AUC},1-\mathrm{AUC}) \ge 0.8$.}
  \label{tab:CRC_pairwise_auc}
\end{table}

\begin{table}[h!]
  \centering
  \begin{tabular}{lccc}
    \hline
    Pair & $r_{\text{target}}=0.7$ & $r_{\text{target}}=0.2$ & $r_{\text{target}}=0.8$ \\
    \hline
    Healthy vs IBS-C &
      $0.966$ (Yes) &
      $0.525$ (No)  &
      $0.995$ (Yes) \\
    Healthy vs IBS-D &
      $0.739$ (No)  &
      $0.658$ (No)  &
      $0.919$ (Yes) \\
    IBS-C vs IBS-D &
      $0.095$ (Yes) &
      $0.602$ (No)  &
      $0.099$ (Yes) \\
    \hline
  \end{tabular}
  \caption{Pairwise AUCs between IBS subgroups for different
  target ratios $r_{\text{target}}$ (with $p=0.5$). ``Yes'' indicates
  strong separation, defined as $\max(\mathrm{AUC},1-\mathrm{AUC}) \ge 0.8$.
  Note that AUC values far below $0.5$ (e.g.\ $0.095$) also correspond
  to strong separation but with the ordering of the groups reversed.}
  \label{tab:IBS_pairwise_auc}
\end{table}

\paragraph{Results for $p = 0.7$}
Overall, the pattern at $p = 0.7$ is very similar to that observed for $p = 0.5$.
Healthy vs.\ pooled non-healthy comparisons (Table~\ref{tab:auc_summary_rp_p07})
show excellent separability for IBD, CRC and CDI across all $r_{\text{target}}$, with
$\mathrm{AUC} \approx 1.0$ in every case. For IBS, the separation is again strong at
$r_{\text{target}} = 0.7$ and $0.8$ ($\mathrm{AUC} \approx 0.99$ and $0.98$,
respectively), but essentially lost at $r_{\text{target}} = 0.2$
($\mathrm{AUC} \approx 0.39$), consistent with the degradation observed for
$p = 0.5$ when $r_{\text{target}}$ is very small.

The pairwise subgroup AUCs in Tables~\ref{tab:IBD_pairwise_auc_p07}--%
\ref{tab:IBS_pairwise_auc_p07} further refine this picture. In IBD, healthy vs.\ CD
and healthy vs.\ UC remain perfectly separated for all $r_{\text{target}}$, whereas
the CD vs.\ UC contrast does not reach the separability threshold and is only moderate
at best. In CRC, the healthy group is cleanly separated from all tumor stages for all
$r_{\text{target}}$, while stage--stage comparisons involving Stage~III--IV typically
show strong separation, in line with the $p = 0.5$ analysis, despite the limited
effective sample size for individual stages. For IBS, both healthy vs.\ IBS-C and
healthy vs.\ IBS-D display very strong separation at $r_{\text{target}} = 0.7$ and
$0.8$, and the IBS-C vs.\ IBS-D comparison yields AUCs very close to zero in these
settings, indicating a clear but reversed ordering of the scores between subtypes.
By contrast, at $r_{\text{target}} = 0.2$ all IBS contrasts fall in the non-separable
range, reinforcing the conclusion that too aggressive a reduction in effective taxa
($r$ too small) compromises the ability of our methodology to distinguish IBS groups.

\begin{figure}[h!]
    \centering
    \includegraphics[width=\linewidth]{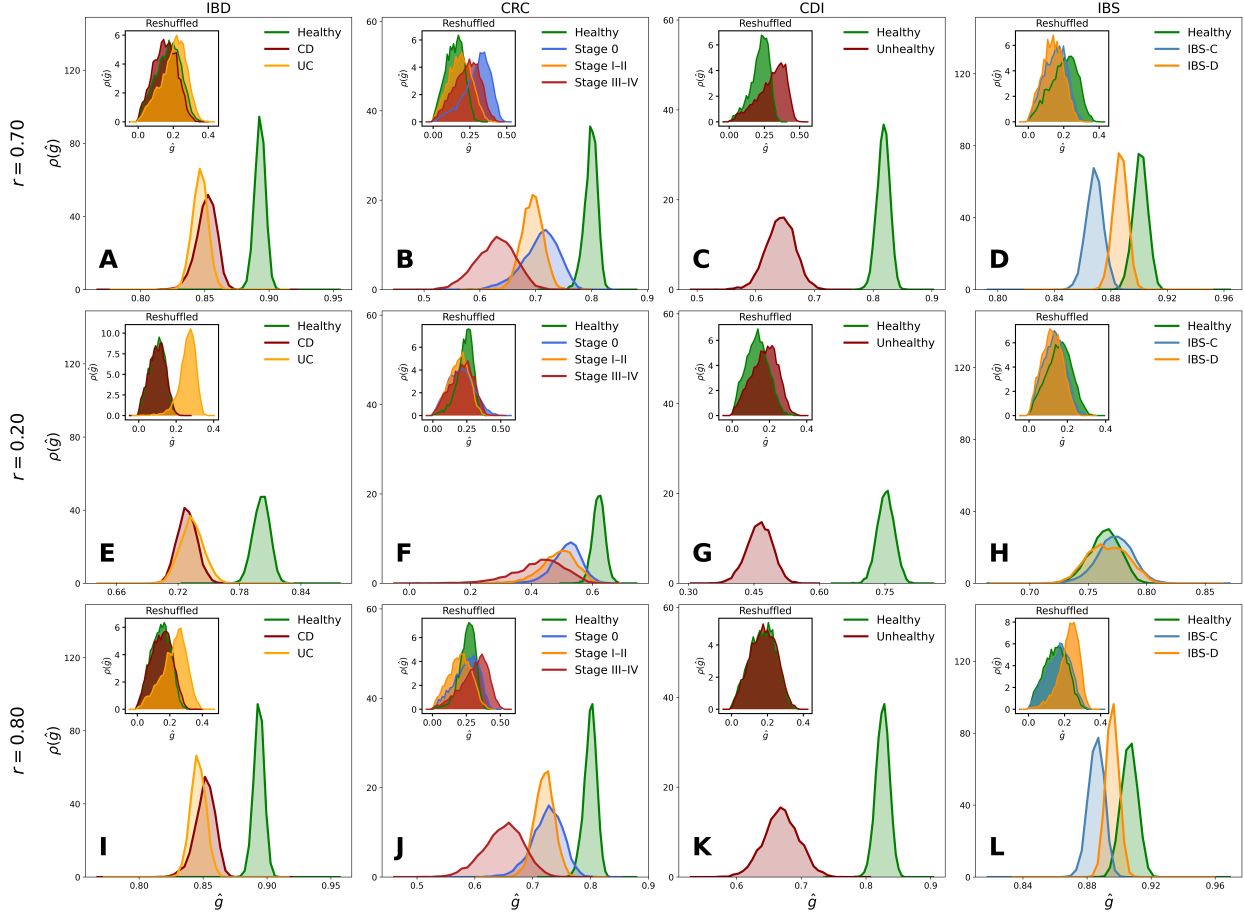}
    \caption{
        \textbf{Distributions of inferred distance to criticality for different disease cohorts at
        fixed subset fraction $p = 0.7$ and varying target species--samples ratio
        $r_{\text{target}} \in \{0.2, 0.7, 0.8\}$}. Rows correspond to different values of
        $r_{\text{target}}$, and columns to IBD, CRC, CDI and IBS, respectively. Within each
        main panel, we show $\rho(\hat{g})$ for healthy and diseased subgroups estimated from
        the original abundance tables, using the $(r,p)$ subsampling scheme. Insets display
        the corresponding distributions obtained after reshuffling the abundance data (random
        permutation of entries), which largely collapses the group-specific curves and
        removes the strong separations observed in the original data. Together with the AUC
        values reported in Tables~\ref{tab:auc_summary_rp_p07}--\ref{tab:IBS_pairwise_auc_p07},
        this figure shows that the separation in distance to criticality between healthy and
        diseased microbiomes is robust for $p = 0.7$ across a broad range of $r_{\text{target}}$,
        and is not reproduced by simple reshuffling controls.}
    \label{fig:SI4_p07_resh}
\end{figure}

These conclusions are visually supported by Fig.~\ref{fig:SI4_p07_resh}. As in the
$p = 0.5$ case, each row corresponds to a different $r_{\text{target}}$ and each
column to a disease cohort. For IBD, CRC and CDI, the healthy and diseased
distributions $\rho(\hat{g})$ in the main panels remain well separated across all
$r_{\text{target}}$, consistent with the near-perfect AUCs in
Table~\ref{tab:auc_summary_rp_p07}. In IBS, the separation is clear for
$r_{\text{target}} = 0.7$ and $0.8$, while at $r_{\text{target}} = 0.2$ the curves
for healthy and pooled IBS groups strongly overlap, matching the poor AUC
observed in that regime. In all cohorts, the reshuffled insets show markedly
reduced or absent separation between groups, confirming that the observed
differences in $\hat{g}$ rely on the genuine correlation structure of the data
rather than on marginal abundance distributions or sample size alone.

\begin{table}[h!]
    \centering
    \begin{tabular}{lcccc}
        \hline
        $r_{\text{target}}$ & IBD & CRC & CDI & IBS \\
        \hline
        0.7 & 
        AUC $= 1.000$ (Yes) & 
        AUC $= 1.000$ (Yes) & 
        AUC $= 1.000$ (Yes) & 
        AUC $= 0.989$ (Yes) \\
        0.2 & 
        AUC $= 1.000$ (Yes) & 
        AUC $= 0.992$ (Yes) & 
        AUC $= 1.000$ (Yes) & 
        AUC $= 0.392$ (No)  \\
        0.8 & 
        AUC $= 1.000$ (Yes) & 
        AUC $= 0.999$ (Yes) & 
        AUC $= 1.000$ (Yes) & 
        AUC $= 0.975$ (Yes) \\
        \hline
    \end{tabular}
    \caption{
        Area under the ROC curve (AUC) for the comparison between healthy and
        pooled non-healthy samples, for different target species--samples ratios
        $r_{\text{target}}$ (with $p = 0.7$ fixed). ``Yes'' indicates successful
        separation according to the same criterion used for $p = 0.5$. For IBS at
        $r_{\text{target}} = 0.2$, the separation is weak (AUC close to $0.5$ and
        opposite ordering).
    }
    \label{tab:auc_summary_rp_p07}
\end{table}

\begin{table}[h!]
  \centering
  \begin{tabular}{lccc}
    \hline
    Pair & $r_{\text{target}}=0.7$ & $r_{\text{target}}=0.2$ & $r_{\text{target}}=0.8$ \\
    \hline
    Healthy vs CD &
      $1.000$ (Yes) &
      $1.000$ (Yes) &
      $1.000$ (Yes) \\
    Healthy vs UC &
      $1.000$ (Yes) &
      $1.000$ (Yes) &
      $1.000$ (Yes) \\
    CD vs UC &
      $0.711$ (No)  &
      $0.374$ (No)  &
      $0.721$ (No)  \\
    \hline
  \end{tabular}
  \caption{Pairwise AUCs between IBD subgroups for
  $p = 0.7$ and different target ratios $r_{\text{target}}$.
  ``Yes'' indicates strong separation according to the same
  criterion as in Table~\ref{tab:IBD_pairwise_auc}.}
  \label{tab:IBD_pairwise_auc_p07}
\end{table}

\begin{table}[h!]
  \centering
  \begin{tabular}{lccc}
    \hline
    Pair & $r_{\text{target}}=0.7$ & $r_{\text{target}}=0.2$ & $r_{\text{target}}=0.8$ \\
    \hline
    Healthy vs Stage 0 &
      $0.999$ (Yes) &
      $0.984$ (Yes) &
      $0.998$ (Yes) \\
    Healthy vs Stage I--II &
      $1.000$ (Yes) &
      $0.993$ (Yes) &
      $1.000$ (Yes) \\
    Healthy vs Stage III--IV &
      $1.000$ (Yes) &
      $0.998$ (Yes) &
      $1.000$ (Yes) \\
    Stage 0 vs Stage I--II &
      $0.692$ (No)  &
      $0.647$ (No)  &
      $0.594$ (No)  \\
    Stage 0 vs Stage III--IV &
      $0.953$ (Yes) &
      $0.825$ (Yes) &
      $0.960$ (Yes) \\
    Stage I--II vs Stage III--IV &
      $0.954$ (Yes) &
      $0.720$ (No)  &
      $0.972$ (Yes) \\
    \hline
  \end{tabular}
  \caption{Pairwise AUCs between CRC subgroups for
  $p = 0.7$ and different target ratios $r_{\text{target}}$.
  As before, ``Yes'' indicates strong separation. Note that
  stage--stage separations are most robust for comparisons
  involving Stage~III--IV.}
  \label{tab:CRC_pairwise_auc_p07}
\end{table}

\begin{table}[h!]
  \centering
  \begin{tabular}{lccc}
    \hline
    Pair & $r_{\text{target}}=0.7$ & $r_{\text{target}}=0.2$ & $r_{\text{target}}=0.8$ \\
    \hline
    Healthy vs IBS-C &
      $1.000$ (Yes) &
      $0.329$ (No)  &
      $0.999$ (Yes) \\
    Healthy vs IBS-D &
      $0.979$ (Yes) &
      $0.454$ (No)  &
      $0.952$ (Yes) \\
    IBS-C vs IBS-D &
      $0.009$ (Yes) &
      $0.606$ (No)  &
      $0.050$ (Yes) \\
    \hline
  \end{tabular}
  \caption{Pairwise AUCs between IBS subgroups for
  $p = 0.7$ and different target ratios $r_{\text{target}}$.
  AUC values far below $0.5$ (e.g.\ $0.009$ or $0.050$) indicate
  strong separation with reversed ordering of the scores.}
  \label{tab:IBS_pairwise_auc_p07}
\end{table}

\subsection{Negative-control analysis with pooled healthy cohorts}

As a negative control, we pooled all healthy subjects from the two independent
cohorts (Healthy 1 and Healthy 2 in \ref{tab:disease_samples_taxa}) into a single abundance table. From this pooled dataset we then constructed two
pseudo-cohorts, denoted ``Healthy 1'' and ``Healthy 2'', by repeatedly drawing
random subsamples of samples without replacement so that both groups contained
the same number of individuals and were statistically equivalent by
construction. For each pseudo-cohort, we applied exactly the same subsampling
and DMFT-based estimation pipeline used in the main text: given a choice of
$(r,p)$ we drew $M=5\times 10^{3}$ subsamples, computed the corresponding
distance-to-criticality estimates $\hat{g}$, and then quantified the separation
between the resulting $\hat{g}$-distributions via the AUC of the
Mann--Whitney--Wilcoxon statistic.

Table~\ref{tab:healthy_split_auc} reports the AUC values obtained when comparing
``Healthy~1'' and ``Healthy~2'' for a grid of species--samples ratios
$r \in \{0.4, 0.6, 0.8\}$ and subset fractions $p \in \{0.1, 0.3, 0.5\}$. Across
all parameter combinations, the AUC remains in the range
$\mathrm{AUC} \approx 0.60$--$0.73$. According to the separability criterion
used in this work, none of these comparisons qualify
as strongly separable: the two pseudo-cohorts are only mildly shifted relative
to each other, as expected when splitting a single homogeneous healthy
population into two arbitrary halves. The associated $p$-values are extremely
small due to the very large number of subsamples ($M=5000$), but the effect
sizes (AUC values close to $0.5$) are modest.

\begin{table}[h!]
    \centering
    \begin{tabular}{cccccc}
        \hline
        $r$ & $p$ & $\langle r_{\mathrm{eff}}\rangle$ &
        AUC(H$_1$ vs H$_2$) & $p$-value & Separable? \\
        \hline
        0.40 & 0.10 & 0.400 & 0.660 & $2.2\times 10^{-168}$ & No \\
        0.40 & 0.30 & 0.397 & 0.700 & $1.4\times 10^{-262}$ & No \\
        0.40 & 0.50 & 0.398 & 0.689 & $6.2\times 10^{-234}$ & No \\
        \hline
        0.60 & 0.10 & 0.600 & 0.725 & $<10^{-300}$          & No \\
        0.60 & 0.30 & 0.598 & 0.688 & $2.9\times 10^{-233}$ & No \\
        0.60 & 0.50 & 0.599 & 0.669 & $4.8\times 10^{-189}$ & No \\
        \hline
        0.80 & 0.10 & 0.800 & 0.704 & $1.1\times 10^{-273}$ & No \\
        0.80 & 0.30 & 0.799 & 0.673 & $6.3\times 10^{-197}$ & No \\
        0.80 & 0.50 & 0.799 & 0.606 & $6.4\times 10^{-75}$  & No \\
        \hline
    \end{tabular}
    \caption{
        Negative-control test for the distance-to-criticality estimator using
        pooled healthy cohorts. We first pooled all healthy samples from the two
        independent control datasets into a single abundance matrix, and then constructed two disjoint pseudo-cohorts
        H$_1$ and H$_2$ by randomly splitting the columns of said abundance table
        into two equally sized subsets. For each choice of $(r,p)$ we then ran
        the same subsampling pipeline used in Figure 5 of the main text on H$_1$
        and H$_2$, obtaining distributions of $\hat{g}$ and computing the AUC
        for the contrast H$_1$ vs.\ H$_2$. The column
        $\langle r_{\mathrm{eff}}\rangle$ reports the mean effective
        species--samples ratio over subsamples, and ``Separable?'' follows the
        same criterion used elsewhere in the manuscript,
        $\max(\mathrm{AUC}, 1-\mathrm{AUC}) \ge 0.8$.
    }
    \label{tab:healthy_split_auc}
\end{table}

\begin{figure}[h!]
    \centering
    \includegraphics[width=\linewidth]{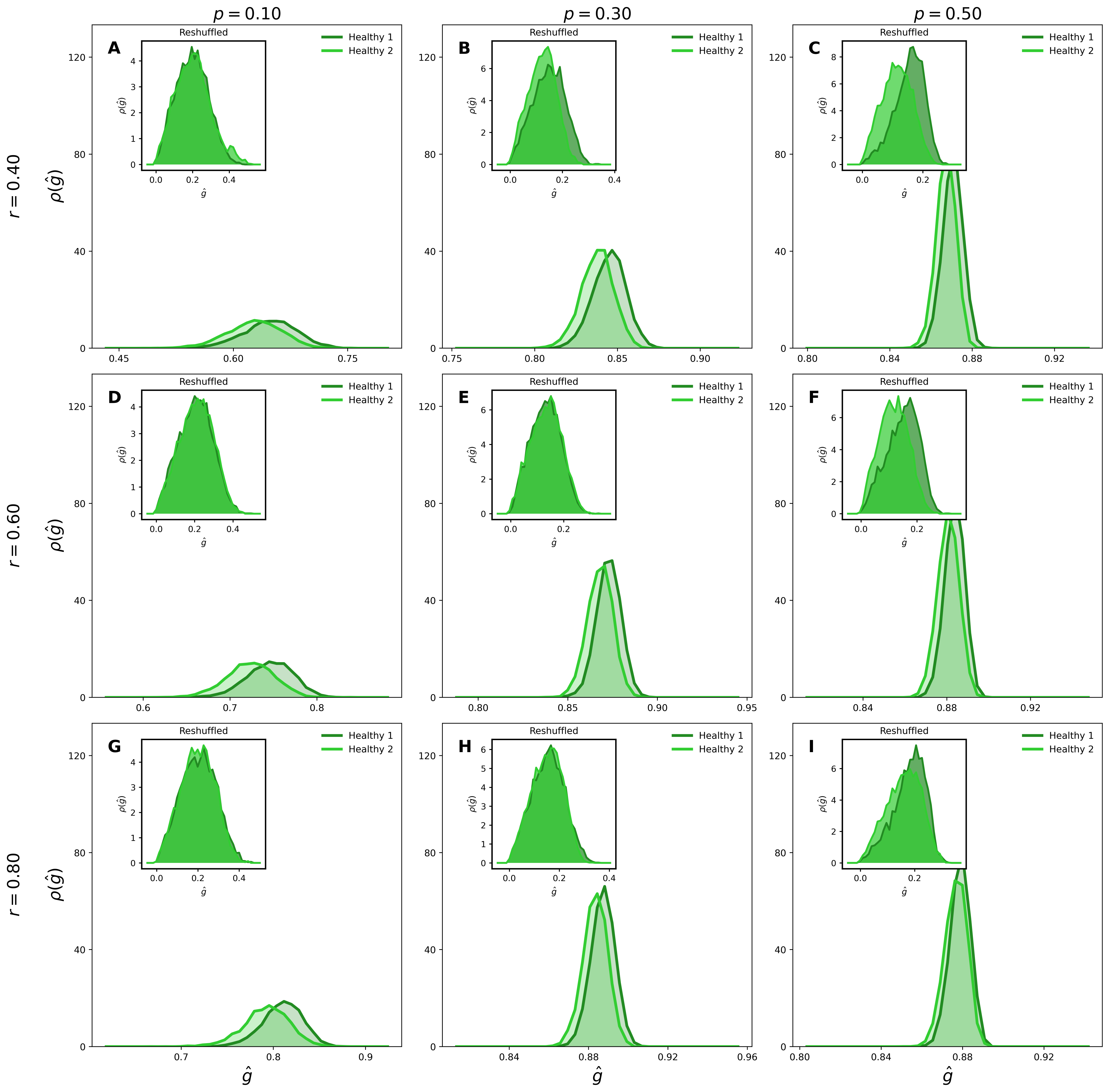}
    \caption{
        \textbf{Negative-control comparison between two pseudo-cohorts of healthy subjects.}
        All healthy samples from the two independent healthy cohorts are pooled
        into a single dataset and then randomly split into two pseudo-cohorts, denoted
        ``Healthy 1'' and ``Healthy 2'', of equal size. For each combination of
        species--samples ratio $r \in \{0.4, 0.6, 0.8\}$ (rows) and subset fraction
        $p \in \{0.1, 0.3, 0.5\}$ (columns), we draw $M = 5\times 10^{3}$ subsamples using
        the $(r,p)$ scheme and compute the corresponding distance-to-criticality estimates
        $\hat{g}$. The main panels show the empirical distributions
        $\rho(\hat{g})$ for ``Healthy 1'' and ``Healthy 2''; in all cases, the two curves
        overlap strongly and exhibit only mild shifts, in agreement with the moderate AUC
        values reported in Table \ref{tab:healthy_split_auc}. Insets display the same
        analysis applied to reshuffled abundance tables, in which the entries are randomly
        permuted within each taxon. In this null model, the two distributions collapse
        almost perfectly for all $(r,p)$, indicating that our estimator and subsampling
        pipeline do not spuriously generate strong separations when comparing two
        statistically equivalent healthy cohorts.
    }
    \label{fig:SI5_resh}
\end{figure}

Figure \ref{fig:SI5_resh} provides a visual representation of the numerical
results shown in Table \ref{tab:healthy_split_auc}. For each combination of
species--samples ratio $r \in \{0.4, 0.6, 0.8\}$ and subset fraction
$p \in \{0.1, 0.3, 0.5\}$, the main panels display the empirical
distributions $\rho(\hat{g})$ obtained for the two pseudo-cohorts
``Healthy 1'' and ``Healthy 2'' constructed by randomly splitting the pooled
healthy dataset. Across all $(r,p)$ pairs, the two curves largely overlap,
with only modest shifts in their central tendency and shape, consistent with
the moderate AUC values ($\mathrm{AUC} \approx 0.60$--$0.73$) reported in
Table \ref{tab:healthy_split_auc}. In particular, none of the panels shows
the kind of pronounced, systematic displacement between distributions that we
observe when comparing healthy and diseased groups in the main text.

The insets show the same analysis applied to reshuffled versions of the data,
where the entries of the abundance table are randomly permuted so as to
destroy the empirical correlation structure while preserving marginal
distributions. In this null model, the $\rho(\hat{g})$ curves for
``Healthy 1'' and ``Healthy 2'' almost perfectly collapse for all $(r,p)$,
indicating that any residual differences seen in the main panels arise from
finite-sample fluctuations rather than systematic biases in the estimator or
the subsampling scheme. Taken together, Table \ref{tab:healthy_split_auc} and
Figure \ref{fig:SI5_resh} confirm that our pipeline does not spuriously
produce strong between-group separations when comparing two statistically
equivalent healthy cohorts, and therefore has a low propensity for false
positives.

\bibliography{Biomes+Edge}